%% file: ms.tex
\documentclass[preprint,12pt]{aastex62}

\shorttitle{HOPS-370 Disk}
\shortauthors{Tobin et al.}

\newcommand{\ntdp}{\mbox{N$_2$D$^+$}}

\newcommand{\cateo}{\mbox{C$^{18}$O}}
\newcommand{\thco}{\mbox{$^{13}$CO}}
\newcommand{\twco}{\mbox{$^{12}$CO}}

\newcommand{\kms}{\mbox{km s$^{-1}$}}

\newcommand{\lsun}{\mbox{L$_{\sun}$}}
\newcommand{\msun}{\mbox{M$_{\sun}$}}

\newcommand{\tbol}{\mbox{T$_{bol}$}}
\newcommand{\lbol}{\mbox{L$_{bol}$}}
\begin{document}

\title{The VLA/ALMA Nascent Disk and Multiplicity (VANDAM) Survey of Orion Protostars IV. Unveiling the 
Embedded Intermediate-Mass Protostar and Disk within OMC2-FIR3/HOPS-370}
\author[0000-0002-6195-0152]{John J. Tobin}
\affiliation{National Radio Astronomy Observatory, 520 Edgemont Rd., Charlottesville,VA 22903, USA}
\author{Patrick D. Sheehan}
\affiliation{Center
for Interdisciplinary Exploration and Research in Astronomy, 1800 Sherman Rd., Evanston, IL 60202, USA}
\author{Nickalas Reynolds}
\affiliation{Homer L. Dodge Department of Physics and Astronomy, University of Oklahoma, 440 W. Brooks Street, Norman, OK 73019, USA}
\author{S. Thomas Megeath}
\affiliation{Department of Physics and Astronomy, University of Toledo, Toledo, OH}
\author[0000-0002-6737-5267]{Mayra Osorio}
\affiliation{Instituto de Astrof\'{\i}sica de Andaluc\'{\i}a, CSIC, Glorieta de la Astronom\'{\i}a s/n, E-18008 Granada, Spain}
\author[0000-0002-7506-5429]{Guillem Anglada}
\affiliation{Instituto de Astrof\'{\i}sica de Andaluc\'{\i}a, CSIC, Glorieta de la Astronom\'{\i}a s/n, E-18008 Granada, Spain}
\author[0000-0001-9112-6474]{Ana Karla D{\'i}az-Rodr{\'i}guez }
\affiliation{Instituto de Astrof\'{\i}sica de Andaluc\'{\i}a, CSIC, Glorieta de la Astronom\'{\i}a s/n, E-18008 Granada, Spain}
\author[0000-0001-9800-6248]{Elise Furlan}
\affiliation{IPAC, Mail Code 100-22, Caltech, 1200 E. California Blvd., Pasadena,CA 91125, USA}
\author[0000-0001-5253-1338]{Kaitlin M. Kratter}
\affiliation{University of Arizona, Steward Observatory, Tucson, AZ 85721}
\author{Stella S. R. Offner}
\affiliation{The University of Texas at Austin, 2500 Speedway, Austin, TX USA}
\author[0000-0002-4540-6587]{Leslie W. Looney}
\affiliation{Department of Astronomy, University of Illinois, Urbana, IL 61801}
\author{Mihkel Kama}
\affiliation{Tartu Observatory, Observatooriumi 1, T\~{o}ravere 61602, Tartu, Estonia}
\affiliation{University College London, Dept. of Physics and Astronomy, Gower Street, London, WC1E6BT, UK}
\author{Zhi-Yun Li}
\affiliation{Department of Astronomy, University of Virginia, Charlottesville, VA 22903}
\author[0000-0002-2555-9869]{Merel L.R. van 't Hoff}
\affiliation{Department of Astronomy, University of Michigan, 1085 S. University Ave, Ann Arbor, MI 48109, USA}
\author{Sarah I. Sadavoy}
\affiliation{Department of Physics, Engineering Physics and Astronomy, Queen’s University, Kingston, ON, K7L 3N6, Canada}
\author[0000-0003-3682-854X]{Nicole Karnath}
\affiliation{SOFIA-USRA, NASA Ames Research Center, MS 232-12, Moffett Field, CA 94035, USA }

\begin{abstract}
We present ALMA (0.87~mm and 1.3~mm) and VLA (9~mm) observations toward the
candidate intermediate-mass protostar OMC2-FIR3 (HOPS-370; \lbol$\sim$314~\lsun) 
at $\sim$0\farcs1 (40~au) resolution for the continuum emission and $\sim$0\farcs25 (100 au) 
resolution of nine molecular lines. The dust
continuum observed with ALMA at 0.87~mm and 1.3~mm resolve a near edge-on 
disk toward HOPS-370 with an apparent radius of $\sim$100~au. The VLA observations detect 
both the disk in dust continuum and free-free emission extended along the jet direction. 
The ALMA observations of molecular 
lines (H$_2$CO, SO, CH$_3$OH, \thco, \cateo, NS, and H$^{13}$CN) reveal rotation of the apparent 
disk surrounding HOPS-370 orthogonal to the jet/outflow direction. 
We fit radiative transfer models to both the dust continuum structure
of the disk and molecular line kinematics of the inner envelope and disk 
for the H$_2$CO, CH$_3$OH, NS, and SO lines. The central protostar mass is determined to be 
$\sim$2.5~\msun\ with a disk radius of $\sim$94~au, when fit using combinations of 
the H$_2$CO, CH$_3$OH, NS, and SO lines, consistent with an intermediate-mass protostar.
Modeling of the dust continuum and spectral energy distribution (SED)
yields a disk mass of 0.035~\msun\ (inferred dust+gas) and a 
dust disk radius of 62~au, thus the dust disk 
may have a smaller radius than the gas disk, similar to Class II disks.
In order to explain the observed luminosity with the measured protostar mass, 
HOPS-370 must be accreting at a rate between 1.7 and 3.2$\times$10$^{-5}$~\msun~yr$^{-1}$.
\end{abstract}

\section{Introduction}

The formation of stars and planets is governed by the collapse of dense clouds
of gas and dust under the force of gravity and conservation of angular momentum.
A rotating disk of gas and dust forms around a nascent protostar due to the conservation of
angular momentum, and material is accreted through the disk onto the protostar. 
However, there are major uncertainties in our understanding of disk formation and
the processes that set their mass and radii. For example, during the collapse process,
magnetic fields must not be strong enough or strongly coupled to the gas on $\lesssim$ 1000~au
scales; otherwise, they could prevent the spin-up of infalling material \citep{allen2003,
mellon2008}. Non-ideal magneto-hydrodynamic (MHD) effects
can also dissipate the magnetic flux and enable the formation of disks 
to proceed during the star formation process \citep[e.g.,][]{dapp2010,li2014,masson2016}.
Additionally, turbulence of the infalling material and misaligned magnetic fields
have also been shown to enable disk formation \citep{seifried2013,joos2012}.

The evolutionary state of a protostar system is typically classified by the properties of their
spectral energy distributions (SEDs), which approximately (but not directly) relate
to the physical evolution \citep[e.g.,][]{robitaille2006,offner2012}.
Observationally, the youngest protostars identified are those in the Class 0 
phase, which is characterized by a dense
infalling envelope of gas and dust surrounding the protostar(s) \citep{andre1993}. 
Following the Class 0 phase is the Class I phase, in which the protostar is 
less deeply embedded, but still surrounded by an infalling envelope,
and by the end of the Class I phase the envelope will be largely dissipated. 
The bolometric temperature (\tbol) is a diagnostic of the evolutionary state
utilizing the SED of a protostar \citep[e.g.,][]{chen1995}, and \tbol=70~K 
is the a canonical dividing line between Class 0 
and Class I protostars \citep{dunham2014}. This border is an observational distinction in 
what is otherwise considered to be a gradual evolution in envelope properties. 
However, the measured \tbol\ can 
vary depending on viewing inclination angle and sampling of the SED at wavelengths longer 
than $\sim$70~\micron\ \citep{furlan2016,tobin2008}; thus,
protostars with \tbol\ near 70~K could belong to either class. 

One such borderline protostar is HOPS-370, 
also known as OMC2-FIR3 \citep{chini1997} and VLA 11 \citep{reipurth1999}, 
located in the northern part of the
integral-shaped filament within the Orion A molecular cloud. 
Recent measurements from the \textit{Herschel} Orion Protostar Survey 
\citep[HOPS; ][]{furlan2016} found that HOPS-370 has a \tbol\ of 71.5~K and a 
bolometric luminosity (\lbol) of 314~\lsun. Model fitting to its spectral 
energy distribution (SED) in the aforementioned paper 
indicate an internal luminosity of 511~\lsun\ (values are adjusted to account for the adopted
distance of 392~pc versus the previously adopted 420~pc)\footnote{The
revised distance is estimated using \textit{Gaia} parallaxes measured
toward more-evolved young stars throughout the Orion region, see the Appendix
of \citet{tobin2020} and \citet{kounkel2018} for more detail.}. 
Thus, HOPS-370 is one of the most luminous protostars forming north of the Orion Nebula
in the OMC2 and OMC3 regions \citep[e.g.,][]{tobin2019}. 

HOPS-370 is also driving a strong jet and
outflow that is seen in radio continuum \citep{osorio2017}, 
[OI] 63~\micron, far-infrared CO, and H$_2$O lines \citep{gonzalez2016}, 
and low-J CO \citep{williams2003,shimajiri2008,takahashi2008,tobin2019}. Furthermore,
observations by the NSF's Karl G. Jansky Very Large Array (VLA) 
and the Atacama Large Millimeter/submillimeter
Array (ALMA) have observed the region with $\sim$0\farcs1 resolution, 
resolving an apparent disk in the dust continuum and indications of 
rotation in methanol, H$^{13}$CN, and NS \citep{tobin2019}. The combination
of observational results toward this region from \citet{tobin2019}, \citet{furlan2014}, and 
\citet{osorio2017} indicate that HOPS-370 is a candidate intermediate-mass
protostar with similar spectral properties to hot corinos \citep{taquet2015, 
ceccarelli2004,jacobsen2018,drozdovskaya2016,lee2018}.

The aforementioned work has resulted in HOPS-370 being regarded as a potential
prototype intermediate-mass protostar given
its well-organized nature. We present new data and further 
analyze the ALMA and VLA molecular line and continuum observations toward this source
at a resolution of $\sim$0\farcs1 (40~au) in the continuum and in molecular lines observed
at $\sim$0\farcs25 (100 au) resolution. Using these data probing $<$500~au scales, 
we examine the structure of the forming disk and its
gas kinematics. We use these molecular line data to measure the mass of the central protostar and
confirm its intermediate-mass status. Finally, we also present near-infrared
spectroscopy toward the protostar. The paper is structured as follows. 
Section 2 describes the observations and data reduction, Section 3
provides and overview of the region around HOPS-370, Section 4 describes the
dust continuum and molecular line kinematics, 
Section 4 presents the radiative transfer modeling results.
We discuss our results in Section 6 and present our conclusions in Section 7.

\section{Observations and Data Reduction}

We make use of data from two ALMA bands, a single band of VLA data, and near-infrared spectroscopy
in our study of HOPS-370. The ALMA 0.87~mm observations and the VLA 9~mm 
observations have already been detailed in \citet{tobin2019} and \citet{tobin2020}; we only
briefly describe those observations. The new ALMA 1.3~mm observations and near-infrared spectroscopic observations
are described in more detail, along with their reduction procedures.

\subsection{ALMA 1.3~mm Observations}
The ALMA observatory is located on the Chajnantor plateau in northern Chile at an elevation
of $\sim$5000~m. HOPS-370 was observed with ALMA at 1.3~mm (Band 6) 
on 2018 January 07 with 43 antennas operating and sampling baselines from 15~m to 2500~m. 
The observations were executed within an 87 minute observation block, 
and HOPS-370 was observed along with 
19 other Orion protostars. The total time spent on HOPS-370 was $\sim$2.42 minutes and
the precipitable water vapor was $\sim$2.3 mm.
The complex gain, bandpass, and absolute flux calibrator was J0510+1800. 
The absolute flux calibration accuracy is expected to be better than 10\%.
The correlator was configured with the first baseband containing a 1.875 GHz continuum band 
centered at 232.5 GHz and observed in time Division Mode (TDM) with 128 channels; the remaining
three basebands were configured in Frequency Division Mode (FDM). The second baseband was
split into two 58.6 MHz spectral windows with 1960 channels each (0.083~\kms\ velocity resolution)
and centered on $^{13}$CO ($J=2\rightarrow1$) and C$^{18}$O ($J=2\rightarrow1$). The third baseband was
split into four 58.6 MHz spectral windows with 980 channels each (0.168~\kms\ velocity resolution)
and centered on SO ($J_N = 6_5\rightarrow5_4$), H$_2$CO ($J=3_{0,3}\rightarrow2_{0,2}$), 
and  H$_2$CO ($J=3_{2,2}\rightarrow2_{2,1}$); the final window was centered between 
H$_2$CO ($J=3_{2,1}\rightarrow2_{2,0}$) and CH$_3$OH ($J=4_{2,2}\rightarrow3_{1,2}$),
enabling both lines to be observed. Finally, the fourth baseband was configured with 
two 234 MHz spectral windows (980 channels, 0.367~\kms\ resolution) and one centered 
on $^{12}$CO ($J=2\rightarrow1$) and the other centered 
between \ntdp\ ($J=3\rightarrow2$) and $^{13}$CS ($J=5\rightarrow4$).

The data were reduced using the ALMA calibration pipeline within CASA version 4.7.2. In order to 
increase the signal-to-noise ratio of the continuum and spectral lines, we performed
self-calibration on the continuum. 
We performed 3 rounds of phase-only self-calibration, the first round used solution
intervals that encompassed the length of an entire on-source scan, then the second 
round utilized 12.1~s solution intervals, and the third round used at 6.05~s solution interval,
corresponding to a single integration.
The phase solutions from the continuum self-calibration were also applied to the 
spectral line bands. The resultant RMS noise in the 1.3~mm continuum 
was $\sim$0.22~mJy/beam and $\sim$12~mJy/beam in 0.33~\kms\ channels for the spectral line observations.
The continuum and spectral line data were imaged using the \textit{clean} task within
CASA version 4.7.2. The continuum image was deconvolved using Briggs weighting and 
a robust parameter of 0.5, while the spectral line observations were deconvolved 
using Natural weighting. The continuum image only uses uv-points $>$25~k$\lambda$ to 
mitigate striping resulting from bright large-scale emission that is not well sampled. 
The typical beam sizes of the continuum and molecular line images are
0\farcs23$\times$0\farcs13 (90~au~$\times$51~au) and 0\farcs32$\times$0\farcs18 
(125~au~$\times$~71~au), respectively. 
The observational setups are also detailed in Table 1 and
the reduced data are available from the Harvard Dataverse \citep{hops_370_spectral_line}\footnote{https://dataverse.harvard.edu/dataverse/VANDAMOrion}.

\subsection{ALMA 0.87~mm Observations}
 The ALMA 0.87~mm observations were taken as three executions of the
scheduling block with two executions on 2016 September 6 and the third on 2017 July 19.
The time on-source during each execution was $\sim$0.3~minutes for a total of $\sim$0.9~minutes
on HOPS-370 at 0.87~mm. The correlator was setup with two basebands observed in TDM mode, 
each with 1.875 GHz of bandwidth and 128 channels, 
centered at 333 GHz and 344 GHz. The other two basebands were observed in FDM mode, centered on \twco\ ($J=3\rightarrow2$) 
at 345.79599 GHz and \thco\ ($J=3\rightarrow2$) at 330.58797~GHz. The bandwidth and spectral resolution
of the spectral windows were 937.5~MHz with 0.489~\kms\ channels and 234.375~MHz with 0.128~\kms\ channels.
Additional detail of the data reduction and imaging is provided in \citet{tobin2019} and \citet{tobin2020}.
The data are available from the Harvard Dataverse \citep{alma_images}.

In addition to the continuum data, in this paper we make use of integrated intensity maps from
the $^{12}$CO ($J=3\rightarrow2$) data cubes observed as part of this work toward HOPS-370 and HOPS-66. 
The $^{12}$CO data cubes were generated with 1~\kms channels using the CASA 4.7.2 \textit{clean}
task with \textit{robust=2} weighting, uv-distances $>$50~k$\lambda$ to avoid artifacts from
large-scale emission, and tapering at 500~k$\lambda$ to increase sensitivity to extended emission.
Masks were created manually through interactive execution of the \textit{clean} task.
The integrated intensity maps were generated using the CASA task \textit{immoments} selecting
the channel ranges where $^{12}$CO emission was detected. The reduced data cubes are also available
from the Harvard Dataverse \citep{alma_12CO}.

\subsection{VLA 9~mm Observations}
The VLA is located on the Plains of San Agustin in New Mexico, USA at an elevation of 2100~m.
The VLA observations of HOPS-370 were conducted on 2016 October 26 while the VLA was in
A configuration with 26 antennas operating. The entire observation lasted 2.5 hours with
$\sim$1 hour on-source. We used the Ka-band receivers with 3-bit samplers, providing a 4~GHz 
baseband centered at 36.9~GHz (8.1~mm) and the other baseband centered at 29~GHz (1.05~cm).
Additional details of the data reduction and imaging are provided in \citet{tobin2019} and \citet{tobin2020};
the reduced data are also available for download through the Harvard Dataverse \citep{vla_images}.

\subsection{Near infrared Spectroscopy}

Near-infrared spectroscopy was obtained from the Astrophysics
Research Corporation (ARC) 3.5~m telescope at Apache Point
 Observatory in New Mexico, USA. HOPS-370 was
observed on 2017 October 13 using the the TripleSpec spectrograph \citep{wilson2004}. 
TripleSpec simultaneously records spectra from $\sim$0.9~\micron\ to 2.5~\micron\ with a 
resolution of R$\sim$3000 with a 1\farcs1$\times$45\arcsec\ slit.

The slit was centered on the base of the near-infrared scattered light
nebula associated with HOPS-370, oriented in the East-West direction. This slit
orientation minimized contamination from a near-infrared point source (MIR 22) located
$\sim$3\arcsec\ to the south. HOPS-370 was observed in an ABBA pattern, and the integration
time was 2.5 minutes at each nod position, with a total on-source time of 30 minutes. Nodding
was done along the slit and the separation of nod positions was $\sim$20\arcsec. The average
airmass during the observation was 1.5. The telluric standard used was the A0 star HD 37887
with a magnitude of 7.74 in Ks-band,
and it was observed in an ABBA pattern with 30 seconds
in each nod position with a total time on source of 4 minutes.

The data were reduced using the IDL package Tspectool, which is a modified version of 
Spextool \citep{vacca2003, cushing2004}. Wavelength calibration was performed 
using OH sky emission lines. Flat fielding was done using exposures of quartz lamps
on the telescope truss. The flat field was constructed by subtracting exposures with
the lamps on from exposures with the lamps off. Flux calibration of the spectrum was performed
using the telluric standard with its cataloged magnitude relative to Vega.
The reduced spectrum is available from the Harvard Dataverse \citep{hops_370_spectral_line}.

\section{Overview of HOPS-370/OMC2-FIR3 Region}

\begin{figure}
\begin{center}
\includegraphics[scale=1.1]{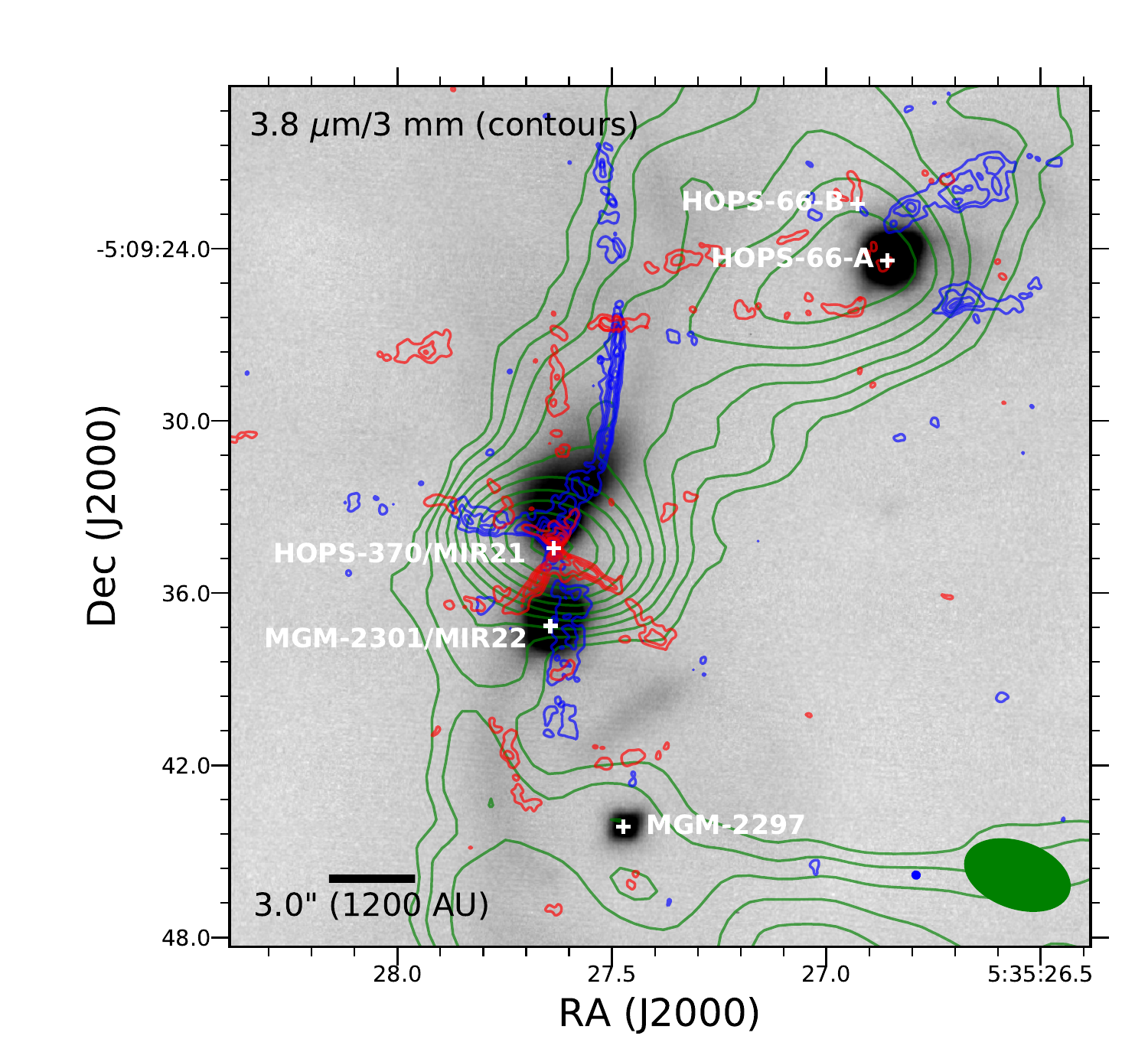}
\end{center}
\caption{
Overview of the HOPS-370/OMC2-FIR3 region. 
The grayscale image is a 3.8~\micron\ image (L$^{\prime}$-band) from \citet{kounkel2016}, the green contours 
are ALMA 3~mm combined 12m and ALMA Compact Array data from \citet{kainulainen2017}.
The red and blue contours are the low-velocity $^{12}$CO ($J=3\rightarrow2$) integrated intensity
toward HOPS-370 (blue: -26 to 3~\kms, red: 14 to 23~\kms) from \citet{tobin2019} 
and HOPS-66 (blue: -9 to 7~\kms, red: 14 to 28~\kms).
The positions shown for HOPS-370 and HOPS-66 are from \citet{tobin2020} and
MIR 22/MGM 2301 and MGM 2297 are from \citet{megeath2012}. 
The ALMA 3~mm contours start at 3$\sigma$ and increase on logarithmically spaced intervals to 100$\sigma$, where
$\sigma$=0.0024~mJy~beam$^{-1}$.  The blue contours
toward HOPS-370 (HOPS-66) start at 6$\sigma$ (3$\sigma$) and increase on 3$\sigma$ (3$\sigma$) 
intervals where $\sigma$=0.12~Jy~beam$^{-1}$ ($\sigma$=0.18~Jy~beam$^{-1}$). The red contours
toward HOPS-370 (HOPS-66) start at 5$\sigma$ (3$\sigma$) and increase on 3$\sigma$ (3$\sigma$) 
intervals where $\sigma$=0.12~Jy~beam$^{-1}$ ($\sigma$=0.15~Jy~beam$^{-1}$).
The 3~mm beam is 3.75\arcsec$\times$2.27\arcsec\ and is shown as the green ellipse the lower right corner,
while the beam for the $^{12}$CO moment maps is 0.25\arcsec$\times$0.24\arcsec\ and is the small 
blue ellipse left of the green ellipse.
}
\label{overview}
\end{figure}

We show an overview of HOPS-370/OMC2-FIR3 and its surroundings
at 3.8~\micron\ and 3~mm in Figure \ref{overview}; the 3.8~\micron\ data
originally appeared in \citet{kounkel2016} and the 3~mm data are 
from \citet{kainulainen2017}. HOPS-370 and HOPS-66
are the protostars present within the 30\arcsec\ region shown,
and two additional sources are also detected: MGM 2297 
and MIR 22 (MGM 2301), both of which
appear to be more-evolved young stellar
objects (YSOs) \citep{megeath2012,nielbock2003}.
HOPS-370 was also detected by \citet{nielbock2003} as MIR 21 in
close proximity to MIR 22. Unlike the other
sources identified in the image, HOPS-370/MIR 21
does not have a corresponding point source at 3.8~\micron\ due to its 
deeply embedded nature. The emission north of HOPS-370 is, however, scattered
light with its outflow cavities illuminated by the central protostar
and disk (e.g., Habel et al. 2020). The
$^{12}$CO ($J=3\rightarrow2$) integrated intensity maps shown in Figure \ref{overview}
illustrate the correspondence of the scattered light emission with the 
low-velocity outflow emission from HOPS-370. HOPS-66 also has some scattered light 
to the west of its position and $^{12}$CO emission along its edges.

At first glance it appears that HOPS-370 could be a multiple system given the
proximity of the other YSOs detected in the infrared; MIR 22 has
a projected separation of just $\sim$3\arcsec\ ($\sim$1200~au). However, 
it was argued in \citet{tobin2019} that MGM 2297 and MIR 22 are likely 
foreground YSOs and not embedded within an envelope like HOPS-370. The lack of 3~mm emission
at their positions is further evidence for lacking an envelope; only HOPS-370 and HOPS-66
have strong 3~mm emission toward their positions.
MIR 22 is detected and resolved from HOPS-370 (MIR 21) at 1 to 18 \micron, and its SED
is consistent with a Class II YSO \citep{nielbock2003}.
Longer-wavelength emission in the mid- to far-infrared is centered on 
HOPS-370, which is also brighter than MIR 22 beyond 12~\micron\ \citep{furlan2014}, and
MIR 22 is not detected at 870 micron nor at 3 mm \citep{tobin2019,kainulainen2017}. 
MIR 22 was, however, detected at centimeter wavelengths as a thermal free-free source, 
$\sim$100$\times$ weaker than HOPS-370/MIR 21
\citep{osorio2017}. Finally, the $^{12}$CO maps show no evidence of an outflow
associated with MIR 22. Taken together, the evidence indicates that MIR 22 
is not likely to be embedded within the envelope of HOPS-370 and is most likely
a foreground YSO and not a true companion.
Thus, the nearest protostar that is likely to be physically associated with HOPS-370 is
HOPS-66 at a projected separation of 6600~au; no additional candidate companions were 
detected down to 30~au separations \citep{tobin2020}.

\section{Results}

The ALMA and VLA observations of HOPS-370 (OMC2-FIR3) offer an unprecedented
view of this protostar and its immediate environment in terms of resolution, sensitivity, 
and breadth of molecular lines/continuum wavelengths observed at $\la$0\farcs25 (100~au) resolution. 
The dust continuum at 0.87~mm and 1.3~mm probes the column density structure toward HOPS-370,
and the kinematics are traced by multiple molecular lines that probe 
complementary physical conditions. The 9~mm continuum
on the other hand traces a combination of dust and free-free emission as described in more detail in the
following section.

\subsection{Continuum Emission}

The ALMA and VLA continuum images of HOPS-370 at 0.87~mm, 1.3~mm, and 9~mm are shown 
in Figure 1. The 0.87~mm and 9~mm images were previously presented in \citet{tobin2019}
and they are reproduced here to emphasize the clear disk emission.
 The 0.87~mm and 1.3~mm images clearly show the presence of a 
disk-like structure in the dust continuum emission that we will
simply refer to as a disk. This disk is orthogonal to the outflow and jet directions
traced both by \textit{Herschel} \citep{gonzalez2016} and the 
continuum between 5~cm and 9~mm \citep{osorio2017}. The 5~cm
contours from \citet{osorio2017} are overlaid on the 1.3~mm map in Figure \ref{continuum-5cm},
further demonstrating the orthogonal relationship between the disk and the jet.

To measure the geometric parameters of the continuum emission and
the integrated flux densities, we fitted Gaussians to the images
using the \textit{imfit} task of CASA 4.7.2.
The single-component Gaussians are not perfect fits to the data, but
the integrated flux densities agree reasonably well with comparison to the flux density
measured within a a polygon surrounding the source.

The results from the Gaussian fits to the continuum emission at 0.87~mm find a deconvolved 
full width at half maximum (FWHM) of 0\farcs34~$\times$~0\farcs11
with a position angle of 109\degr\ (Table 2). 
Adopting the 2$\sigma$ value of the Gaussian fit
as the disk radius \citep{tobin2020}\footnote{The FWHM of a Gaussian is equivalent 
to $\sqrt{8~ln(2)}$$\sigma$~$\simeq$~2.355$\sigma$.}, we find a disk 
radius of $\sim$113~au, and the inclination of HOPS-370 can be estimated 
to be $\sim$71\degr\ by assuming that it is a
geometrically thin disk and then calculating the inverse cosine of 
the deconvolved minor axis divided by the deconvolved major axis.

The VLA 9~mm continuum image in Figure 1 shows a distinctly different morphology 
with respect to the ALMA images. Rather than a simple disk feature, the VLA 9~mm 
image shows a cross-like morphology. The extension in the 
northeast to southwest direction is longer and more prominent than the extension in the southeast to northwest
direction. The longer axis is orthogonal to the major axis of the disk and in the presumed 
direction of the jet/outflow from HOPS-370. We also show the spectral index
map determined from the VLA 9~mm data alone. The spectral index
of the emission ($S_{\nu}~\propto~\nu^\alpha$) indicates that 
the presumed-jet is emitting via the 
free-free emission process, with values between -0.1 and $\sim$1 
\citep[e.g.,][]{anglada1998}. The spectral index 
in the direction of the disk is close to 2 at the edges of the spectral index map
and more consistent with dust emission
than free-free emission. Towards the central region, the spectral index 
is $\sim$1, too low to be dust, and suggesting a significant free-free 
contribution. Therefore, at 9~mm we trace
both the jet and disk emission. The direction of the jet emission at 9~mm is
consistent with the resolved jet reported by \citet{osorio2017} at
longer wavelengths (see Figure \ref{continuum-5cm}) and the molecular 
outflow in \twco\ reported by \citet{tobin2019}.

We fit the VLA data with four Gaussian
components simultaneously to describe the cross-like structure observed at 9~mm. 
The first component is a point-like component to match the peak intensity, the 2nd
and 3rd components model the surface brightness distribution of the extended jet, and 
the 4th component fits the emission along the position angle of the disk. We add 
parameter estimates to help imfit arrive at a solution that can fit the disk component.
The two components are needed for the emission along the jet axis because 
it cannot be well-described with a single Gaussian. The parameters of each component
are given in Table 3; note that the optimization of the 4 components yield one component
with a negative flux. When this negative component is combined with the brighter positive
component the surface brightness distribution is best reproduced.

The extent of the emission from the disk at 9~mm is estimated to be 0\farcs27$\times$0\farcs09
and a position angle of 112\degr, and the flux density of the
disk is $\sim$0.73~mJy. The position angle of the disk is consistent with the fit at 0.87~mm, and the
angular size at 9~mm is modestly smaller than at 0.87~mm, 
which is typical when comparing such short and long wavelength data \citep[e.g.,][]{segura-cox2016,tobin2020}.

We plot the radio spectrum of HOPS-370 in Figure \ref{radio-spectrum}, including the flux
densities reported in this work and those from \citet{osorio2017} 
at longer wavelengths. We fit power-laws to the dust emission between 0.87~mm and 9~mm
and remove this estimated contribution from the longer wavelength data points. Then the 
longer wavelength data points are assumed to only trace the free-free emission from
HOPS-370. The flux density of the dust emission scales $\propto$~$\nu^{2.84\pm0.08}$ 
and the flux density of the free-free emission scales $\propto$~$\nu^{0.19\pm0.08}$.
The spectral index of the free-free emission indicates partially optically-thick emission since
flux density of optically thin free-free emission is expected to scale $\propto$~$\nu^{-0.1}$. This is consistent
with the spectral index derived at the position of HOPS-370 by \citet{osorio2017}, 
while at larger distances from the protostar they found that the jet exhibits non-thermal spectral indices.

\subsection{Estimated Disk Mass}

The mass of the disk traced by dust continuum emission can be estimated under the assumption of
isothermal and optically thin emission using the equation
\begin{equation}
\label{eq:dustm}
M_{dust} = \frac{D^2 F_{\nu} }{ \kappa_{\nu}B_{\nu}(T_{dust}) }.
\end{equation}
$D$ is the distance ($\sim$392~pc), $F_{\nu}$ is the observed flux density, $B_{\nu}(T_{dust})$ is
the Planck function, $T_{dust}$ is the dust temperature, and $\kappa_{\nu}$ is the
dust opacity at the observed wavelength. We adopt $\kappa_{1.3mm}$~=~0.899~cm$^2$~g$^{-1}$ 
and $\kappa_{0.87mm}$~=~1.81~cm$^2$~g$^{-1}$ \citep{ossenkopf1994}, 
appropriate for protostellar envelopes. We adopt
 $\kappa_{9.1mm}$~=~0.13~cm$^2$~g$^{-1}$ by using $\kappa_{1.3mm}$ as a reference
point and extrapolating it to 9~mm assuming a dust opacity power-law index of 1.0.
This adhoc extrapolation from 1.3~mm to 9~mm is necessary because the \citet{ossenkopf1994}
models predict an opacity that is too low at 9~mm to yield consistent dust masses with shorter
wavelength observations \citep{tobin2016a,segura-cox2016, tychoniec2018b}.
$T_{dust}$ is often assumed to be 20 or 30~K \citep{jorgensen2009,tobin2015b, tychoniec2018}
for solar-luminosity protostars, consistent with temperature 
estimates on $\sim$100~au scales \citep{whitney2003a}. 
However, as part of the VANDAM Orion survey \citep{tobin2020} we used radiative
transfer models to estimate the average dust temperature dependence on radius and
luminosity. Using those results, we estimate a dust average temperature
for a 100~au embedded protostellar disk of $\sim$31~K around a 1~\lsun\ protostar. 
Assuming that T$_{dust}$ scales as (L/\lsun)$^{0.25}$, 
the expected average temperature for the HOPS-370 disk is $\sim$131~K, with
\lbol~=~314~\lsun\ \citep{furlan2016}. 
We finally multiply the resulting value of M$_{dust}$ by
100, assuming the typical dust to gas mass ratio of 1:100 \citep{bohlin1978}
to arrive at an estimate of the total disk mass.

The continuum flux density at 1.3~mm is 0.207~Jy, corresponding
to a disk mass of 0.084~\msun. At 0.87~mm, the continuum flux density is 0.533~Jy,
corresponding to a disk mass of 0.048~\msun. Lastly, at 9~mm the continuum
flux density from the disk is  0.732~mJy, corresponding to a disk mass of 0.098~\msun.
The modest discrepancy between 
0.87~mm and 1.3~mm  could result from the dust continuum at 0.87~mm being more 
opaque and/or uncertainty in the relative dust opacity between
0.87~mm and 1.3~mm. Also, the larger beam at 1.3~mm may enable more emission
to be recovered due to more short baselines being included in those
observations. The 9~mm measurement agrees well with the others considering
the uncertainty in dust opacity and the separation of its emission from the
jet using multi-component Gaussian fitting.

\subsection{Molecular line Emission}

The continuum imaging from both ALMA and the VLA strongly indicate the presence of a 
disk surrounding HOPS-370. However, the nature of the disk structure can be 
analyzed further using the kinematics of molecular line emission. To this end,
a suite of molecular lines were observed toward HOPS-370 with ALMA at 0.87~mm and 1.3~mm. 
However, the longer integration of the 1.3~mm data and shorter baseline
coverage leads to those data being more sensitive to molecular
line emission than the few lines covered in the 0.87~mm observations. 
The \cateo, \thco, H$_2$CO, SO, NS, and CH$_3$OH molecular lines all
 trace a signature of rotation from the disk revealed by the
dust continuum. NS and H$^{13}$CN are the only lines highlighted here from the 0.87~mm
observations, but other complex organic molecules are also detected toward the disk at
lower S/N \citep{tobin2019}. Of the other targeted molecular lines, the 
$^{13}$CS line was only weakly detected,
\ntdp\ was not detected, and \twco\ traces the outflow. The non-detection of \ntdp\
is expected because of the warm temperature of the warm temperature of the disk and
inner envelope resulting from the high luminosity of the protostar \citep[e.g.,][]{emprechtinger2009,tobin2013}.

We show the integrated intensity maps of the blue- and 
red-shifted emission of each molecular line (except \twco, \ntdp, and $^{13}$CS) 
in Figure \ref{moment0}. The integrated intensity maps were created using
selected channel ranges where there is spectral line emission using 
the CASA task \textit{immoments}. We selected the channels corresponding 
to blue and redshifted emission using the system velocity of
$\sim$11.0~\kms\ \citep{tobin2019}; the blueshifted maps are integrated between 4 and
11~\kms and the redshifted maps are integrated between 11 and 18~\kms.
These blue and redshifted integrated intensity images are used 
to assess the kinematics traced by the particular molecular lines. We do not show the 
\twco\ moment maps because they were presented \citet{tobin2019}. We also do not show 
\ntdp\ because it is not detected and $^{13}$CS is not shown because it is only marginally detected.

The consistency of the velocity gradient in all well-detected molecular lines 
indicates that the disk around HOPS-370 is clearly rotating and may be rotationally supported.
We note that the peaks of the line emission can be northeast and/or 
southwest with respect to the center of the continuum position,
and these peaks tend to avoid the midplane traced in dust continuum. 
Continuum opacity and/or line opacity are likely to cause
this feature; molecular freeze-out is not likely because of the warm disk temperatures
due to the high luminosity of the protostar.

The integrated intensity maps only show limited spectral information; 
therefore we also show position-velocity (PV) diagrams for each
molecular line in Figure \ref{pvdiagrams}.
The PV diagrams were extracted from the molecular line data cubes using 
a custom Python code to obtain a complementary view of the kinematic structure.
The PV diagrams are extracted
along a 0\farcs6 wide strip (15 pixels) along the major axis of the disk, at a position 
angle 100\degr\ East of North. The emission is summed within the 0\farcs6 strip
to produce a two dimensional spectrum. The PV diagrams bear strong 
resemblance to other protostellar disks that have been detected and 
characterized \citep{tobin2012,sakai2014,oya2015,aso2017}. 
The PV diagrams bear the signature of a rotating Keplerian disk with a finite radius: a linear velocity gradient
transitioning from blue to red-shifted on opposite sides of the protostar and higher-velocity emission
from spatial scales closer to the central protostar following a Keplerian velocity profile.

These features are produced because the outer radius of the disk causes the
 emission from the largest radii of the disk to have a similar velocity profile
to a rotating ring; in this case the ring is the outer disk. The finite radius
of the outer disk means that Keplerian rotation will not continue to spatial
scales that extend beyond the disk. Then radii closer to
the protostar rotate more quickly, producing the higher-velocity emission 
that is typically associated with Keplerian rotation toward embedded protostars.
The linear transition from blue to red is most easily seen in the PV diagrams
for NS and SO molecules. However, the transition is not always
easily detectable toward protostellar disks \citep[e.g.,][]{tobin2012,ohashi2014,murillo2013} 
due to spatial filtering and blending with the infalling envelope and 
the molecular cloud near the system velocity. But, certain molecular transitions 
that specifically trace the disk and not 
the surrounding cloud/envelope show the low-velocity emission of the 
disk with higher fidelity. This is due to a variety of possible 
reasons, but mainly critical density, chemistry, excitation temperature. 
The lines SO, CH$_3$OH, NS, and H$_2$CO ($J=3_{2,2}\rightarrow2_{2,1}$), 
($J=3_{2,1}\rightarrow2_{2,0}$) with T$_{ex}$ $\sim$60~K best trace the disk of HOPS-370.

Previous studies have been able to use point-like line emission in velocity channels
away from the system velocity toward disks around lower-mass
Class 0 and Class I protostars to fit the emission centroids from PV diagrams, data cubes, and/or
the uv-data themselves to map the rotation curves in one dimension. In these cases, the emission
centroids systematically changed with each velocity channel, high velocities near the
continuum source and lower velocities centered farther from the continuum source \citep{yen2013,
tobin2012,ohashi2014}. However, the clear non-Gaussian 
emission in the images prevents such analyses from being viable for HOPS-370. Thus, the 
rotation curve must be examined using modeling of the molecular line emission which will be
discussed in Section 5.

\subsection{Near Infrared Spectrum}
The near-infrared spectrum of HOPS-370 is shown in Figure \ref{near-ir-spectrum} 
from 2 to 2.4~\micron. HOPS-370 is also detected shortward of 2~\micron\, but with lower
S/N and is not shown. The main spectral features 
are the prominent H$_2$ emission lines detected throughout the band.
We also detect weaker emission from the Brackett $\gamma$ (Br $\gamma$) atomic
hydrogen recombination line and CO band head emission. Continuum emission 
is detected, but there are no obvious photospheric absorption features detected 
in this medium resolution spectrum. The H$_2$ emission lines are likely associated 
with shocks in the outflow from HOPS-370 given that most of the near-infrared 
emission detected in this spectrum is from the
outflow cavity due to the central protostar being too highly extincted. 
The properties of the Br $\gamma$ emission are of the greatest interest to extract 
due to their frequent association with accretion processes in young stars 
\citep[e.g.,][]{muzerolle1998,connelley2009}. The equivalent width of the Br $\gamma$
line is -1.42~\AA, with an integrated flux of 2.9$\times$10$^{-15}$~erg~s$^{-1}$~cm$^{-2}$.
This line flux translates to a Br $\gamma$ luminosity of 
5.6$\times$10$^{28}$~erg~s$^{-1}$ or 1.5$\times$10$^{-5}$~\lsun. The equivalent
width and line luminosity of the Br $\gamma$ line are within the ranges typically
observed toward Class I protostars by \citet{connelley2010}. However, these values are
on the low end for a protostar that is expected to be accreting rapidly. The interpretation of
the line emission will be further discussed in Section 6.3.

\section{Radiative Transfer Modeling}

In order to further interpret our dust continuum and molecular line observations,
radiative transfer modeling is necessary. We use molecular line radiative transfer
modeling to fit the kinematics of the system, primarily constraining the protostar mass
and rotating disk radius. The continuum radiative transfer modeling on the other hand 
enables more detailed constraints on the disk structure to be derived. We make
use of the software packages \textit{pdspy}\footnote{https://github.com/psheehan/pdspy} 
\citep{sheehan2017,sheehan2019} and RADMC-3D \citep{dullemond2012} for our modeling efforts.

\subsection{Molecular Line Modeling}
The molecular line images and PV diagrams show strong rotation signatures 
(Figures \ref{moment0} and \ref{pvdiagrams}). To quantitatively determine 
if the rotation is tracing a Keplerian disk and, if so, to measure 
the protostar mass we must make use of radiative transfer 
modeling to fully utilize the constraints offered by the 
channel maps for multiple molecular lines.
The \textit{pdspy} package has distinct modes for fitting molecular line kinematic data
and continuum data. The basis for both models is an analytic physical model for a protostellar
system with a surrounding disk embedded within an infalling envelope.

\subsubsection{Physical Model}

The disk structure for the molecular line models uses an 
exponentially-tapered disk density profile \citep{lyndenbell1974}.
The exponentially-tapered density profile is
described by
\begin{equation}
\Sigma(r) = \, \Sigma_0 \left(\frac{r}{r_c}\right)^{-\gamma} \, \exp\left[-\left(\frac{r}{r_c}\right)^{(2-\gamma)}\right],
\end{equation}
where $r$ is the disk radius defined in cylindrical coordinates. The 
power-law index of the surface density profile is defined by $\gamma$. 
The normalization constant, $\Sigma_0$ is given by
\begin{equation}
\Sigma_0 = \frac{(2-\gamma) M_d}{2 \pi r_c^2},
\end{equation}
where $M_d$ is the disk mass and the other parameters are as defined above.
The vertical structure of the disk is set by hydrostatic equilibrium, assuming 
that the disk is vertically isothermal such that
\begin{equation}
h(r) = \left(\frac{k_b \, r^3 \, T_g(r)}{G \, M_* \, \mu_m \, m_H}\right)^{1/2},
\end{equation}
where $k_b$ is the Boltzmann constant, $G$ is the gravitation constant, $M_*$ is the
central stellar mass, $\mu_m$ is the mean molecular weight of 2.37 \citep{lodders2003},
$m_H$ is the mass of a hydrogen atom, and
$T_g(r)$ is the temperature profile defined as
\begin{equation}
T_g(r) = T_0 \, \left(\frac{r}{1 \, \mathrm{au}}\right)^{-q}
\end{equation}
where both $T_0$ and $q$ can be free-parameters.

The rotating, infalling envelope is described by the density profile of a
rotating collapse model \citep{ulrich1976,cassen1981,terebey1984}. The envelope density
profile is defined as
\begin{equation}
\footnotesize
\rho = \frac{\dot{M_{env}}}{4\pi}\left(G M_* r^3\right)^{-\frac{1}{2}} \left(1+\frac{\mu}{\mu_0} \right)^{-\frac{1}{2}} \left(\frac{\mu}{\mu_0}+2\mu_0^2\frac{R_c}{r}\right)^{-1}
\end{equation}
where $\dot{M}_{env}$ is the mass infall rate of the envelope onto the disk,
$R_c$ is the centrifugal radius where the infalling material has enough angular momentum to
orbit the star, $\mu$~=~cos $\theta$, and $\mu_0$ is the cosine polar angle
of a streamline out to r $\rightarrow$ $\infty$. The density profile
inside of $R_c$ will be $\rho_{env} \propto r^{-1/2}$, and outside $R_c$,
it will be $\rho_{env} \propto r^{-3/2}$. 
For the sake of our modeling, we consider $R_c$ to be equivalent 
to the outer disk radius for a truncated disk model and the critical
radius for an exponentially-tapered disk. Thus, $R_c$ = $r_c$ from Equations 2, 3, and 4.

The envelope model includes outflow cavities with reduced envelope density. The
width of the outflow cavities are parameterized as
\begin{equation}
z~>~1~au + r^{\xi},
\end{equation}
and the envelope density is reduced by a factor of $f_{cav}$.
The outflow cavity opening angle will be less than 45\degr\ for $\xi$~$<$~1 and greater
than 45\degr\ for $\xi$~$>$~1. The outflow cavity opening angle, $\psi$, 
can be directly calculated from
\begin{equation}
\psi=2tan^{-1}(\xi).
\end{equation}

The velocity profile of the infalling envelope is also adopted from the rotating collapse model \citep{ulrich1976} where
\begin{equation}
\scriptsize
v_r(r, \theta) = - \left(\frac{G M_*}{r}\right)^{1/2} \, \left(1 + \frac{\cos\theta}{\cos\theta_0}\right)^{1/2},
\end{equation}
\begin{equation}
\scriptsize
v_{\theta}(r, \theta) = - \left(\frac{G M_*}{r}\right)^{1/2} \, (\cos\theta_0 - \cos\theta) \, \left(\frac{\cos\theta_0 + \cos\theta}{\cos\theta_0\sin^2\theta}\right)^{1/2},
\end{equation}
\begin{equation}
\scriptsize
v_{\phi}(r, \theta) = - \left(\frac{G M_*}{r}\right)^{1/2} \, \left(\frac{\sin\theta_0}{\sin\theta}\right) \, \left(1 - \frac{\cos\theta}{\cos\theta_0}\right)^{1/2}.
\end{equation}
The velocity profile of this equation results in rotational velocity in the 
equatorial plane of the envelope that is equivalent to the Keplerian 
orbital velocity at a radius of $R_c$. As such, material located within the
disk at radii smaller than $R_c$, have velocities described
by Keplerian rotation with 
\begin{equation}
v_{\phi}(r) = \left(\frac{G M_*}{r}\right)^{1/2}.
\end{equation}
The $v_r$ and $v_{\theta}$ components are expected to cancel out upon
incorporation into the disk.

\subsubsection{Parameters}
We are principally interested in fitting the protostar mass and disk radius with the
molecular line models. However, we also fit disk mass, the system velocity, position angle, 
central position, power law index of the surface density profile $\gamma$, and temperature at 1~au.
In addition to these parameters, we computed a second set of models that 
also fit the envelope mass and radius that are presented in the Appendix.
We do not regard $\gamma$ with high confidence given that it
may not truly reflect the underlying surface density profile, rather a convolution of the radial
abundance profile, surface density profile, and dust continuum opacity. Furthermore,
the envelope mass and radius will also not be robust because we do not
include uv-data from scales $<$50~k$\lambda$ (4\arcsec) in our fitting, and the
extended emission from the envelope is weak for the molecular lines shown in Figures \ref{moment0} and \ref{pvdiagrams}.
Also, we do not account for a radial variation in the abundance profile of molecules
in our modeling. Moreover,
the disk and envelope masses are degenerate with the assumed molecular abundances (which are uncertain). The full range of
parameters fixed and varied in the line modeling are provided in Table 4.

To
limit the parameter space, we fix several parameters that do not strongly impact
the modeling results or have constraints from other data.   
The inclination is fixed at 72.2\degr, as determined from an earlier
model fit to the continuum data (Section 5.2); the small ($\sim$2\degr) difference
has an insignificant effect on the fit. Finally, we fix the power-law index of the
temperature profile, $q$, to be 0.35, appropriate for protostellar disks \citep{vanthoff2018}.
We adopt gas phase abundances as follows:
H$_2$CO abundance of 1.0$\times$10$^{-9}$ per H$_2$, SO abundance of 3.14$\times$10$^{-9}$,
CH$_3$OH abundance of 1.0$\times$10$^{-8}$, and NS abundance of 3.14$\times$10$^{-9}$. The CH$_3$OH abundance is 
adopted from the estimate made toward HOPS-370 and HOPS-108 in \citet{tobin2019}, 
while the SO and H$_2$CO abundances are adopted to be consistent with the 
range of abundances reported in \citet{schoier2002,gerner2014,feng2016}, and the NS abundance is within the ranges
found by \citet{crockett2014} and \cite{xu2013}.  When two different molecules are modeled simultaneously, 
the abundance of one molecule is allow to vary such that a single disk mass can fit the data well.

\subsubsection{Model Fitting}

To fit models with the underlying physical structure
described in the previous section, we employ a Markov-Chain Monte Carlo (MCMC)
modeling framework to sample the parameter space and fit radiative transfer
models. We use the software package \textit{pdspy} \citep{sheehan2017,sheehan2019},
which uses \textit{emcee} \citep{foreman2013} to conduct the MCMC and sample the 
parameter space, RADMC-3D \citep{dullemond2012} is used to compute the
molecular line radiative transfer in the limit of local thermodynamic 
equilibrium (LTE) for each sample of the parameter space, and finally 
the GPU Accelerated Library for Analysing Radio Interferometer 
Observations \citep[GALARIO;][]{tazzari2018} is used to Fourier transform the
synthetic datacubes output by RADMC-3D for comparison with the visibility data.
Our molecular line modeling is based on the implementation presented 
in \citet{sheehan2019} to fit the protostar mass and disk radius. 
We restrict fitting to uv-data at baselines longer than 50~k$\lambda$ to limit
the contribution of emission more extended than 4\arcsec, 
which is not recovered in our observations.

We calculate the goodness of fit for each model using the visibility data
from the observations and model at each velocity channel
 between 0 and 19.8~\kms\ with 0.33~\kms\ channels. 
The MCMC uses 200 walkers to explore the multi-dimensional
parameter space. During a single iteration, all walkers are advanced by running a model
and the goodness of fit is calculated for each model.
After the completion of a single iteration, each walker takes its next step 
by comparing its likelihood with the likelihood of the other walkers, and moving 
towards or away from them based on the comparison.

We used the model to fit the molecular line emission, in the uv-plane,
toward HOPS-370 that best traces the disk without significant contamination
from the molecular cloud/envelope. Thus, the molecular lines fit are 
SO, CH$_3$OH, NS, and H$_2$CO ($J=3_{0,3}\rightarrow2_{0,2}$), ($J=3_{2,2}\rightarrow2_{2,1}$), 
and ($J=3_{2,1}\rightarrow2_{2,0}$). We excluded \thco\ and \cateo\ from the
fitting due to their low S/N. The modeling includes the
dust continuum emission produced by the disk parameters of a particular model 
run during the course of the MCMC fitting. 
The continuum emission is generated
using dust grains with $a_{min}$=0.005~\micron, 
$a_{max}$=1~\micron\ and $p$=3.5 with optical properties from
\citet{pollack1994}; this dust grain size distribution and composition
reproduces the features of a typical protostellar SED well \citep{sheehan2014,sheehan2017} and
provides opacities as a function of wavelength similar to \citet{ossenkopf1994}.
The continuum emission is included in the radiative transfer calculation
to approximate an attenuation of the line emission by the continuum opacity. 
Then, the continuum is subtracted from the model to 
to approximate the continuum subtraction that has been applied to the observed data. Note that
the continuum emission that is calculated and subtracted corresponds to the continuum from the molecular
line model setup and \textit{not} from the
fit to the dust continuum data described in Section 5.2.

In addition to fitting individual spectral lines, we also performed simultaneous 
fits to those species that had multiple transitions. The capability exists in \textit{pdspy}
to fit both two species simultaneously and multiple 
transitions of the same molecular species.  
We computed the following simultaneous fits:
all three observed H$_2$CO lines, only the H$_2$CO ($J=3_{2,2}\rightarrow2_{2,1}$) and 
($J=3_{2,1}\rightarrow2_{2,0}$) lines, SO 
and H$_2$CO, SO and CH$_3$OH, SO and NS, NS and CH$_3$OH, and NS and H$_2$CO.
In the case of fitting two different molecules, we allowed the abundance of one
molecule to vary in order to enable fitting to converge with a single disk mass.

\subsubsection{Modeling Results}

We present the results from the model fitting without including the envelope in Figure \ref{linefit}.
The disk-only models find similar protostar mass and disk radii as the models that include an
envelope;  the model results that include an envelope are presented in the Appendix.
Figure \ref{linefit} shows the results from fitting three H$_2$CO lines
and the SO line; we also show the results from CH$_3$OH and NS when 
simultaneously fit with H$_2$CO. The fit parameters for each line or combination
of lines are given in Table 5. The best fitting values are determined
from the median of the posterior distribution for each parameter and the uncertainties
reflect the standard deviation of the posterior distribution. In calculating the
best fitting values and their uncertainties, we filtered outliers
from the posterior distributions by rejecting walkers that did not conform 
to a $\xi^2$ distribution. These outliers were typically walkers that never converged to the 
best fitting values and typically represent less than 5\% of the 200 walkers used. To avoid
filtering too aggressively, we relaxed the filtering criteria such that we included at 
least 95\% of the walkers in the final statistics.

As can be seen in Table 5, there is variation 
in the best fitting protostar mass and
disk radius, depending on the molecule(s) being fit.
We note that the uncertainties listed in Table 5 are smaller than the differences between
best-fit values for the different molecules fit; 1$\sigma$ uncertainties
are shown, but differences are even in excess of 3$\sigma$ uncertainties. 
The variation of best fit parameters can reflect both the adopted model 
not being fully representative of the prototstellar disk and envelope, 
as well as the molecular line emission from the various
molecules tracing different spatial extents. Differences in the spatial distribution of
molecular line emission have been found toward several protostars on comparable spatial
scales \citep{sakai2014,yen2014}. The models assume a constant molecular abundance with
radius and if this is not true, systematic differences between modeling of different
molecules can arise.

Averaging the 12 independent fits 
we find an average protostar mass of 2.5$\pm$0.15~\msun.
The full range of best fitting protostar masses is
between 1.8 and 3.6~\msun, with the two highest 
and lowest masses being quite different from the other 10 fits, which
are much closer to the average value.
The best fitting disk radii are between 70 and 121~au. Averaging the 12 
independent fits we find an average radius of $\sim$94$\pm$13~au. The
uncertainties of the average are calculated using the median absolute deviation (MAD) from the
collection of 12 model fits, scaled such that the MAD would correspond 
to one standard deviation of a Normal distribution. Other average fitting parameters 
are V$_{lsr}$~=~11.1$\pm$0.04~\kms, PA~=~352.7$\pm$1.4\degr, 
$\gamma$~=~0.92$\pm$0.14, and the position offsets are also very small, typically $\sim$0\farcs01.
We discuss the other fitted parameters and the possible ramifications of 
various assumptions for the modeling in the Appendix.

We compare our best fitting protostar mass of $\sim$2.5~\msun\ to the PV diagrams
in Figure \ref{pvdiagrams}. We overlaid a Keplerian rotation curve for a mass of 2.5~\msun\ at an 
inclination of 72.2\degr (see Section 5.2). The Keplerian rotation curves 
encompass the high-velocity emission, indicating that the average fitted mass
of 2.5~\msun\ is quite consistent with the observed data. Thus, 
the measured protostar mass clearly confirms that HOPS-370
is an intermediate-mass protostar.

\subsection{Dust Continuum Modeling}

To determine the physical structure of the disk around HOPS-370, the dust continuum
emission must also be modeled using radiative transfer, but using a more realistic treatment
of the disk and envelope temperature structure than the molecular line modeling employed.
 We model the disk and envelope around HOPS-370,
also with \textit{pdspy} and RADMC-3D, but taking advantage of the radiative equilibrium mode where the photons are 
propagated from a central luminosity source and the temperature structure of the disk
and envelope are calculated self-consistently. This mode is much more computationally intensive that using
prescribed temperature profiles, as such we could not make use of this mode for the molecular line modeling.
We follow the methodology
employed by \citet{sheehan2017}, which utilizes the \textit{pdspy} package, employing the 
same underlying MCMC sampling of the parameter space as the molecular line 
fitting, using 200 walkers.

The dust continuum and SED modeling within \textit{pdspy} adopts the
same envelope physical model as the molecular line modeling. The principal
differences, however, are in the temperature and density structure of the disk. 
The volume density structure of the disk is defined as 
\begin{equation}
\rho(r) = \left(\frac{r}{r_c}\right)^{-\alpha}\frac{\Sigma(r)}{\sqrt{2\pi}\ h(r)} \, \exp\left(-\frac{1}{2}\left[\frac{z}{h(r)}\right]^2\right),
\end{equation}
where $\Sigma(r)$ is defined in Equation 3, $\alpha$ is the power-law index of the disk
volume density profile, $z$ is the height above the disk
midplane in cylindrical coordinates, and $h(r)$.
The vertical density structure for the disk in the continuum model
is not defined by hydrostatic equilibrium, but parameterized as
\begin{equation}
h(r) = h_0 \left(\frac{r}{1 \text{ AU}}\right)^{\beta}.
\end{equation} 

Here $\beta$ refers to the power-law index that defines how the vertical 
scale height of the disk varies with radius in the disk, not the
power law index of the dust opacity curve. The power-law index of the
disk surface density profile is equivalent to $\gamma$~=~$\alpha$~-~$\beta$, resulting
from the multiplication of the volume density profile and the vertical density profile.
The temperature structure for the continuum model is not prescribed as it 
is for the molecular line model. The temperature of the dust is set by the 
radiative equilibrium calculation performed by the RADMC-3D code.
The disk dust properties are taken from \citet{woitke2016} where the
 maximum size of the dust grains is parameterized as $a_{max}$ 
and the power-law index of the dust grain size
distribution, $p$, ($n(a)~\propto~n^{-p}$), both of which are free parameters; 
$a_{min}$ is fixed to be 0.05~\micron. The envelope dust properties
are taken from \citet{pollack1994} with $a_{min}$=0.005~\micron, 
$a_{max}$=1~\micron\ and $p$=3.5.

We simultaneously fit the 0.87~mm continuum emission 
in the uv-plane and the SED from the near-infrared to the millimeter. 
The parameter space explored for the dust continuum is significantly larger than that of
the molecular line kinematic modeling because now we fit
the emission of the envelope, disk, and the overall SED of the system.
Our goodness of fit metric is a weighted $\chi$$^2$ where
\begin{equation}
\chi^2 = \omega_{0.87mm,vis}\chi_{0.87mm,vis}^2 + \omega_{SED}\chi_{vis}^2.
\end{equation}
The terms $\omega_{0.87mm,vis}$
and $\omega_{SED}$ are determined empirically
to be 0.2 and 1.0. The $\chi$$_{vis}$ are calculated by directly comparing the real and 
imaginary visibility components between the data and the model; the comparison is done with the
two dimensional visibility data and not using azimuthally averaged one dimensional 
profiles, such profiles are only used for visual comparison of the models and data.

The best-fitting models compared to the data are shown in 
Figures \ref{contvis} and \ref{contimage}. The circularly
averaged visibility amplitude profile at 0.87~mm demonstrates that the model fits the 
0.87~mm visibility data quite well at uv-distances greater than $\sim$50~$k\lambda$.
Also plotted in Figure \ref{contvis} is the estimated contribution of just the disk alone to
the visibility amplitude profile, this shows that the envelope contributes significantly
to the emission at uv-distances less than 300~k$\lambda$. The fit to the SED, also shown 
in Figure \ref{contvis}, is not perfect, but is a close approximation to the
the shape and flux densities of the SED. Some flux density points are over or under predicted,
but the relatively low angular resolution of \textit{Spitzer} and \textit{Herschel} at wavelengths
longer than 10~\micron\ makes it difficult to construct a SED that only includes emission from
HOPS-370. However, the fact that it is the most luminous protostar within an arcminute means that
contributions from other sources will not have an extremely negative impact on the SED. But,
extended emission at wavelengths longer than 100~\micron can cause the luminosity to be over
estimated at those wavelengths.

We show an image-plane comparison of the data, model, and residuals in Figure \ref{contimage}.
The residual images are generated from the residual visibility amplitudes and not an 
image-plane subtraction. The main disk feature in the continuum is well-modeled and removed
from the residual image, but the 0.87~mm residual image does show structure
around the protostar that is not captured in the model. There are clear over-subtractions
in the outer disk and center, while there are under-subtractions within the disk as well.
This residual emission may represent complexities in the disk and inner envelope density structure
that are not reflected in our physical model.
 We note that in the 0.87~mm residual map there is compact emission southeast of the protostar 
(-0\farcs4, -0\farcs3), but its nature is unclear and could stem from heating along
the outflow cavity wall.

We list the parameters and their best fitting values for the disk of 
HOPS-370 in Table 6, but discuss the most relevant parameters here, which are
the disk radius, disk mass, 
surface density power-law index, and luminosity. The best fitting disk mass is
0.035~\msun, which is somewhat lower than the mass calculated under the assumption of
optically thin emission and an average temperature of 131~K. However, the maximum
dust grain size fit by the modeling is 440~\micron, so the dust of the model will
emit much more efficiently at 0.87~mm due to its opacity than the \citet{ossenkopf1994} 
dust adopted for the simple mass estimate. We do note
that the temperature of the disk in the model fit at a radius of 100~au is $\sim$122~K 
(Figure \ref{sigmaTQ}), which implies that the disk average temperature is
comparable to the value estimated from the bolometric luminosity.
The luminosity of the protostar is fit to be 276~L$_{\sun}$, which is a bit less than 
L$_{bol}$ of 314~L$_{\sun}$. The more extended emission at wavelengths longer than
100~\micron, which is not fit well by the model SED, could result from warm dust surrounding the protostar and 
may cause the bolometric luminosity to be over-estimated.

The best fitting envelope mass is 0.12~\msun\ with a radius of $\sim$1900~AU.
However, the poor sampling of short uv-distances will limit the 
robustness of the envelope model fit to the ALMA visibility data, despite the
additional constraints from the SED. This is because the SED shortward of 100~\micron\ 
is most sensitive to the inner envelope density and not the overall mass or radius. The overall
mass and radius can affect the longer wavelength data more, but there is a degeneracy between
dust temperature and mass. Thus, it is possible
that the total mass and radius of the envelope is not accounted for in our model fit. 
The full outflow cavity opening angle is fit to be $\sim$98\degr, as computed from
the value of $\xi$ in Table 6. While this may seem large at first glance, it appears
comparable to the width of the outflow cavities near the protostar  viewed in low-velocity 
$^{12}$CO emission and shown in \citet{tobin2019}. However, the shape of the full outflow cavity
extent may be more parabolic, meaning that the apparent opening angle
at larger radii will appear smaller.

The disk radius from the continuum fit, 62.1~au, is smaller than our estimate
of the radius from Gaussian fitting; however, Figure \ref{sigmaTQ} does show that
the fitted disk surface density is $\sim$1.3~g~cm$^{-2}$ at 100~au because the 
exponentially-tapered disk extends beyond the critical radius $r_c$. The continuum
disk radius is also comparable to the range of exponentially-tapered disk radii 
fit with the molecular line modeling. However, it is known
that the gas disk tends to be larger than the dust disk from Class II 
disks \citep{ansdell2018}. This is thought to be caused by radial drift of dust particles due to 
gas drag experienced because the gas orbits the star at slightly sub-Keplerian velocities \citep{weidenschilling1977}.
Thus, the dust in protostellar disks may also experience radial drift \citep{birnstiel2010}, which
would cause a disagreement between the dust and gas disk radii.

While the disk outer radius is reasonable, there are some peculiarities with the
disk structure. The surface density profile 
increases with radius as $\Sigma$~$\propto$~$r^{0.47}$. However, this may result 
from the high opacity of the disk and its high inclination, leading to a sub-optimal
model fit. Moreover, given that the steepness
of the exponential cutoff depends on $\gamma$, the smaller $\gamma$
leads to the disk surface density to fall off more quickly. Thus, the best fitting $\gamma$ 
may tell us more about the sharpness of the disk's `edge' than the surface density profile.
The negative residuals from over-subtraction in Figure \ref{contvis} could indicate that the
disk needs a sharper cutoff than the exponential taper provides.
In addition to $\Sigma$ increasing with radius, the inner disk radius is fit to be 0.51~au.
This radius slightly smaller than the radius at which dust is expected to be destroyed, where T$\sim$1400~K
which occurs at $\sim$1~au in our model.

The disk vertical height is also not highly flared, the best fitting in disk scale height
with radius is $h(r)$~$\propto$~$r^{0.66}$, normalized to 0.131~au at 1~au, which
indicates that the disk height increases slowly with radius. This fit likely reflects 
the large grain population of the disk and may not accurately reflect the 
disk properties of the smaller grains and/or gas disk.

\section{Discussion}

The ALMA and VLA continuum and molecular line emission are reshaping our 
understanding of HOPS-370/OMC2-FIR3 by providing a detailed view of the disk and 
jet toward this candidate intermediate-mass protostar.
While the \tbol\ of 71~K measured from the SED suggests that this is a Class I 
protostar \citep{furlan2016}, its location near the canonical border of 70~K between 
Class 0 and I indicates that HOPS-370 is very much a young, embedded protostar.

The continuum images 
indicate that there are no resolved companions within 1000~AU, and the nearby 
infrared sources mentioned in Section 3 do not appear to be embedded within
the envelope of HOPS-370. Thus, HOPS-370 may have been the 
result of a single core collapsing within
the OMC2 region, and its seemingly well-ordered disk and outflow morphology makes
it an ideal system for characterizing the early evolution of an intermediate-mass protostar.

HOPS-370 also appears to have a profound influence on the
environment of the surrounding protostars. For instance, there is strong evidence 
that its outflow is interacting with the ambient cloud and the 
OMC2-FIR4 clump, which is associated with HOPS-108 and at least six other
protostars \citep{shimajiri2008,lopez-sepulcre2013,furlan2014,gonzalez2016,osorio2017,tobin2019}.
The shocks associated with the interaction are strong enough to emit non-thermal
synchrotron emission \citep{osorio2017}, and it is the brightest known far-infrared
line emitter in Orion outside of the Orion Nebula itself \citep{manoj2013}.

The dust continuum emission toward HOPS-370 indicates a large disk around the protostar.
The inferred disk radius from a Gaussian fit to the dust continuum of 113~au 
is in excess of the disk radii around most Class 0 and I 
protostars \citep{harsono2014,yen2017,tobin2020}. Even if one only considers the 
62~au radius from modeling, it is still larger than the mean disk radii for protostars in Orion; 
the mean disk radii for Class 0 and I protostars in Orion are 
$\sim$45~au and $\sim$37~au, respectively \citep{tobin2020}. However, it is not clear
if intermediate-mass protostars typically have larger disks given that \citet{tobin2020}
observed no correlation between disk radius and bolometric luminosity.

\subsection{Protostellar Mass and Accretion Rate}

Stellar mass is the most fundamental property of a stellar system, 
given that the entire evolution of a star is determined by its mass. 
Thus, the protostar mass measurement of $\sim$2.5~\msun\ for HOPS-370
solidifies its status as an intermediate-mass protostar
still in the early stages of formation. While there is some uncertainty 
in the mass when fitting the molecular line emission for kinematics
of different molecules, the differences are most likely
systematic because SO, CH$_3$OH, NS, and H$_2$CO do not trace the same material
in the disk and inner envelope, meaning that their abundance profiles in the radial and
vertical directions are not equivalent. 

We can see this particularly in the SO and NS line emission, which are
more compact than the H$_2$CO ($J=3_{0,3}\rightarrow2_{0,2}$) line
emission (Figures \ref{moment0} and \ref{pvdiagrams}), with 
H$_2$CO ($J=3_{0,3}\rightarrow2_{0,2}$) emission extending
to slightly higher velocities on the blue-shifted side. The masses from fitting are
between 1.8 to 3.6~\msun;  however, considering the full range of masses fit the mean is $\sim$2.5~\msun\ with
a fractional uncertainty of $\sim$7\%.

To determine the mass accretion rate, we use the equation for accretion luminosity
\begin{equation}
L_{acc}~=~\frac{GM_{ps}\dot{M}}{R_{ps}},
\end{equation}
where $G$ is the gravitational constant, $M_{ps}$ is the protostar mass, $\dot{M}$ is the mass
accretion rate from the disk to the protostar, and $R_{ps}$ is the protostar radius. Thus, 
$M_{ps}$ is now the most highly constrained of the parameters needed to determine $\dot{M}$.
\citet{palla1993} calculated pre-main sequence evolution of intermediate-mass stars, indicating
that the luminosity of the protostellar object itself ($L_{*}$) will be less than $\sim$10~\lsun\, meaning
that the vast majority of the \lbol\ = 314~\lsun\ will be dominated by luminosity from 
mass accretion. We assume simplistically that 
\begin{equation}
L_{acc}~\approx~L_{tot} - L_{*}.
\end{equation}
\citet{palla1993} also calculated that the likely stellar radius for a 
protostar like HOPS-370 is $\sim$5~R$_{\odot}$. We note that these protostellar 
stellar birthline models require assumptions that are still subject to debate, the effect
of accretion in particular, so there is uncertainty in the most appropriate stellar 
radius to assume for a given mass \citep{baraffe2009,hosokawa2010}.

 If assume \lbol\ $\approx$ $L_{tot}$, then the mass accretion rate
from the disk to the protostar is likely 2.25$\times$10$^{-5}$~\msun~yr$^{-1}$.
This value compares well to the outflow rate of $\sim$2.3$\times$10$^{-6}$~\msun~yr$^{-1}$
calculated from the [OI] jet emission by \citet{gonzalez2016} if one assumes that the outflow
rate is $\sim$10\% of the accretion rate \citep{shu1994,frank2014}. 
Due to beaming of the luminosity and foreground extinction, the actual $L_{tot}$ may differ
from \lbol\ \citep{whitney2003a,offner2012}. Our model fit to the continuum visibilities and SED provides a
luminosity of 276~\lsun, which is slightly lower than \lbol, implying an accretion rate
of $\sim$1.7$\times$10$^{-5}$~\msun~yr$^{-1}$. The SED fit by \citet{furlan2016} provides
a much higher estimate of the luminosity at 511~\lsun, implying an accretion rate of 
$\sim$3.2$\times$10$^{-5}$~\msun~yr$^{-1}$.
Another SED modeling effort by \citet{adams2012} inferred a luminosity of 300~\lsun, 
which would imply that the accretion luminosity is comparable to the bolometric luminosity.
In summary, the mass accretion rate appears to be well constrained to be
between 1.7$\times$10$^{-5}$~\msun~yr$^{-1}$ and 3.2$\times$10$^{-5}$~\msun~yr$^{-1}$; 
for comparison, these accretion rates are $\sim$1000$\times$ the typical accretion
rates found in low-mass T Tauri stars \citep[e.g.,][]{ingleby2013,acala2017}.

The inferred infall rate from the envelope to the disk is
3.2$\times$$10^{-5}$~\msun~yr$^{-1}$ from our best fitting model. This
is very comparable to previous estimates derived from SED fitting of 
2.2$\times$10$^{-5}$~\msun~yr$^{-1}$ \citep{furlan2016}
and 4.4$\times$10$^{-5}$~\msun~yr$^{-1}$ \citep{adams2012}, both of which 
are scaled to reflect the protostar mass of 2.5~\msun.
The inferred range of accretion rates from the disk to the protostar
are comparable to the envelope to disk infall rates derived from SED fitting and our best-fitting
model. This indicates that the envelope is supplying the disk
with mass at a comparable or higher rate as compared to how rapidly the disk material
drains onto the protostar.

All of the aforementioned analytic models, however, assume simplified geometries 
for the envelope, disk, and outflow cavity. The differences in the parameters from different models
suggests that the underlying physical models may not accurately describe 
the true structure of the protostellar system. But, the analytic models 
ignore the potential effects of turbulence and magnetic pressure support, which
may also play a role in regulating the infall from the envelope to disk \citep[e.g.,][]{li2014,seifried2013}.
For this reason, the mismatches in the infall and accretion rates from different model fits 
could arise from physical model inadequacies, and degeneracies due to only fitting the SED in
some cases. Finally, while we are quoting the infall rates derived from these models, the precise
numbers should be regarded with caution for the reasons outlined in this paragraph.

\subsection{Importance of Disk Self-gravity}

Based on our knowledge of the disk mass and protostar mass from modeling, we can estimate
how important self-gravity is to the HOPS-370 disk using Toomre's $Q$ and its
relationship to the disk and protostar mass
\begin{equation}
\label{eq:qapprox}
Q \approx 2\frac{M_{ps}}{M_d}\frac{H}{r}.
\end{equation}
Note that this is not the conventional representation for Q, but is rewritten for
a rotationally supported disk around a central gravitating body, in our case
a protostar \citep{kratter2016,tobin2016b}.
The disk scale height is $H$ (equivalent to $c_s/\Omega$ where 
$c_s$ the disk sound speed and $\Omega$ the Keplerian angular velocity),
$M_{ps}$ is the protostar mass, and $M_d$ is the disk mass. We calculate $Q$ at a radius of 50~AU and
find that $Q$$\sim$6, assuming the inferred disk gas mass from modeling (0.035~\msun), protostar mass ($\sim$2.5~\msun),
and a typical disk temperature of $131$~K ($c_s$~=~0.56~\kms) at a 
radius of 50~AU ($\Omega$~$\sim$~8.9$\times$10$^{-10}$ for M$_*$=2.5~\msun). 
Thus, the disk around HOPS-370
is not expected to be highly self-gravitating. We note that Q could be lower if 
the disk mass was higher, but the disk mass would have to be significantly underestimated
for Q to approach 1.

We also calculated $Q$ using the temperature and density structure
derived from the best fitting radiative transfer model to the continuum data.
We extracted the disk surface density and midplane temperature profile 
from the best fitting radiative transfer model (plotted
in Figure \ref{sigmaTQ}). Then using the surface
density and temperature, we calculated $Q$ as a function of radius using
\begin{equation}
\label{eq:qexact}
Q = \frac{\Omega(r) c_s(r)}{\pi G \Sigma(r)}.
\end{equation}
where $\Sigma$ is the radial surface density profile and $G$ is the gravitation constant.
Figure \ref{sigmaTQ} shows $Q$ as a function of radius. The radial distribution of Q
shows that the disk does not approach instability at any radius and the minimum
value of $Q$ is $\sim$14, even larger than the approximate calculation
from Equation \ref{eq:qapprox}. This further demonstrates that 
self-gravity is not likely important in the disk of 
HOPS-370 at the present time.

\subsection{Constraints from the Near-infrared Spectrum}

The near-infrared spectral features were presented in Section 4.4, where the principal
features of importance with respect to the accretion rate and luminosity of the protostar
are the Br~$\gamma$ line emission and CO band head emission. \citet{connelley2010} found
a strong correlation between Br~$\gamma$ emission, veiling, and CO band head emission. The
authors inferred that when Br~$\gamma$, CO band heads are in emission, and veiling is
high, that mass accretion is also high. Modeling of the CO band head spectra for different
stellar types and accretion rates by \citet{calvet1991} predicted when CO band head emission 
should appear in absorption or emission for a given mass accretion rate and stellar
effective temperature. For example, very high accretion rate systems, like FU Ori systems,
have CO band head absorption due to the hot accreting disk midplane and cooler 
disk atmosphere.

For the accretion rate needed to produce the observed
bolometric luminosity with the expected stellar temperature
of $\sim$4500~K \citep{palla1993} for HOPS-370, the models of \cite{calvet1991} indicate
that CO band head absorption should be expected rather than the observed
emission. On the other hand, \citet{najita1996}
suggested that high accretion rates could still lead to CO band head emission because
accretion through the disk could cause a temperature inversion in the inner disk, making the surface hotter 
than the midplane, leading to emission. In addition, disk accretion is not the only
possible mechanism for producing CO band head emission; other studies suggest
that winds could also produce the emission \citep[e.g.,][]{chandler1995}, 
but higher spectral resolution and S/N is required to differentiate between a wind 
and disk origin.

While the exact origin of the CO band head emission in HOPS-370 is uncertain, 
the CO band head and Br $\gamma$ emission demonstrate that despite the high inferred
accretion rate from the luminosity, the spectrum of HOPS-370 is decidedly not FU Ori-like
since FU Ori-type spectra have no detectable Br~$\gamma$ emission and
the CO band heads are in absorption \citep{connelley2010,fischer2012}. Br~$\gamma$
line luminosity has been used to infer the accretion luminosity of young stars, including
protostars with relationships defined by \citet[][log$_{10}$($L_{acc}$) = 1.26~log$_{10}$($L_{Br\gamma}$) + 4.43]{muzerolle1998}. 
Applying this relationship, our inferred $L_{acc}$ for HOPS-370 from Br~$\gamma$ emission is 2.2$\times$10$^{-2}$~\lsun,
which is at odds with the accretion luminosity inferred in section 6.1
by a factor of 10000. The Br~$\gamma$ line luminosity 
has not been corrected for attenuation by dust extinction, and correction will only lead to 
higher accretion luminosities. Furthermore, the Br~$\gamma$ emission we detect is
from scattered light, and radiative transfer models from 
\citet{whitney2003b} indicate that the amount of emergent flux 
at 2~\micron\ in the outflow cavities can be between 2 to 4 
orders of magnitude lower than the input stellar spectrum. Therefore, it is plausible
that the Br $\gamma$ emission we observe is originating from accretion, but it is difficult
to accurately infer an accretion rate from the line emission due to the combined effects of
extinction and observing the spectrum in scattered light.

\subsection{Comparison to Other Protostars with Measured Masses}

The most current compilation of protostar masses is found in \citet{yen2017}, which contains
protostar masses measured by ALMA, the Plateau de Bure Interferometer, and the Submillimeter Array. 
HOPS-370 is one of the most massive
protostars to have a kinematic mass measurement. Comparable protostars
are HH111 MMS (1.8~\msun), Elias 29 (2.5~\msun), R CrA IRS7B (2.3~\msun), 
Oph IRS 43 (1.9~\msun), and L1489 IRS (1.6~\msun). However, these are all Class I
protostars, except for R CrA IRS7B, which is a borderline Class 0/I protostar, and 
HOPS-370 is the only one with \lbol~$>$~100~\lsun. Of these protostars with comparable
masses, the highest luminosity system is HH111 MMS at 17.4~\lsun. 
Thus, HOPS-370 is unique and requires a very high accretion luminosity 
to explain its \lbol.
The accretion rate for HOPS-370 inferred from the luminosity is greater than an order of 
magnitude larger than the other protostars with a similar
 mass, and it has a higher inferred accretion rate that all other protostars
listed in \citet{yen2017}.

\subsection{The Nature of Accretion in HOPS-370}

The high rate of accretion in HOPS-370 begs the question, is this an outbursting source
or a higher-mass star being formed through sustained infall from the envelope to disk?
Outbursts and variability seem to be common among protostars
\citep[e.g.,][]{hartmann1996,safron2015,fischer2017,dunham2010,audard2014} which are thought to reflect
higher accretion rates than average for short intervals of time (100s to 1000s of years).
However, the study of intermediate-mass star formation has been complicated by multiplicity for systems
other than HOPS-370. We will discuss the merits of the two possible accretion scenarios for HOPS-370.

\subsubsection{An Outbursting Protostar in a High Accretion State?}

The mass accretion rate from the disk to the protostar needs to be sustained by
some mechanism that transports angular momentum. At different radii in the disk,
different processes may be required to transport angular momentum, which will dictate how
rapidly the disk can transport mass inward. If the disk is sufficiently massive, gravitational
instability can transport angular momentum \citep[e.g.,][]{adams1989,zhu2012}, and
when the disk mass is low, disk winds could transport angular momentum 
\citep{pudritz1983,konigl2000} or turbulent viscosity \citep{ss1973}.
 Thus, mass accretion
may require two mechanisms to be active in different regions of the disk at different
times to produce the estimated high accretion rate from the disk to protostar. Such a scenario has been
proposed by \citet{zhu2009} to explain FU Ori outbursts where gravitational instability
transports mass to the inner disk and mass builds up until the Magneto-Rotational Instability \citep[MRI,][]{balbus1991,gammie1996}
is triggered causing rapid accretion from the inner disk to the star.
In this context, the mass accretion rate of HOPS-370 does not need to be constant at its current rate.

While the disk is currently gravitationally stable, under the assumption that HOPS-370
is in a high-luminosity state, it could have previously had a much lower luminosity. Since
$Q$ scales $\propto$ $c_s$, which is $\propto$~T$^{0.5}$ and T~$\propto$L$^{0.25}$, $Q$ is thus $\propto$ L$^{1/8}$.
Therefore, if HOPS-370 in a low accretion state had a luminosity 100$\times$ lower, $Q$ would be
reduced by a factor of 1.77, reducing the $Q$ in the outer disk from $\sim$14 to $\sim$9. 
Thus, even with a lower temperature, the disk around HOPS-370 does not
have enough mass for gravitational instability to be important. Thus, unless the disk
mass is severely underestimated or the mass was much higher in the past,
 the scenario of an outburst triggered by clump accretion resulting
from disk fragmentation \citep{vorobyov2006,dunham2014b} is not highly compelling.

\subsubsection{Sustained Accretion?}
The alternative is that we are not witnessing an outburst, but sustained high accretion rates
that could be typical for the formation of an intermediate-mass protostar. Models of
intermediate to high mass star formation \citep[e.g.,][]{wuchterl2003,mckeetan2003}
predict the luminosities over time during the formation of such systems. While
these analytic models do not assume accretion through a disk, rather direct infall from the 
envelope on to the protostar, they demonstrate a scenario in which a protostar
system that will ultimately form a high-mass star will have a significantly higher overall luminosity
during its formation, as compared to the stellar luminosity 
of the central protostar due to a sustained high accretion accretion rate. Moreover, competitive
accretion models for protostars forming within clustered environments, not unlike the OMC2 region,
also predict that the protostars that ultimately have higher final masses will accrete
at higher rates \citep{bonnell2001,hsu2010,bate2012}.

The estimates of the accretion rate from the disk to the protostar, based on the bolometric
luminosity and protostar mass, range from 1.7$\times$10$^{-5}$~\msun~yr$^{-1}$ to 
3.2$\times$10$^{-5}$~\msun~yr$^{-1}$. Then the estimates of the infall rate from the 
envelope to the disk range between 2.2$\times$10$^{-5}$~\msun~yr$^{-1}$ to 
4.4$\times$10$^{-5}$~\msun~yr$^{-1}$. Thus, the disk is being fed with material rapidly
enough that accretion could be sustained at its high rate, regardless of the spread in the
estimates. However, the infall rates are model-dependent and their accuracy is more questionable
than the disk to star accretion rates. 

These high accretion rates for HOPS-370 indicate that it could 
gain another solar mass of material within $\sim$3$\times$10$^4$ yr to $\sim$9$\times$10$^4$ yr.
But, the envelope surrounding HOPS-370 may not have sufficient mass to sustain high
accretion indefinitely. The envelope in our model fit only had $\sim$0.12~\msun\ of material
out to $\sim$1900~au. However, our modeling does not account for
the total mass available in the envelope; other observational measurements find $\sim$2.5~\msun\ from 3~mm 
continuum data \citep{kainulainen2017}. Even if the immediate surroundings of HOPS-370 have
limited mass, it is embedded within the dense molecular environment of the northern
integral-shaped filament. This means that the total reservoir that HOPS-370 could
accrete from is larger than what is larger that the envelope mass fit from modeling.

Even though we expect the disk to be gravitationally stable, material must still accrete through
the disk with a sustained high accretion rate that keeps
the luminosity around the present value $\sim$300~\lsun. 
If the disk self-gravity is negligible, then gravitational instability
cannot efficiently transport angular momentum anywhere in the disk
\citep[e.g.,][]{rice2010}. Even if disk self-gravity is not large enough to drive accretion,
there are alternative ways to promote the accretion of material through
the disk toward the protostar. One such way is the excitation of spiral density
waves in the disk due to infalling material \citep{lesur2015}. In this scenario,
the infalling material has lower specific angular momentum than the disk at the
point where material arrives to the disk. This creates an unstable accretion shock
that promotes the formation of spiral arms and outward angular momentum transport,
enabling efficient accretion from the outer disk to the inner disk.

While the MRI or some other viscous process may be responsible for accretion between the inner and outer
disk, the exact mechanism remains debated. However, the challenge for MRI is 
to have enough ionization such that the magnetic field is coupled strongly enough to the gas
to transport angular momentum. Work by \citet{offner2019} has shown that cosmic rays produced
by accretion could increase the ionization enough to enable
MRI in the inner $\sim$10~au of the disk. Assuming that accretion proceeds due to turbulent
viscosity, we can use the observationally inferred accretion rate to constrain the necessary
inner disk properties using the equation
\begin{equation}
\dot{M} \sim 3\pi \alpha c_s H \Sigma,
\end{equation}
following \citet{hartmann2008}. The term $\alpha$ refers to the $\alpha$-viscosity 
of the disk \citep{ss1973},  $\Sigma$ is the surface density of the disk a
the radius in question, and the other parameters are as defined previously. Rearranging these
terms to solve for $\Sigma$ at 1~au with $\dot{M}_{acc}$ = 
1$\times$10$^{-5}$~\msun~yr$^{-1}$, we adopt $\alpha$=0.1 which is 
typical for a high accretion rate \citep{zhu2009}, 
$c_s$ = 1.86$\times$10$^{5}$~cm~s$^{-1}$ (from T~=~2000~K),
$H$=8.98$\times$10$^{11}$~cm (from the continuum model, Table 6). With these values,
we find that $\Sigma$ at 1~au should be $\sim$3700~g~cm$^{-2}$ for the disk to accrete
at 1$\times$10$^{-5}$~\msun~yr$^{-1}$. Our best fitting \textit{pdspy}
continuum model only has a surface density at 1~au of $\sim$4.6~cm$^2$~g$^{-1}$, which is inconsistent with the need
for a high surface density in the inner disk to facilitate accretion.
Even if we extrapolate the maximum disk surface density of 19~g~cm$^{-2}$ at 40~au to 1~au 
assuming $\gamma$=1, we would find a surface density of $\sim$760~g~cm$^{-2}$, several times lower than 
the value needed to sustain accretion at $\sim$1$\times$10$^{-5}$~\msun~yr$^{-1}$. Thus, this can be taken
as further evidence that the surface density profile of our best-fit disk model is inconsistent with 
other characteristics of the system and is likely not well-constrained from the current model. Furthermore, 
the disk surface density depends on the dust grain size distribution and if the dust opacities are too
high (the best fitting a$_{max}$ $\sim$440.4~\micron) then the surface density will be too low. With only a single
wavelength, a$_{max}$ can have a degeneracy with disk mass.  Despite the uncertainties in the 
underlying density structure of the disk model and dust opacity, the surface density would need to be 
an order of magnitude larger than currently observed for self-gravity to be important in the disk of HOPS-370.
It is unlikely that the true surface density could be so large and fit the observed data.

Disk winds have been promoted as an alternative mechanism to promote accretion in 
the inner disks of young stars \citep[e.g.,][]{pudritz2007,bai2013}. However, it is unclear if
disk winds can promote accretion at rates as high as $\sim$10$^{-5}$~\msun~yr$^{-1}$.
This is because disk winds can only be active in a thin, ionized layer in the disk and
the overall surface density of this layer is significantly lower than the overall disk surface
density.

The peculiarities of the inner disk fit aside, it seems plausible that HOPS-370 
could be forming an intermediate-mass star from steady accretion supplied from the 
envelope to disk and the disk to the star, without
an outburst being necessary.

\subsection{Prospects for Planet Formation in the Context of Disks Around Herbig Ae/Be Stars}

The current protostar mass of HOPS-370 is comparable to that of disk-hosting Herbig Ae/Be stars, which 
are intermediate-mass pre-main sequence stars, with masses between $\sim$1 to $\sim$4~\msun. 
Herbig Ae/Be systems typically have disk radii that
are comparable to or larger than that of the HOPS-370 disk. Thus, HOPS-370 could be a 
progenitor of these Herbig Ae/Be systems if the protostar does not 
grow significantly larger in mass and a reasonably massive disk remains for $\sim$few~Myr
once infall from the envelope has stopped.

\citet{kama2020} examined the disk masses of Herbig Ae/Be 15 systems, which have
stellar masses ranging between $\sim$ 1.5 to 3.0~\msun using the gas mass upper limits provided
by HD ($J=1\rightarrow0$) observations from the \textit{Herschel Space Observatory}.
The upper limits from HD were then interpreted in the context of dust disk masses measured from continuum
emission and gas masses measured using CO molecular line mission.
Many of these disks show ring or asymmetric structures in their dust
continuum emission that have been interpreted as signatures of giant planet 
formation \citep[e.g.,][]{isella2018,teague2018,pinte2018,keppler2018}.
However, the upper limits of the gas masses of these disks indicate that they are most likely gravitationally stable
at present. The projected planet masses require conversion of disk mass 
to planet masses at efficiencies $>$10\%, and
these efficiencies are high enough that the presence of these putative giant 
proto-planets are easier to explain if the formed earlier during the embedded 
protostar phase, possibly enabled by gravitational instability
\citep{tychoniec2020}. But, HOPS-370 in its current state with a high 
bolometric luminosity is gravitationally stable and is expected to still be stable
even if its luminosity was 10$\times$ lower.

Between the two possible mechanisms for giant planet formation, core accretion and gravitational instability, 
core accretion seems to be the most favorable in HOPS-370. Gravitational instability in the disk
of HOPS-370 is unlikely because the disk would have to remain massive, while the system luminosity decreased
significantly, presumably due to lower accretion. 
Giant planet formation via core accretion is also plausible; the cores could form during the protostar phase, 
and they would be able to begin gas accretion while the disk is still being fed with infalling material. 
Constant replenishment of the disk with gas and dust would enable proto-planets to grow to 
Jupiter masses. The efficiency of disk mass to planet mass is not required to be high because there will be 
a large mass flux onto and through the disk.

Furthermore, the current disk mass of $\sim$0.035~\msun\ and the current accretion rate from
the disk to the star of $\sim$10$^{-5}~$\msun~yr$^{-1}$ indicate that 
the entire gas disk could be accreted in less than 10$^4$~yr.
However, the properties of the
current disk are still relevant for the further evolution of 
solid material because the gas disk may be accreted, but the
solid material with in the disk may not be completely accreted 
with the gas. This is because pressure
bumps in the disk can trap large dust particles, promoting 
further dust growth \citep[e.g.,][]{pinilla2012}. It has been
demonstrated that infalling material to the disk can trigger 
Rossby Wave Instabilities and provide a mechanism 
for particle trapping \citep{bae2015}. 
Thus, while the disk will continue to evolve through the protostellar phase, its
current properties are not completely removed from the initial conditions of planet formation.

On the other hand, HOPS-370 may
continue to gain mass and become a higher mass star,
which could render it significantly different from the 
Herbig Ae/Be systems discussed here. Moreover, HOPS-370 is currently within
a relatively dense cluster of protostars, while the Herbig Ae/Be 
systems studied by \citet{kama2020} are much more
isolated. Thus, it is unclear if a long-lived disk around HOPS-370 would be able to survive a few Myr in a
clustered environment.

\section{Conclusions}

We have conducted ALMA, VLA, and near-infrared observations
toward HOPS-370 (OMC2-FIR3), obtaining an unprecedented view of the dusty and molecular
disk structure surrounding this intermediate-mass protostar in the Orion A molecular cloud. 
The lack of close multiplicity, the clear disk in continuum and molecular
line emission, and well-developed outflow establish HOPS-370 as a prototype
intermediate-mass protostar. Our specific results are detailed as follows.

\begin{itemize}

\item We resolved a clear disk in the dust continuum at 0.87~mm, 1.3~mm, and 9~mm
toward HOPS-370. From the 870~\micron\ continuum, where the disk is best resolved, 
the disk radius is measured to be $\sim$113~AU using Gaussian fitting, and 
we estimate the mass of the disk to be between 0.048 and 0.098~\msun\
using the dust continuum flux density at 0.87, 1.3, and 9.1~mm, assuming a gas-to-dust 
ratio of 100, and an average dust temperature of 131~K as indicated by its bolometric luminosity.

\item Rotation is detected in the disk around HOPS-370 in 
the \cateo, \thco, H$_2$CO, SO, NS, and CH$_3$OH molecular lines. Fitting 
the H$_2$CO, CH$_3$OH and SO channel maps using \textit{pdspy}
enabled us to measure a protostellar mass of $\sim$2.5~\msun\ and a rotationally-supported 
gas disk radius of $\sim$94~AU. The protostar mass likely has a systematic uncertainty of $\sim$6\%,
and the disk radius likely has a systematic uncertainty of $\sim$14\%. The range of 
gas disk radii fit are comparable to the best fitting disk radius from dust continuum modeling,
it is typical for the gas disk to extend to larger radii than the dust disk.

\item We also used \textit{pdspy} and RADMC3D to fit the 0.87~mm dust continuum emission from the 
disk and envelope to constrain their physical properties. We find a disk mass 0.035~\msun\
(assuming a dust to gas mass ratio of 1:100). The critical radius of
the dust disk is fit to be 62~AU; this is the radius at which the exponential decrease
in density begins. Since the disk extends beyond this radius, it is approximately consistent with the
radius derived from Gaussian fitting.

\item The best fitting protostar mass ($\sim$2.5~\msun) and
bolometric luminosity of 314~\lsun\ 
are used to infer a current disk to protostar accretion rate of 
$\sim$2.25$\times$10$^{-5}$~\msun~yr$^{-1}$. The luminosity of HOPS-370 and its accretion rate are
higher than other protostars with comparable masses by an order of magnitude, indicating
that HOPS-370 is in a phase of rapid build-up.

\item The near infrared spectrum from 2 to 2.4~\micron\ shows no photospheric
absorption in the medium resolution spectrum. However, we do detect
several H$_2$ emission lines, Br $\gamma$ emission, and the CO band heads in emission.
The Br $\gamma$ line emission appears to be highly attenuated given that the 
accretion luminosity inferred from  the line luminosity is orders of magnitude lower
than that inferred from the bolometric luminosity and estimated stellar luminosity.

\end{itemize}

The authors thanks the anonymous referee for useful comments that improved
the quality of th manuscript.
The authors also wish to acknowledge useful discussions with L. Hartmann
and J. Bae about the results.
JJT acknowledges support from grant AST-1814762 
from the National Science Foundation and past support 
from the Homer L. Dodge Endowed Chair at the University of Oklahoma.
GA, MO, and AKD-R acknowledge financial support from the State Agency for Research 
of the Spanish MCIU through the AYA2017-84390-C2-1-R grant (co-funded by FEDER) and 
through the ``Center of Excellence Severo Ochoa'' award for the Instituto de 
Astrof\'\i sica de Andaluc\'\i a (SEV-2017-0709). 
ZYL is supported in part by NASA 80NSSC18K1095 and NSF AST-1716259.
MK gratefully acknowledges funding by the University of Tartu 
ASTRA project 2014-2020.4.01.16-0029 KOMEET "Benefits for Estonian 
Society from Space Research and Application", financed by the 
EU European Regional Development Fund. 
MLRH acknowledges support from the Michigan Society of Fellows.
This paper makes use of the following ALMA data: ADS/JAO.ALMA\#2015.1.00041.S and
ADS/JAO.ALMA\#2017.1.00419.S.
ALMA is a partnership of ESO (representing its member states), NSF (USA) and 
NINS (Japan), together with NRC (Canada), NSC and ASIAA (Taiwan), and 
KASI (Republic of Korea), in cooperation with the Republic of Chile. 
The Joint ALMA Observatory is operated by ESO, AUI/NRAO and NAOJ.
The National Radio Astronomy 
Observatory is a facility of the National Science Foundation 
operated under cooperative agreement by Associated Universities, Inc.
These results
are based on observations obtained with the Apache Point
Observatory 3.5-meter telescope which is owned and operated 
by the Astrophysical Research Consortium (ARC). We wish to
thank the APO 3.5m telescope operators in their assistance
in obtaining these data. Access to the APO 3.5m telescope
is funded by the University of Oklahoma and the Homer L. Dodge Endowed Chair.
This research made use of APLpy, an open-source plotting package for Python 
hosted at http://aplpy.github.com. This research made use of Astropy, 
a community-developed core Python package for 
Astronomy (Astropy Collaboration, 2013) http://www.astropy.org.

 \facility{ALMA, VLA, TripleSpec/ARC 3.5m}

\software{
Astropy \citep[http://www.astropy.org; ][]{astropy2013,astropy2018}, 
APLpy \citep[http://aplpy.github.com; ][]{aplpy}, CASA \citep[http://casa.nrao.edu; ][]{mcmullin2007},
pdspy \citep[https://github.com/psheehan/pdspy; ][]{sheehan2017,sheehan2019}, GALARIO \citep[https://github.com/mtazzari/galario; ][]{tazzari2018},
RADMC-3D \citep[http://www.ita.uni-heidelberg.de/~dullemond/software/radmc-3d/index.php; ][]{dullemond2012}
}

\appendix

\section{Additional Modeling Results}

Here we report the results from a second set of models that
include the envelope in fitting the molecular line emission, in addition
to further discussion on the reliability of the fitted parameters for both
sets of models.

\subsection{Model Fits Including an Envelope Component}

We also fitted the molecular line emission including an envelope component 
whose mass and radius were free parameters in the fitting. The best fitting
parameters from these model fits are presented in Table 7.

The range of protostar
masses fit is comparable to the disk only fits.
The average mass from the 12 independent fits is 2.6$\pm$0.2~\msun, only 
larger than the disk-only fit by 0.1~\msun. The best-fitting protostar mass values 
range between 1.85 and 3.83~\msun. Like the disk-only fits, the largest outliers are driven by
the NS-only fit (3.83~\msun) and the NS and SO fit (1.85~\msun).

The best fitting disk radii are between 14 and 114~au. The average 
disk radius is $\sim$68$\pm$21~au; smaller than the average from the disk-only
fits by 26 au. The smallest disk radius
comes from the NS and SO fit, which also has the smallest protostar mass.
Other average fitting parameters 
are V$_{lsr}$~=~11.2$\pm$0.07~\kms, PA~=~352.6$\pm$1.3\degr, 
$\gamma$~=~1.08$\pm$0.12, and the position offsets are also very small, typically $\sim$0\farcs01.

The protostar masses derived from fitting with an envelope component appear robust and
agree well with the disk-only component. Indeed, the averages are consistent within 
their associated uncertainties. However, the range of disk radii fit points to
degeneracy when attempting to fit both an envelope component and a disk simultaneously.
This is likely due to the fact that the spatial scales of both structures are not well-sampled 
in the observations and the molecular abundance is a constant for both the envelope and disk. 
For these reasons, we regard the range of disk radii derived from the disk-only fits with greater confidence.
It is important to point out that some of the combinations of molecular lines fitted with both
a disk and envelope component do
fit disk radii that are consistent with the disk-only fits. The line fits with the 3 H$_2$CO transitions
and another line in particular performed well.

\subsection{Impact of Assumptions and Unreliable Fitted Parameters}

Several other parameters that were fit during modeling most likely do not reflect the actual 
physical parameters that they correspond to due to model limitations and limitations of the data
themselves. This applies to both the disk-only fits and the disk plus envelope fits.
The disk masses are unreliable in an absolute sense because they depend on the 
assumed abundance of the molecule, which we assume is constant throughout the disk. The envelope
masses and radii are considered unreliable because of the limited sensitivity 
to spatial scales beyond $\sim$4\arcsec\ and the fixed molecular abundances.
The temperature normalization of the disk ($T$(1~au)) is also likely unreliable
because the fitted molecules have little temperature sensitivity leading to a degeneracy
between mass and temperature. Finally, the molecular abundance fit for the cases where
we fit more than one species are also unreliable since these were only varied
to improve fitting with a constant disk mass.

While we fixed the power-law index of the temperature profile $q$ 
 for all the models, we did examine the impact of this assumption. We ran another
fit to all H$_2$CO transitions because their different excitation energies can enable their line ratios to
provide constraints on temperature.
Allowing $q$ to vary resulted in a best fitting protostar mass that was 0.29~\msun\
lower than when $q$ was fixed at 0.35. However, this fit is likely unrealistic because
 $q$ was fit to be consistent with 0, indicating no temperature change with radius, and this was also
the lower limit for the parameter. Thus, the overall impact of fixing $q$ results in $\sim$10\%
higher masses, but the best-fitting value for $q$ was not realistic indicating that
our choice of a fixed $q$ is more reasonable; this particular model is not included in the average
protostar mass and radius.

With regard to the protostar mass and disk radius, the model fits
with the most discrepant fitting parameters were NS alone, NS and SO.
The fit to NS alone resulted in the highest protostar mass of all the fits
for both the disk-only and disk plus envelope models. Then the lowest
protostar masses fit were from the NS and SO fits. Furthermore,
the NS and SO fit with an envelope component resulted in a model fit with a very small disk radius.
The reason for these discrepant fits could be because both NS and SO had the smallest spatial
extent of line emission in our data. This potentially explains the small disk radii.
The more concentrated emission from NS likely resulted in a discrepant protostar mass determination because there was 
less of a rotation curve to model. Thus, we regard the fits to multiple molecular
lines as the most robust, particularly when the majority of lines sample the full range of radii
in the disk.

\subsection{Special Analysis Required to Fit CH$_3$OH}

To perform the fit for CH$_3$OH,
we had to first subtract its extended, off-center emission from the visibilities.
This is because the brightest CH$_3$OH emission
is not centered on HOPS-370, but comes from a spot on the edge of 
the outflow cavity wall several arcseconds south of the disk. This is outside the
field of view in Figure \ref{moment0} and not shown
because it is not relevant for this analysis. This emission
was preventing model convergence because it dominated the imaginary component of the
visibility amplitude data. To remove this emission, we interactively 
cleaned the CH$_3$OH emission, only cleaning
the emission that was not associated with the disk. This created a model for the CH$_3$OH emission
not associated with the disk. We used \textit{clean} to save the model to the measurement set,
and we subtracted it from the data using the CASA task \textit{uvsub}. This made it possible to 
reliably model the CH$_3$OH emission only coming from the vicinity of the disk.

\begin{small}
\bibliographystyle{apj}
\bibliography{ms}
\end{small}

\clearpage

\begin{figure}
\begin{center}
\includegraphics[scale=0.475]{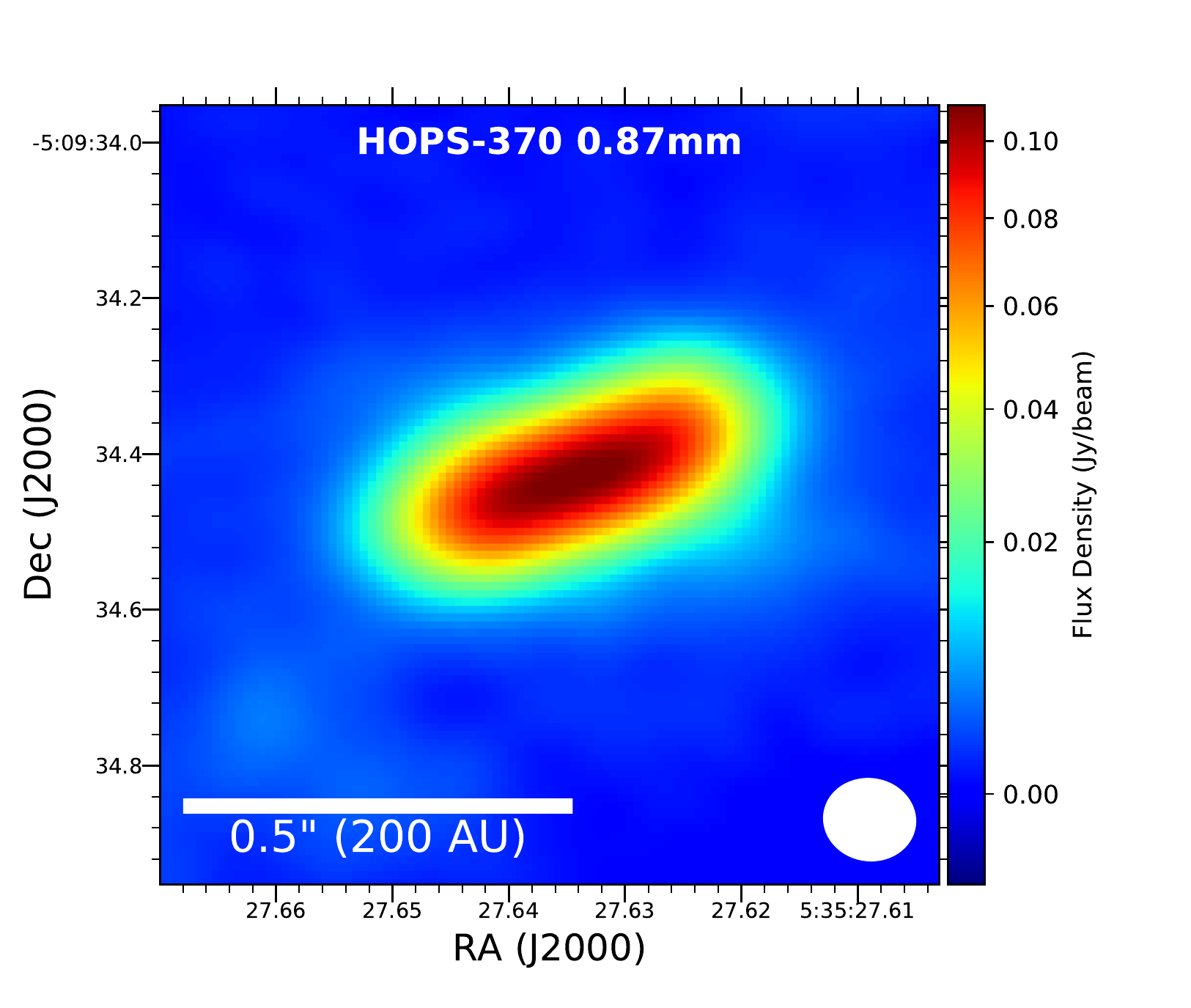}
\includegraphics[scale=0.475]{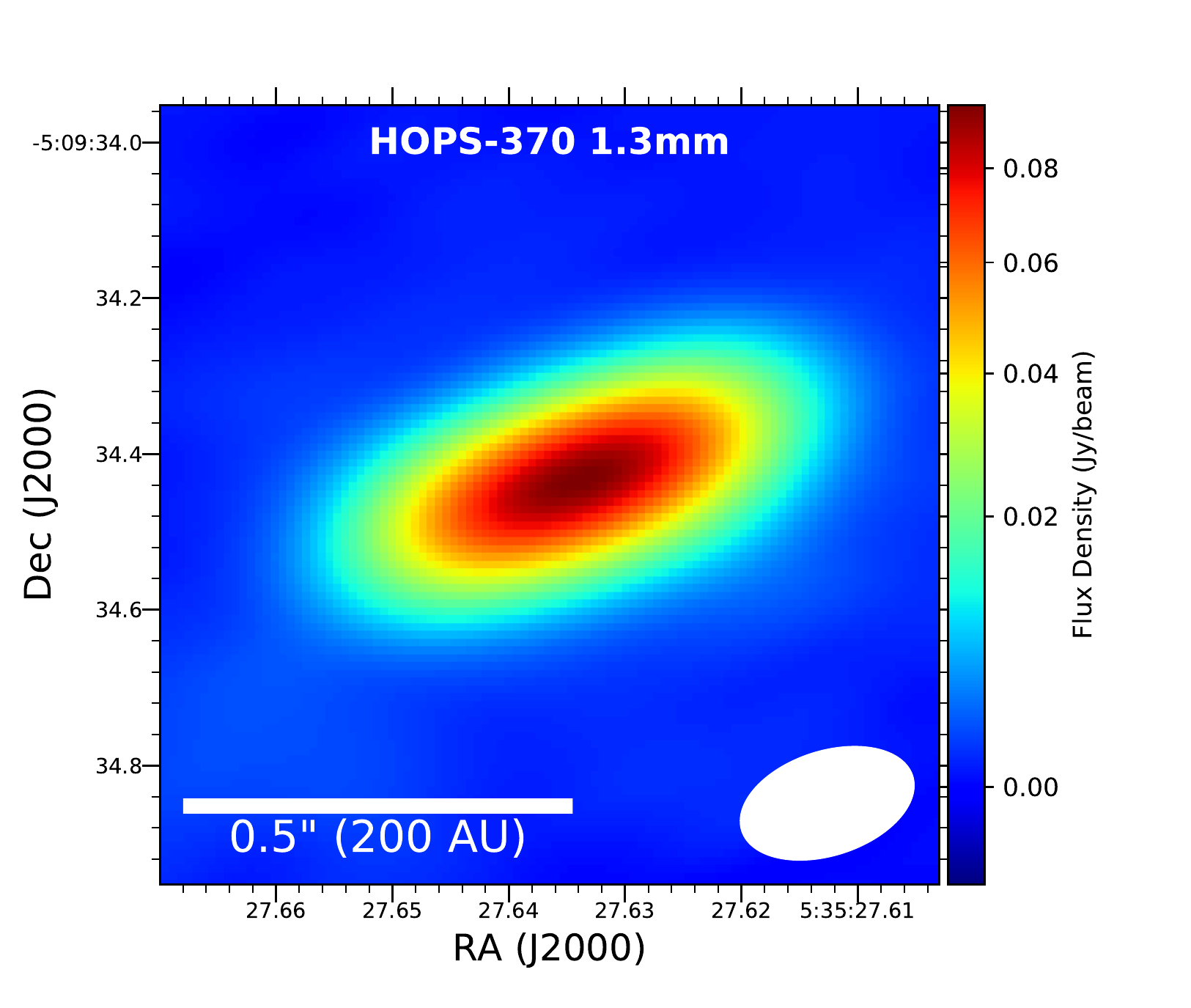}
\includegraphics[scale=0.475]{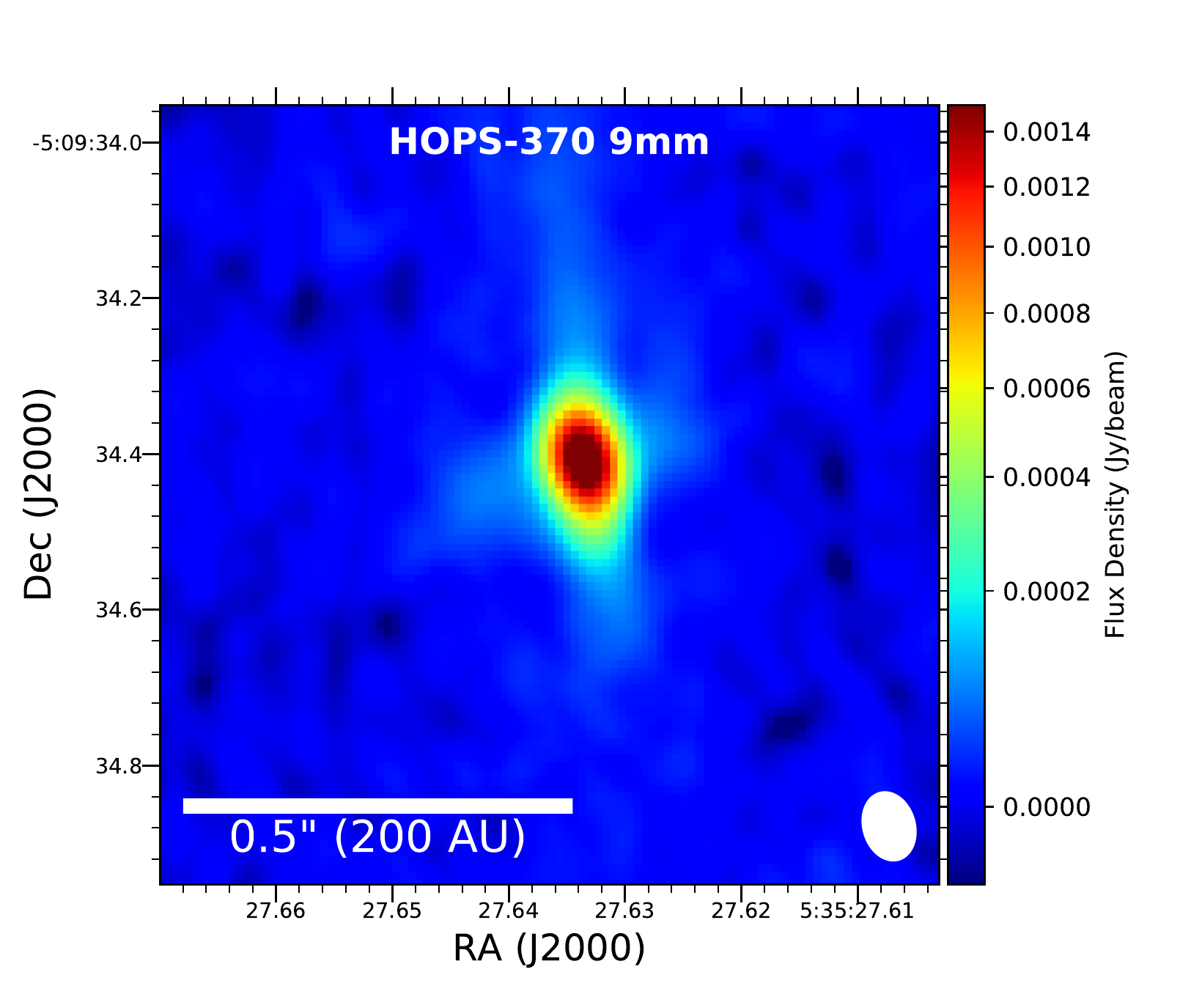}
\includegraphics[scale=0.475]{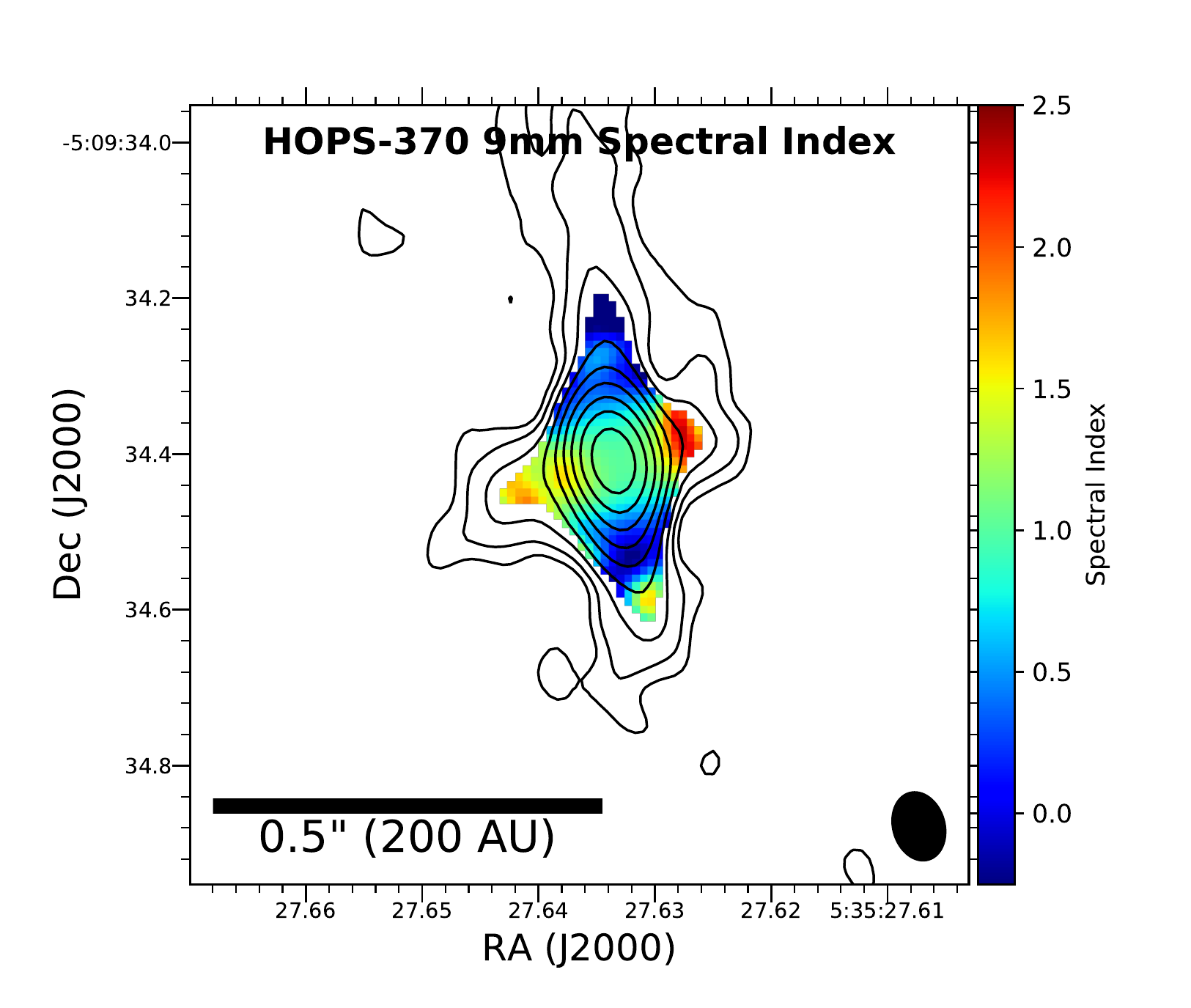}
\end{center}
\caption{ALMA 0.87~mm, 1.3~mm, and VLA 9~mm continuum images of HOPS-370
and the VLA 9~mm in-band spectral index map produced by the CASA task \textit{clean}. The 0.87~mm and
1.3~mm continuum emission traces an obvious disk-like structure. The 9~mm emission 
has extensions in the direction of the disk as well as the outflow tracing 
both the jet and disk emission. The contours on the 9~mm spectral index map  
are the VLA 9~mm continuum and are logarithmically
spaced between $\pm$3$\sigma$ and 300$\sigma$, where $\sigma$=7.2~$\mu$Jy.
The beam size is indicated in the lower right of
 each panel and is 0\farcs11$\times$0\farcs10 (43~au~$\times$~39~au), 
0\farcs23$\times$0\farcs13 (90~au~$\times$51~au), and
0\farcs08$\times$0\farcs07 (32~au~$\times$~28~au) at 0.87~mm, 1.3~mm, and 9~mm, respectively. 
}
\label{continuum}
\end{figure}

\begin{figure}
\begin{center}
\includegraphics[scale=0.65]{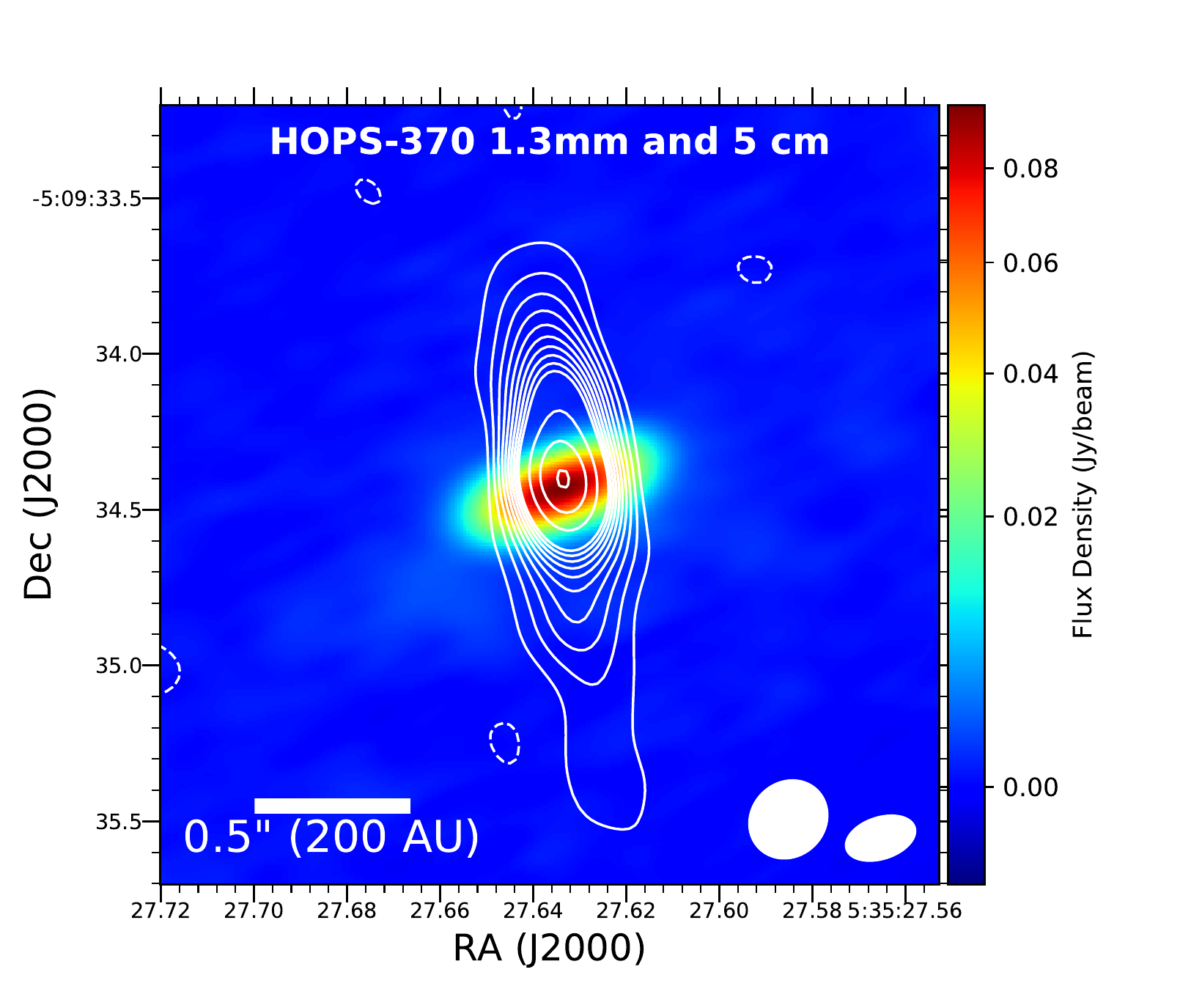}
\end{center}
\caption{ALMA 1.3~mm continuum images of HOPS-370 with the VLA 5~cm data
from \citet{osorio2017} overlaid showing the relation of the jet traced by 5~cm emission
to the disk traced by 1.3~mm emission. The contours start at and increase on $\pm$3$\sigma$
 intervals until 30$\sigma$where contours begin to increase on 15$\sigma$ intervals; $\sigma$=7.2~$\mu$Jy. 
The beam of the 1.3~mm image is 0\farcs23$\times$0\farcs13 (90~au~$\times$51~au) (white ellipse in lower
right) and the beam of the 5~cm image is 0\farcs26$\times$0\farcs23 (101~au~$\times$~90~au) 
(white ellipse located left of the 1.3~mm beam).
}
\label{continuum-5cm}
\end{figure}

\clearpage
\begin{figure}
\begin{center}
\includegraphics[scale=0.5]{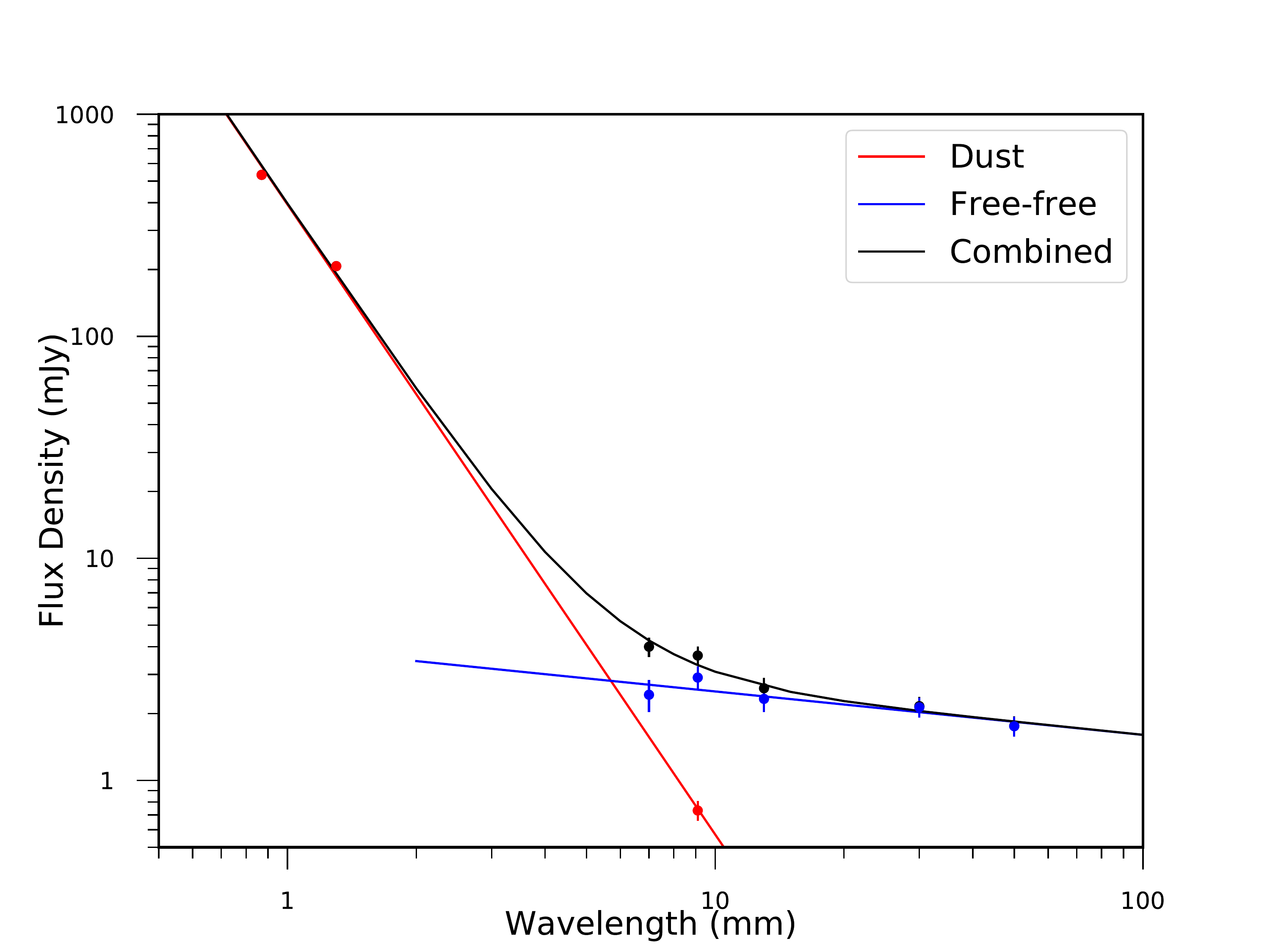}

\end{center}
\caption{Radio spectrum of HOPS-370. The 0.87~mm, 1.3~mm, and 9.1~mm flux density 
measurements are from this work; the other measurements are from \citet{osorio2017} and
are listed in Table 2. The red points are assumed to trace only dust emission; a power-law
fit is performed finding that the flux density from dust emission scales
$\propto$~$\nu^{2.84\pm0.08}$. The blue points are expected to only
trace free-free emission, having their estimated 
dust emission contribution removed using the power-law fit to the dust emission.
A power-law slope is fit to the free-free emission (blue points) 
finding a dependence $\propto$~$\nu^{0.19\pm0.08}$.
The combined spectrum is drawn as a black line and agrees well with the total flux densities
of the points between 7~mm and 20~mm where both emission mechanisms are contributing. The uncertainties
on the points at 0.87~mm, 1.3~mm, and 9.1~mm do not include the absolute flux calibration uncertainty of $\sim$10\%.
The other points from \citep{osorio2017} do include the additional absolute flux calibration uncertainty.
}
\label{radio-spectrum}
\end{figure}

\begin{figure}
\begin{center}
\includegraphics[scale=0.375]{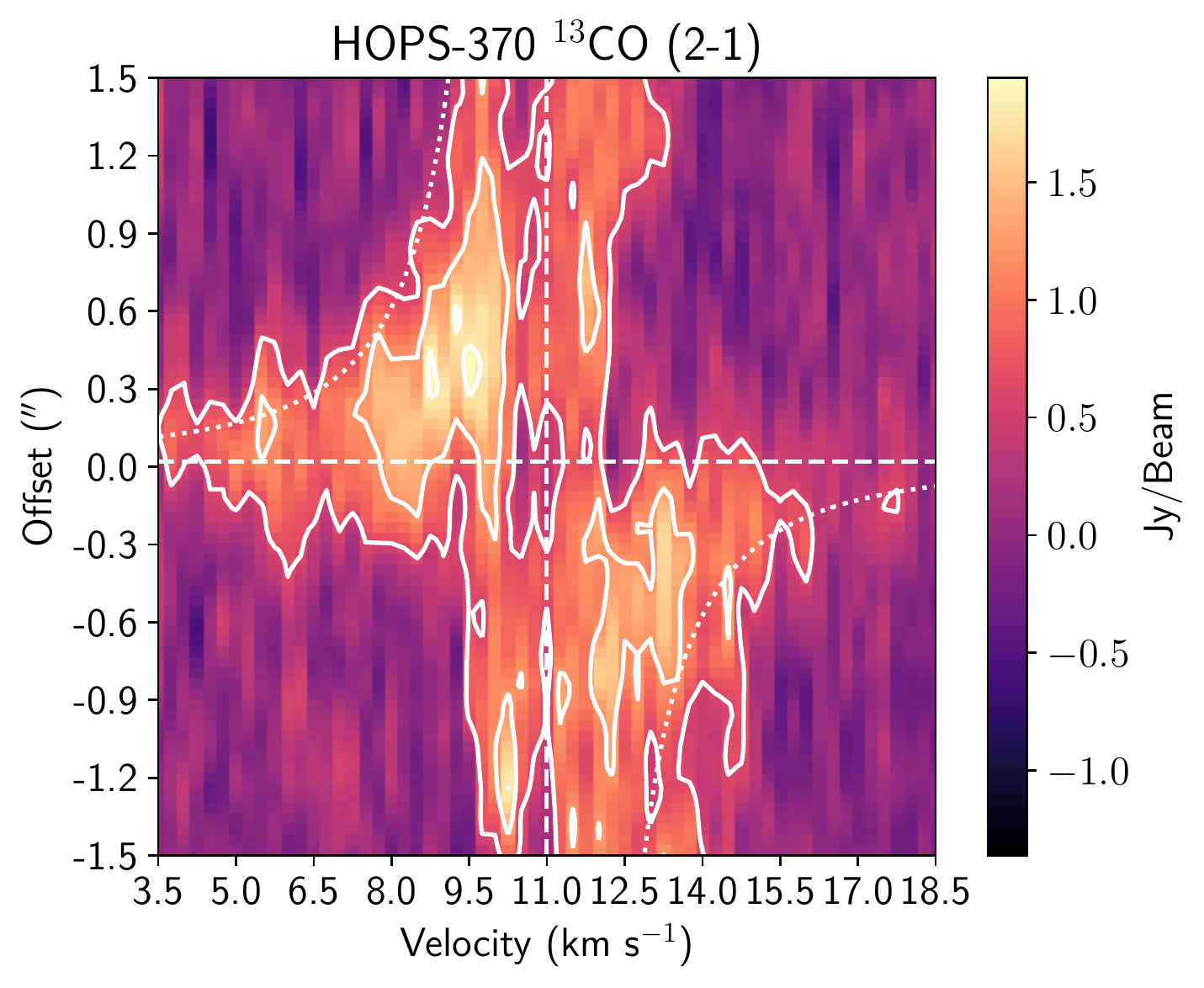}
\includegraphics[scale=0.375]{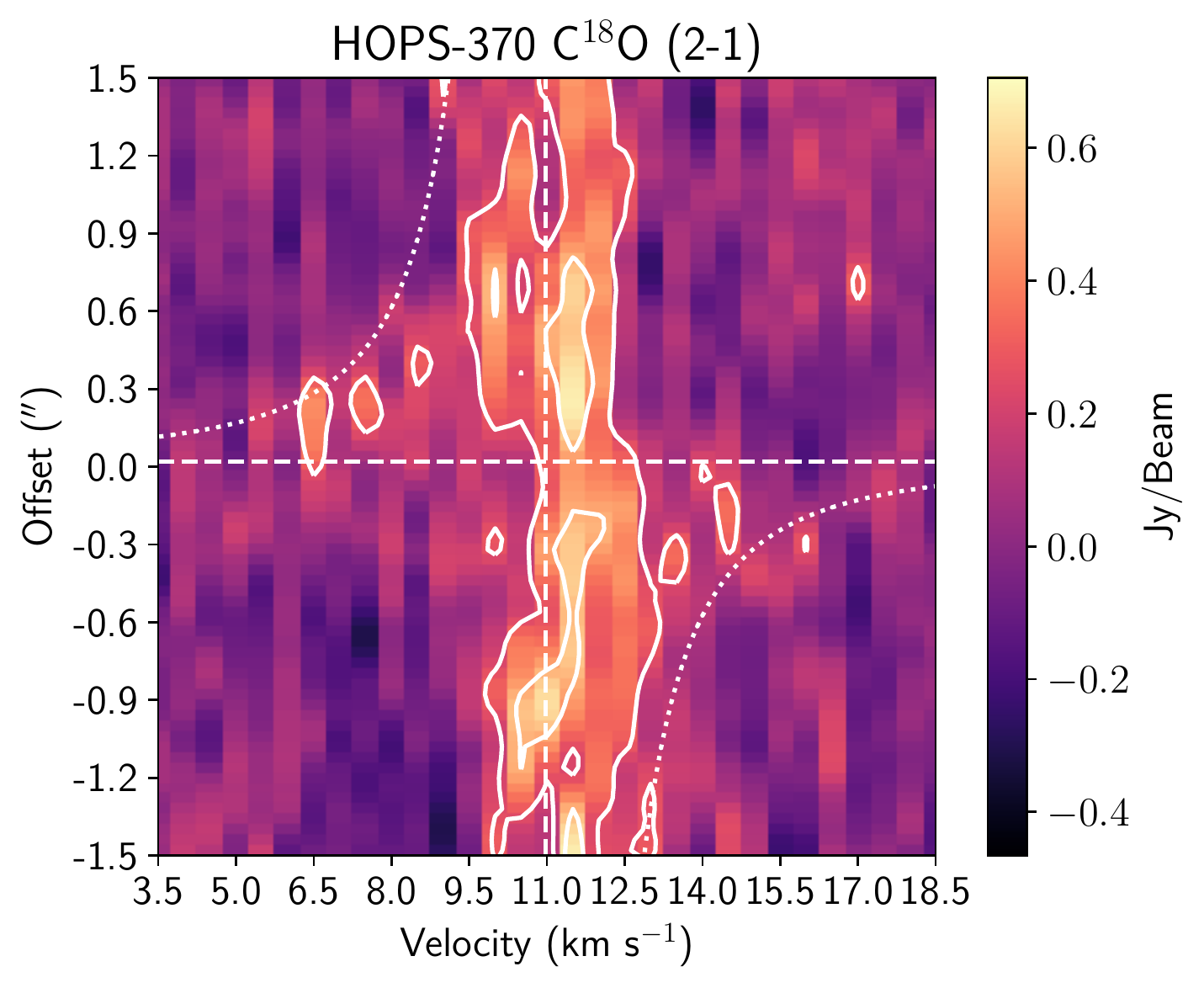}
\includegraphics[scale=0.375]{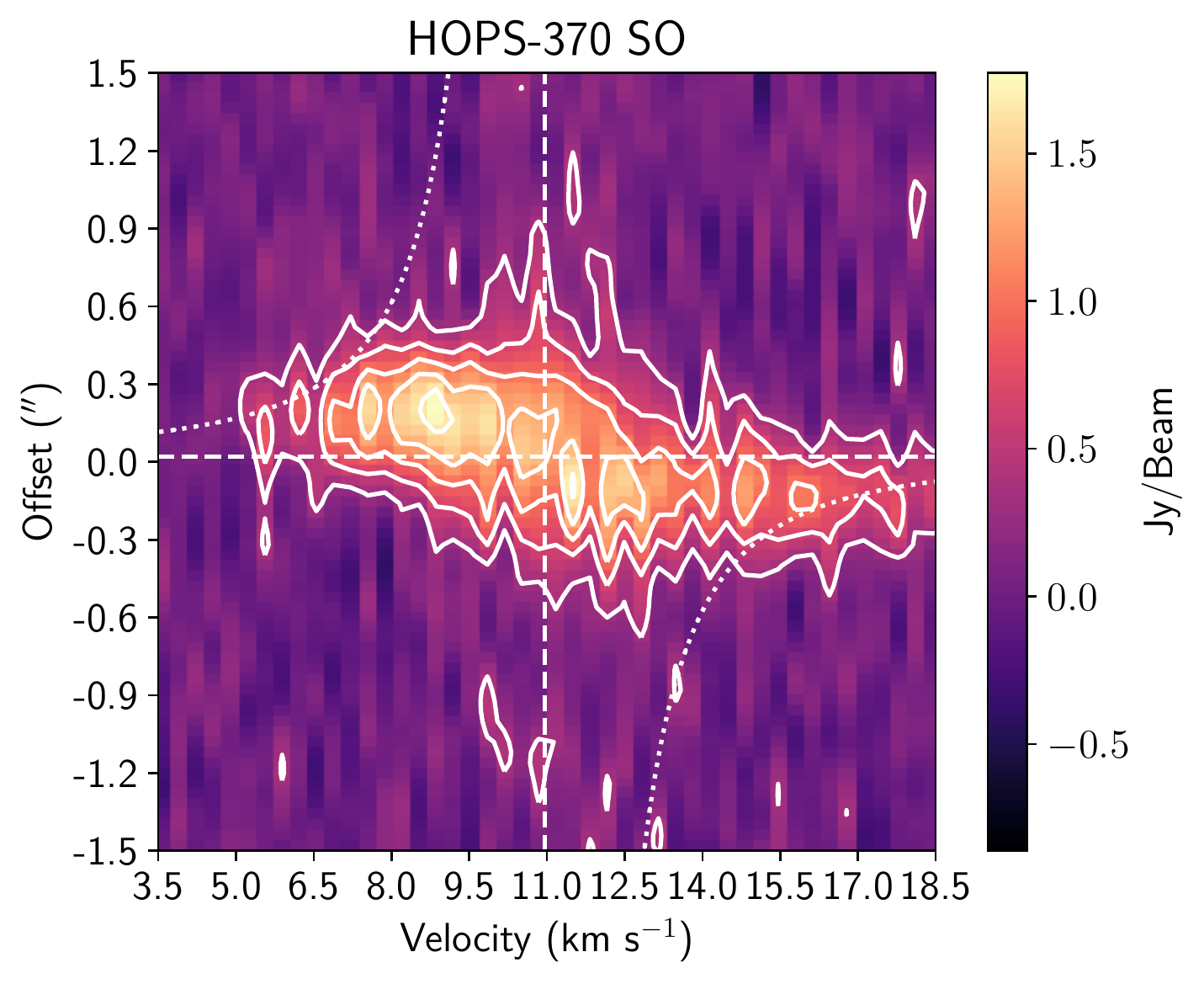}
\includegraphics[scale=0.375]{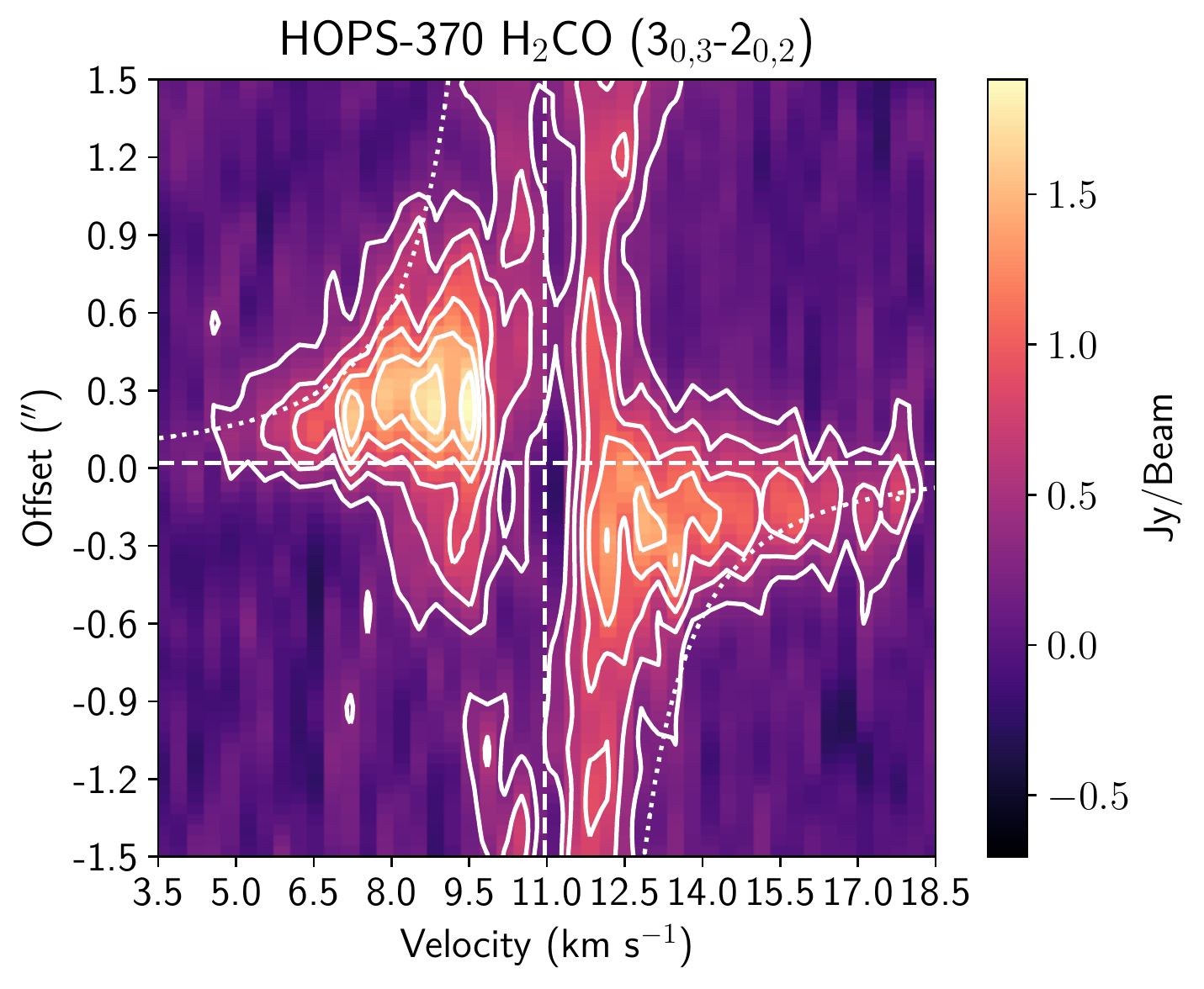}
\includegraphics[scale=0.375]{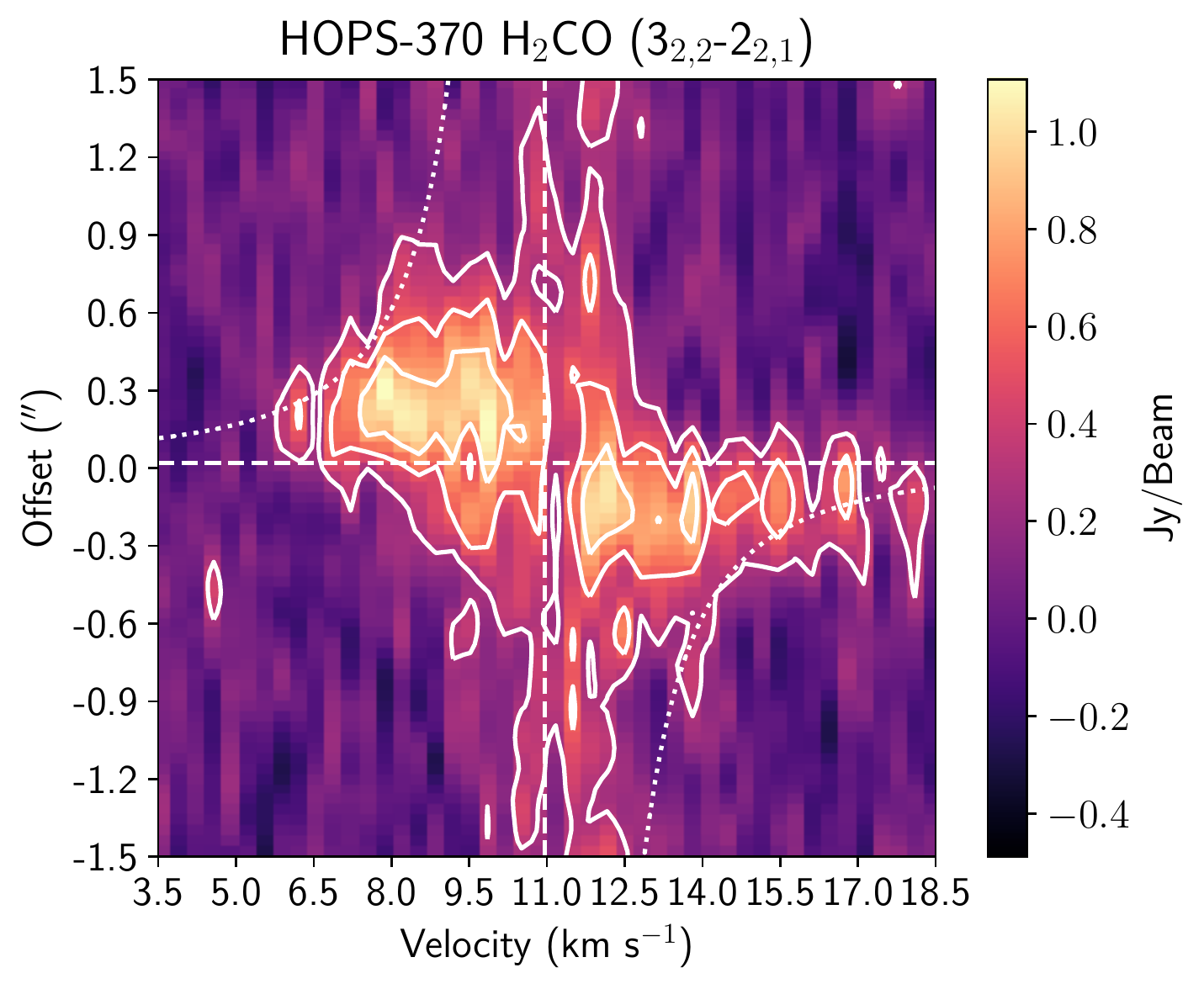}
\includegraphics[scale=0.375]{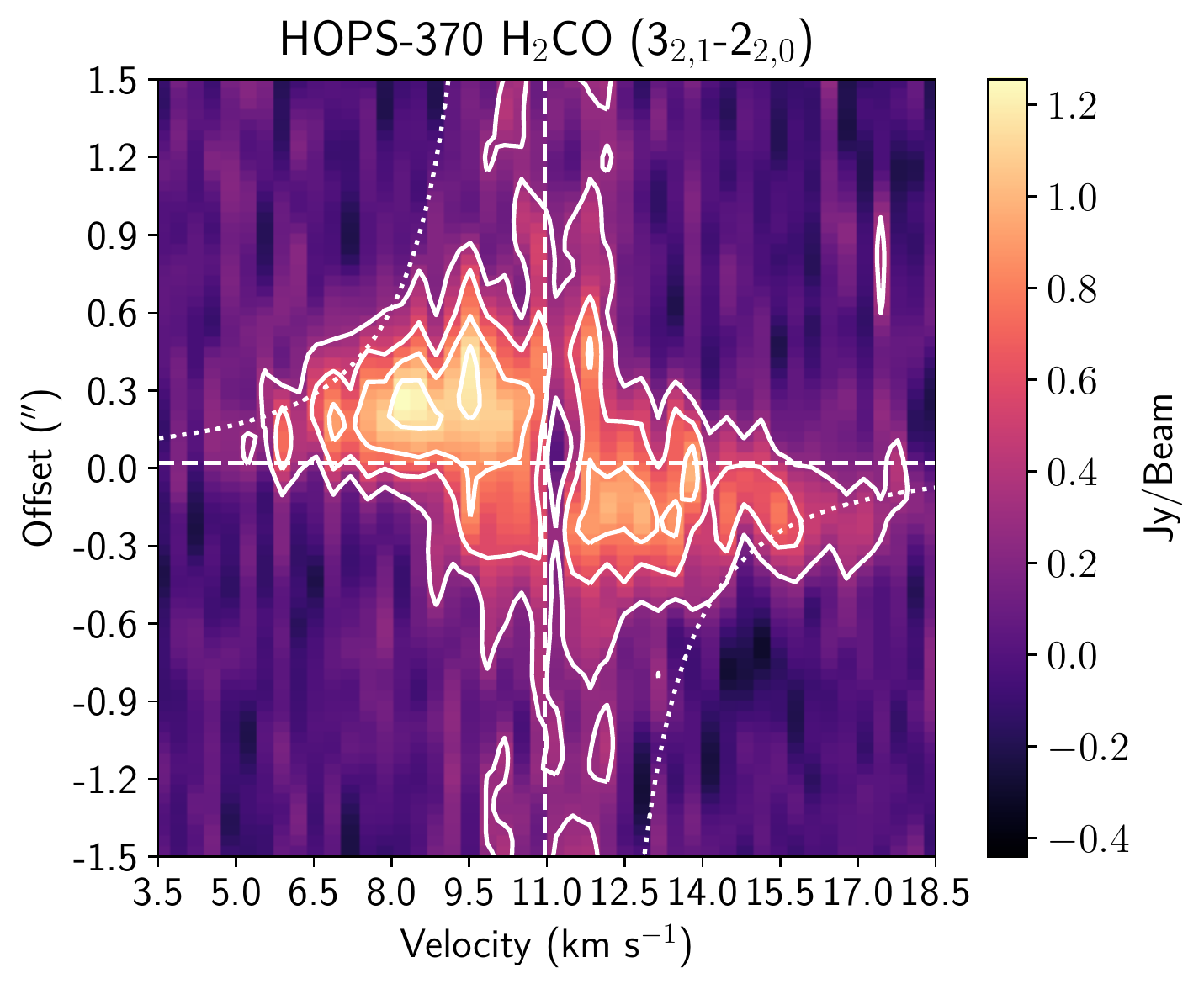}
\includegraphics[scale=0.375]{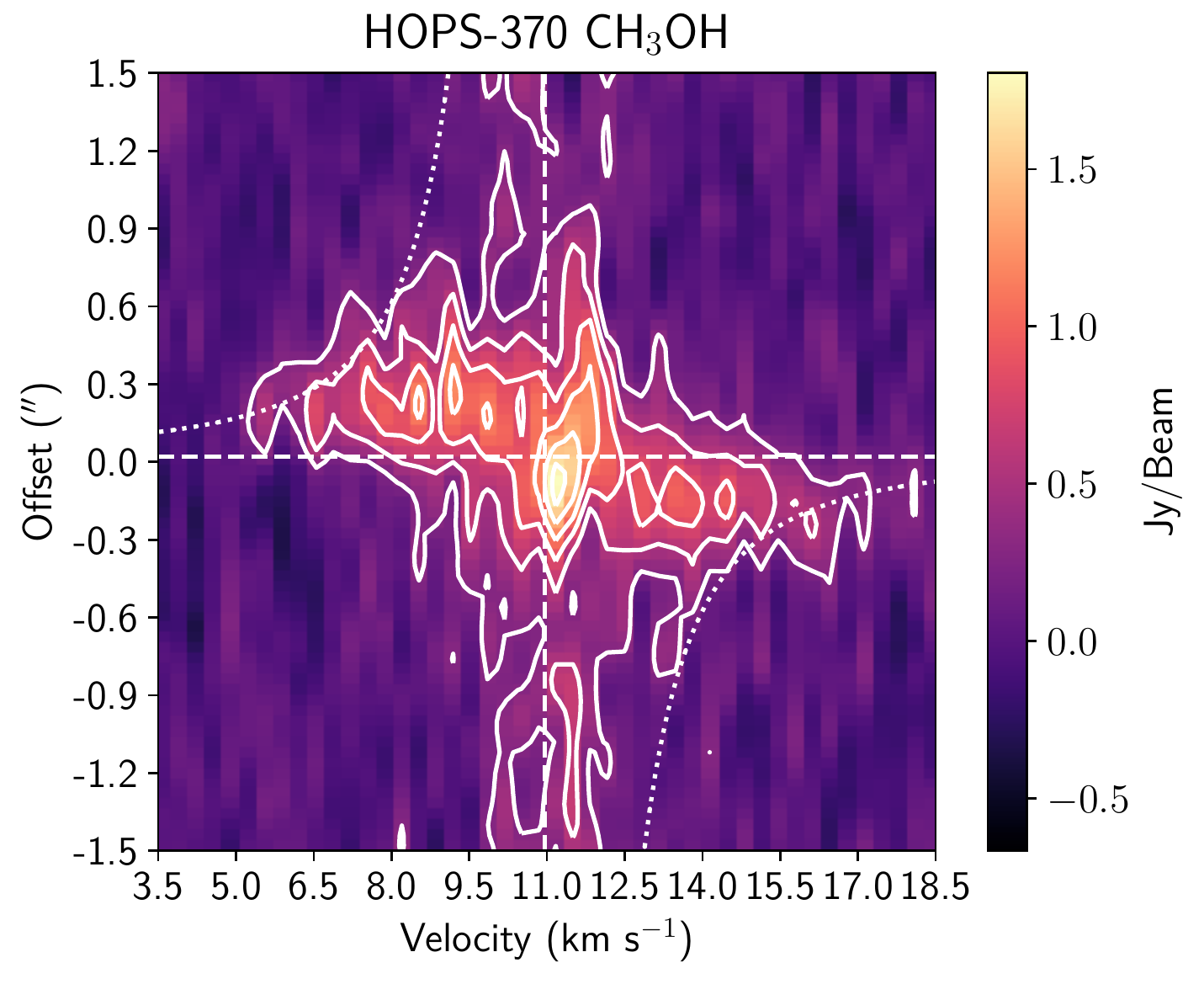}
\includegraphics[scale=0.375]{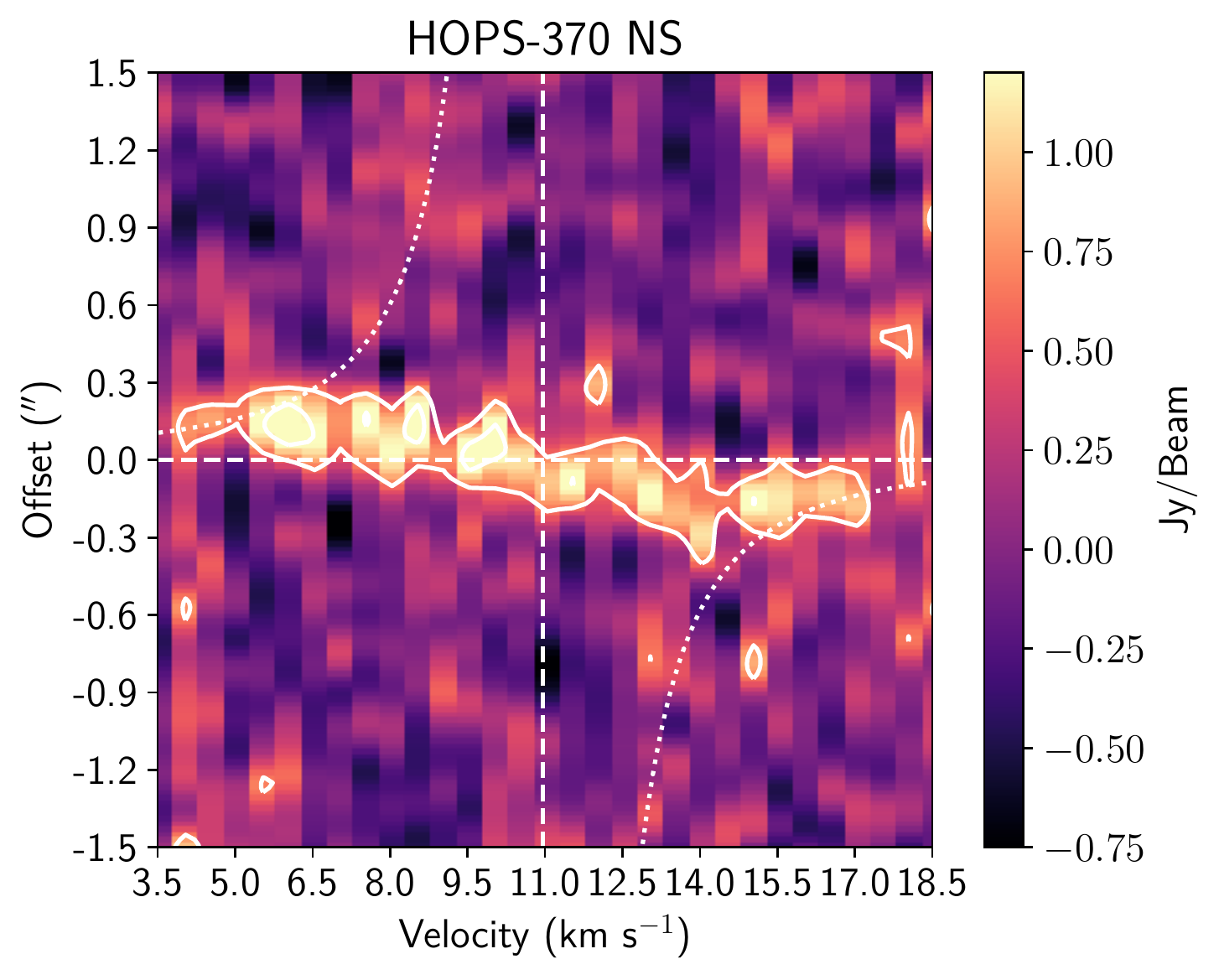}
\includegraphics[scale=0.375]{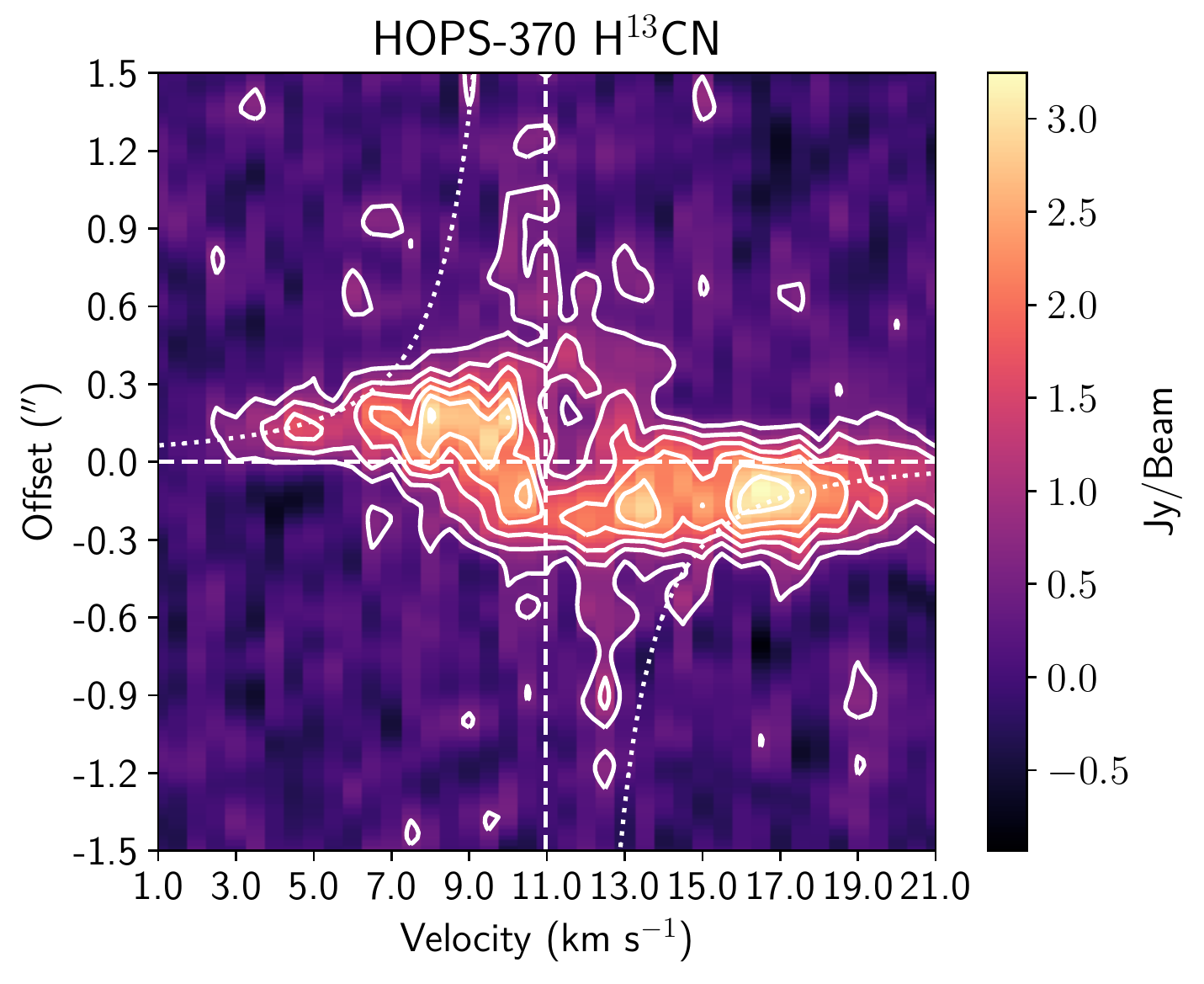}
\end{center}
\caption{ALMA integrated intensity images of the molecular lines at blue- and red-shifted velocities
overlaid with their corresponding contour colors and overlaid on the 0.87~mm continuum (grayscale). 
The red- and blue-shifted 
emission show a clear rotation pattern in the disk that is consistent for all molecules.
The integrated intensity maps for many 
molecules have their emission peaks spatially located above and below the 
continuum emission which could be related
to continuum or line opacity. The NS molecule in contrast does not seem to avoid
the disk midplane. The blue-shifted images are typically integrated between 4 and 11~\kms, while
the red-shifted images are typically integrated between 11 and 18~\kms. The contours plotted
start at and increase on 5$\sigma$ intervals for the three H$_2$CO transitions, SO, and CH$_3$OH 
where $\sigma$=24~mJy~beam$^{-1}$. The contour levels for the other transitions are as follows:
NS starts at 3$\sigma$ and increases on 2$\sigma$ intervals and $\sigma$=38~mJy~beam$^{-1}$,
$^{13}$CO starts at 3$\sigma$ and increases on 2$\sigma$ intervals and $\sigma$=61~mJy~beam$^{-1}$, 
C$^{18}$O starts at 2$\sigma$ and increases on 1$\sigma$ intervals $\sigma$=36~mJy~beam$^{-1}$,
and H$^{13}$CN starts at 4$\sigma$ and increases on 4$\sigma$ intervals and $\sigma$=33~mJy~beam$^{-1}$.
The green ellipse denotes the beam for the molecular line data which is $\sim$0\farcs23$\times$$\sim$0\farcs13 
(90~au~$\times$51~au) for all but NS and H$^{13}$CN whose beams are 0\farcs15$\times$0\farcs14 (59~au~$\times$~55~au).
The black ellipse denotes the beam for the 0.87~mm continuum data 0\farcs11$\times$0\farcs10 (43~au~$\times$~39~au).
}
\label{moment0}
\end{figure}

\begin{figure}
\begin{center}
\includegraphics[scale=0.375]{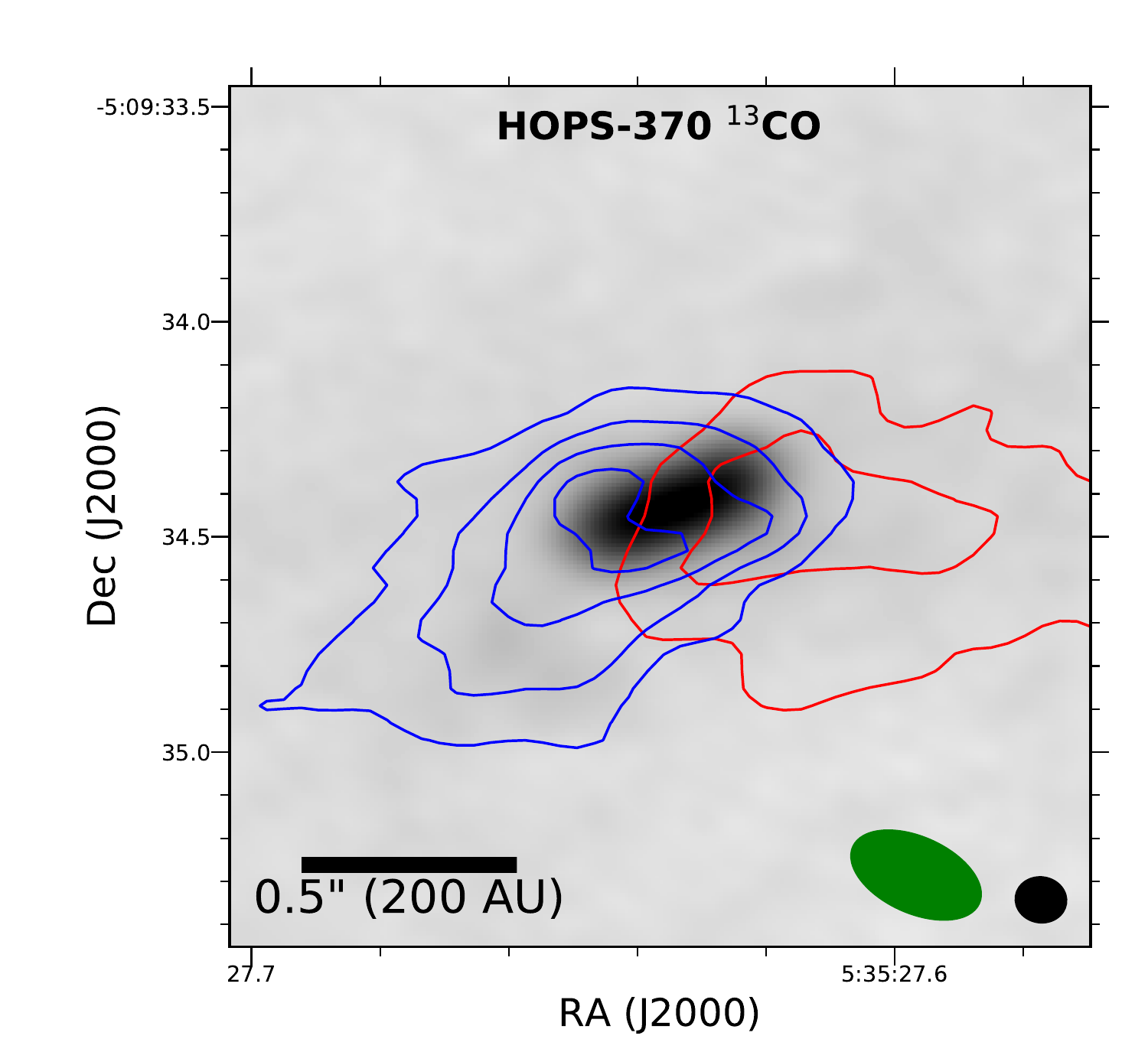}
\includegraphics[scale=0.375]{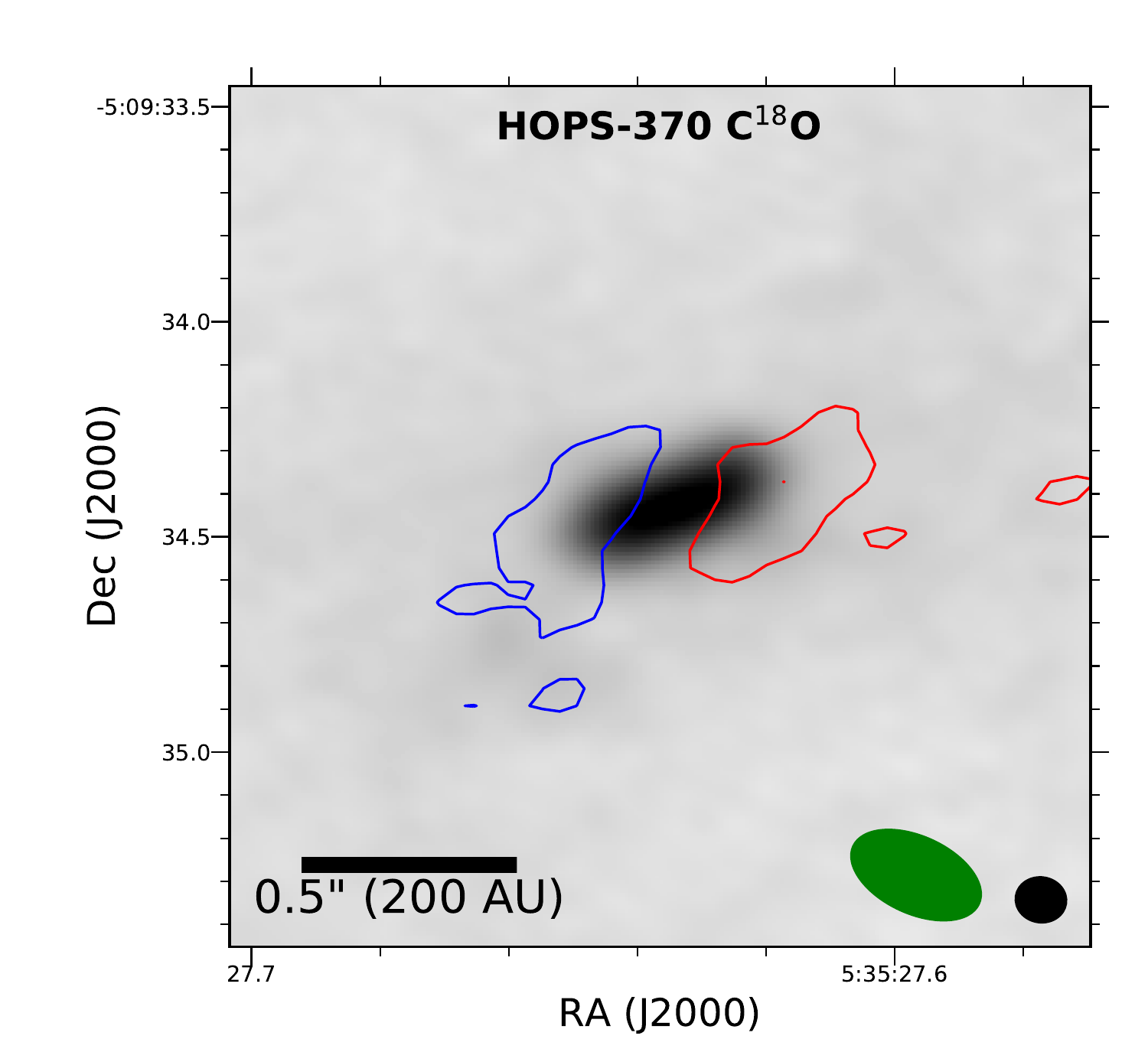}
\includegraphics[scale=0.375]{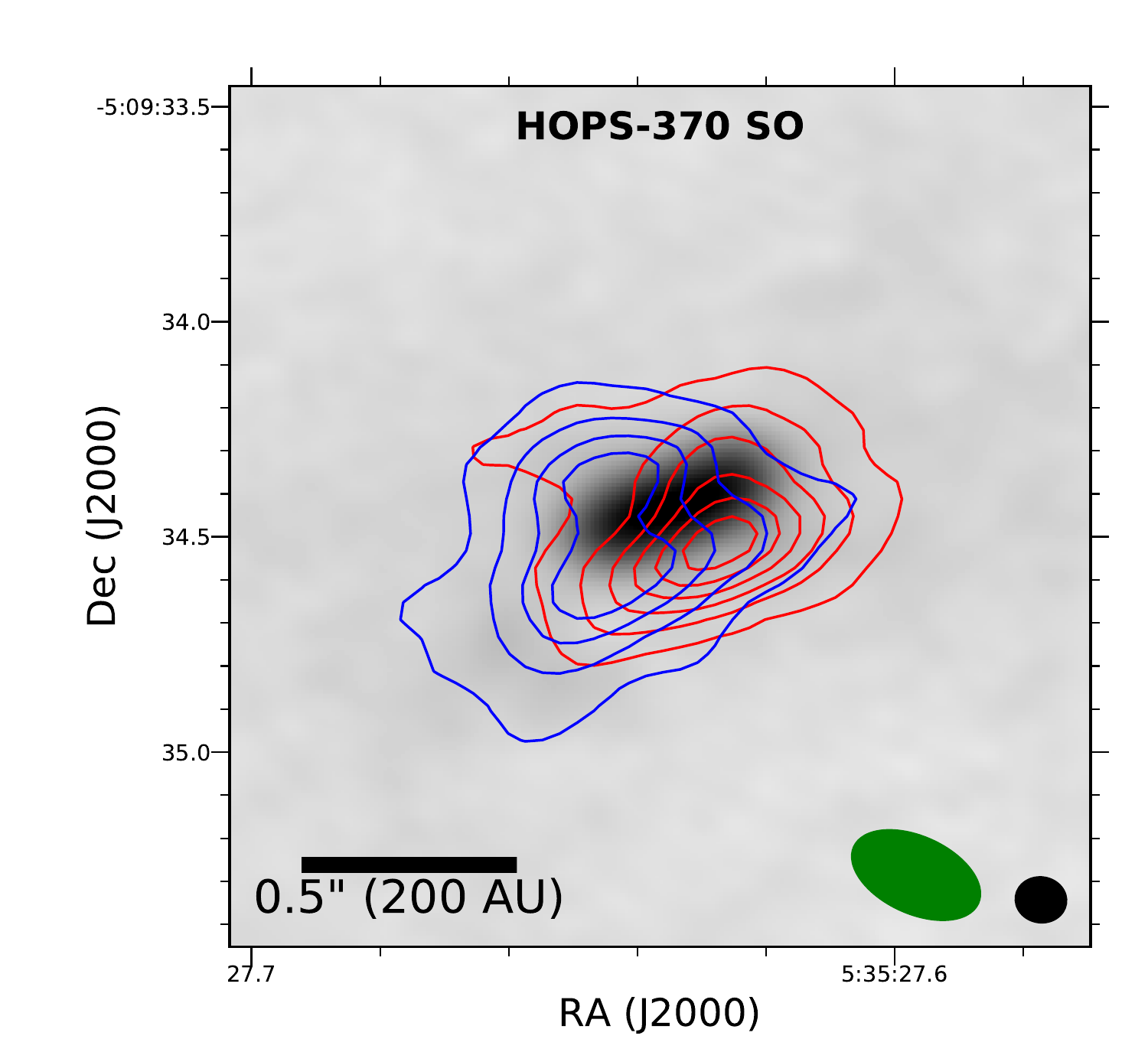}
\includegraphics[scale=0.375]{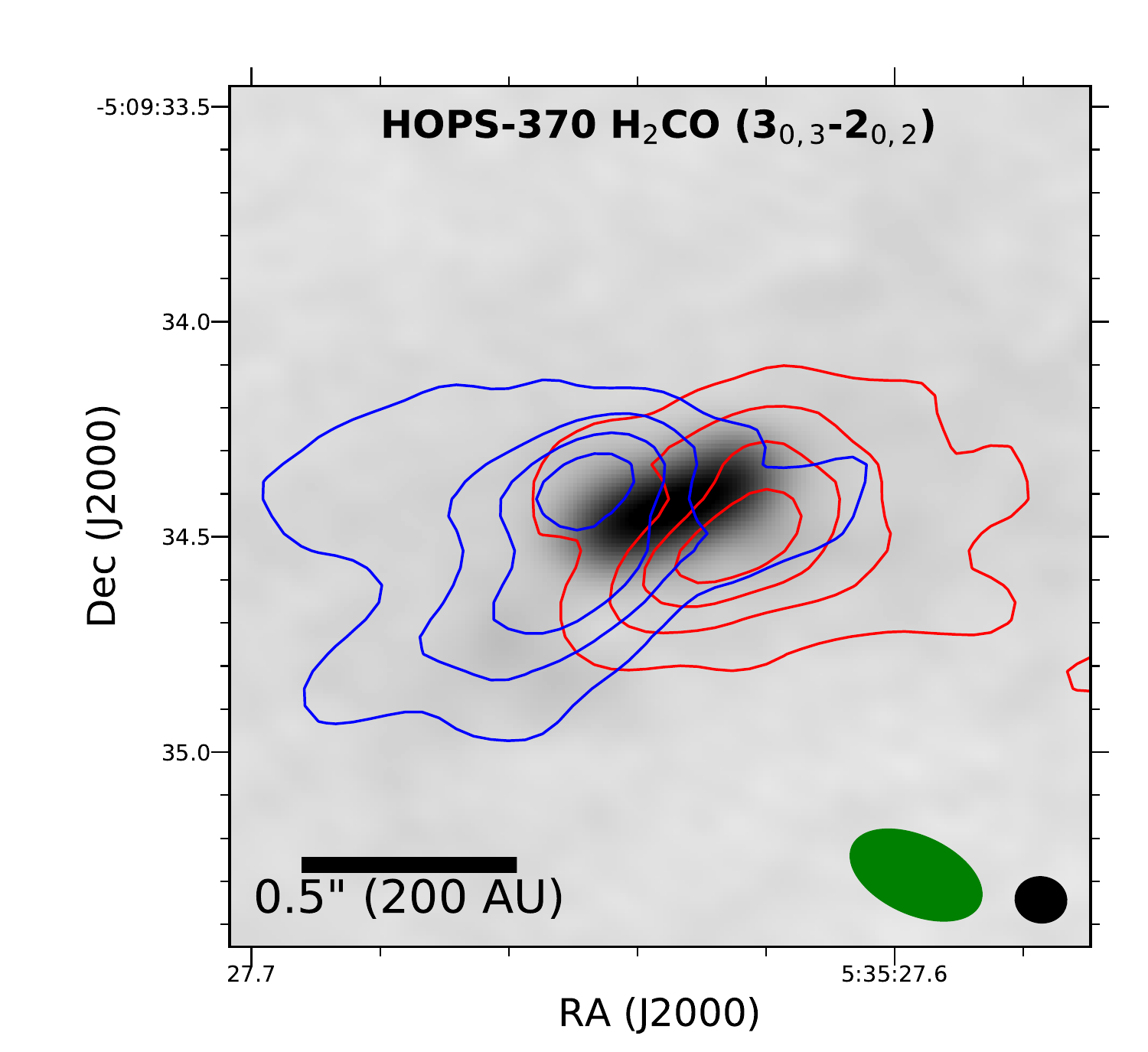}
\includegraphics[scale=0.375]{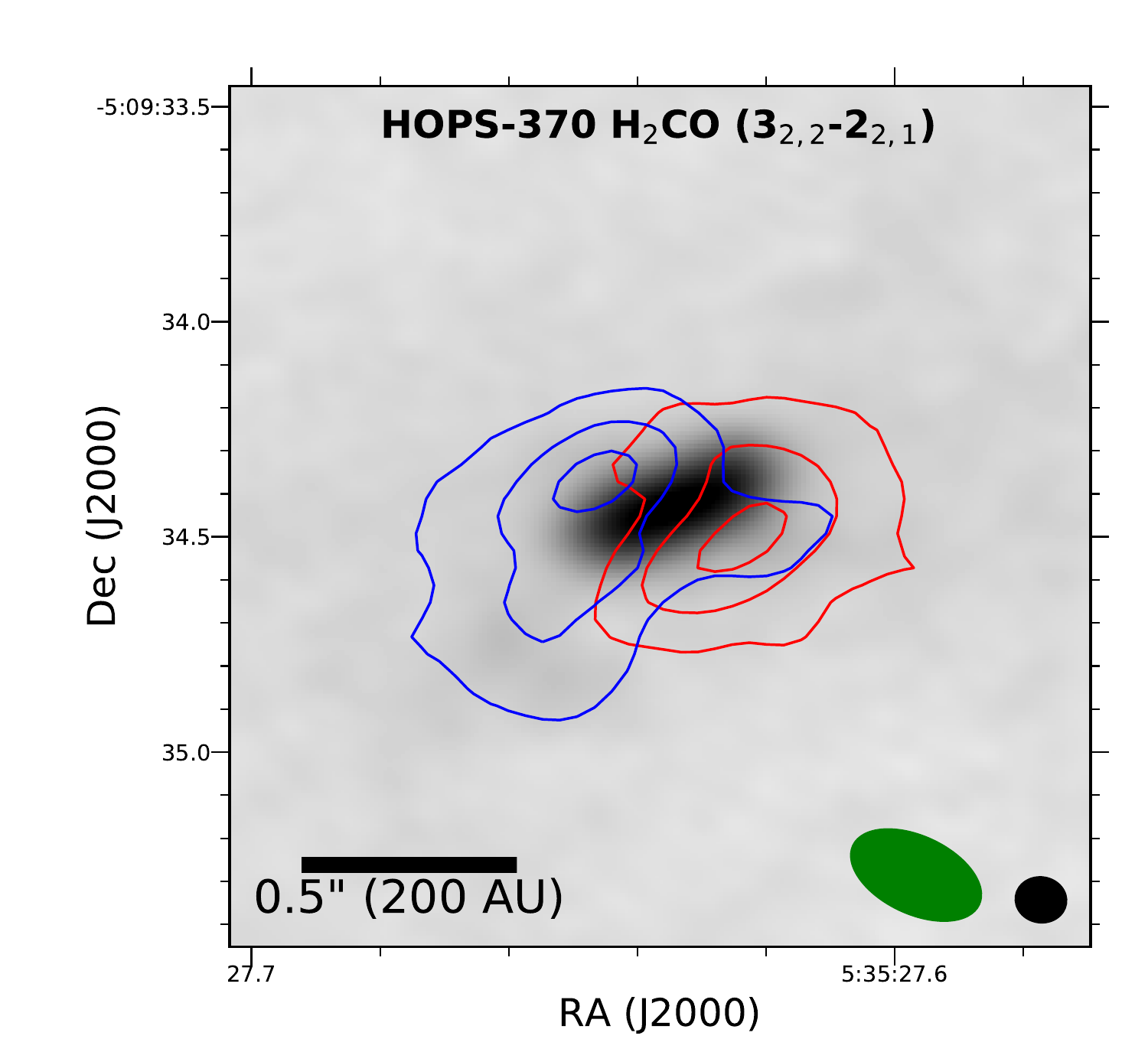}
\includegraphics[scale=0.375]{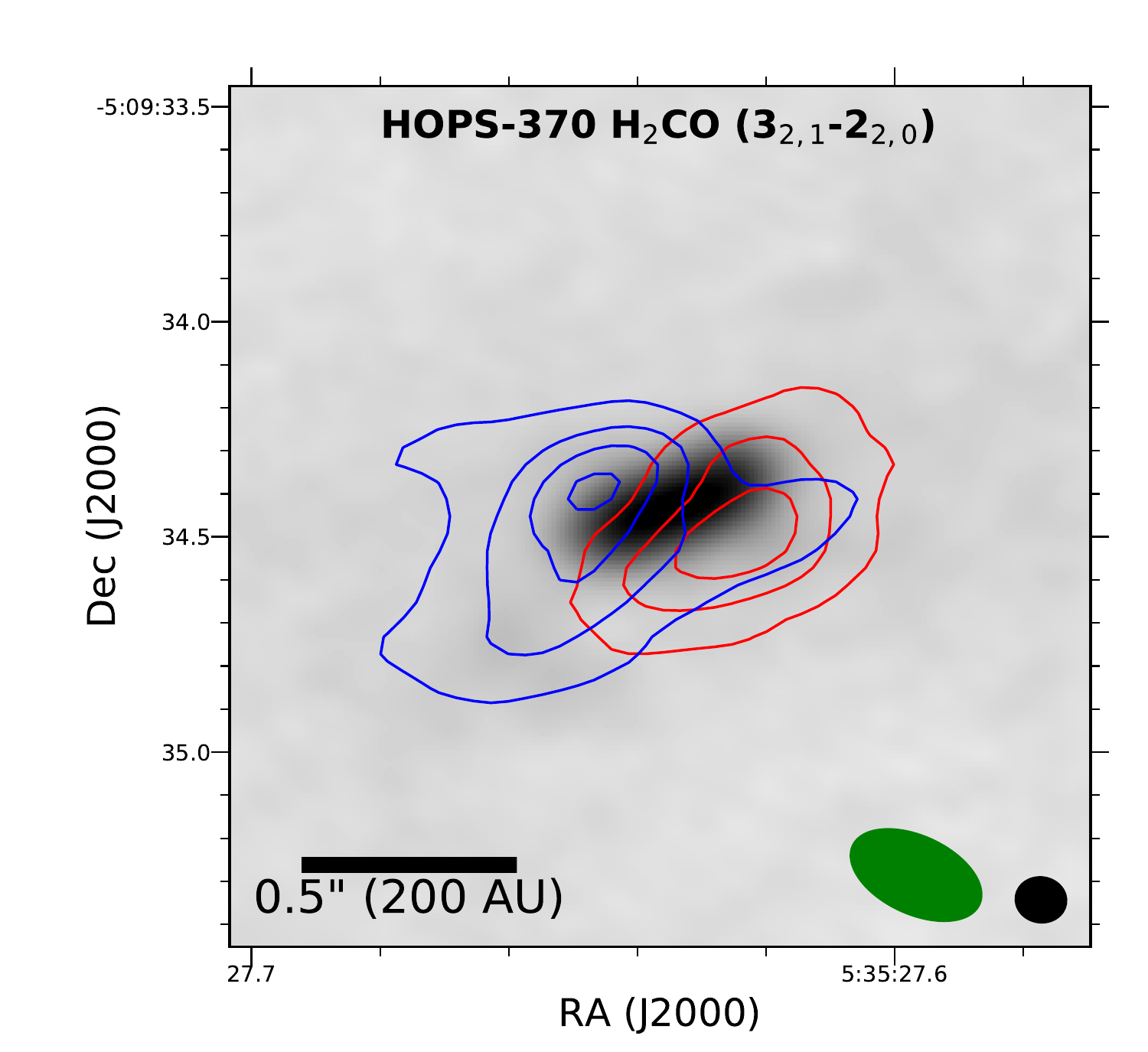}
\includegraphics[scale=0.375]{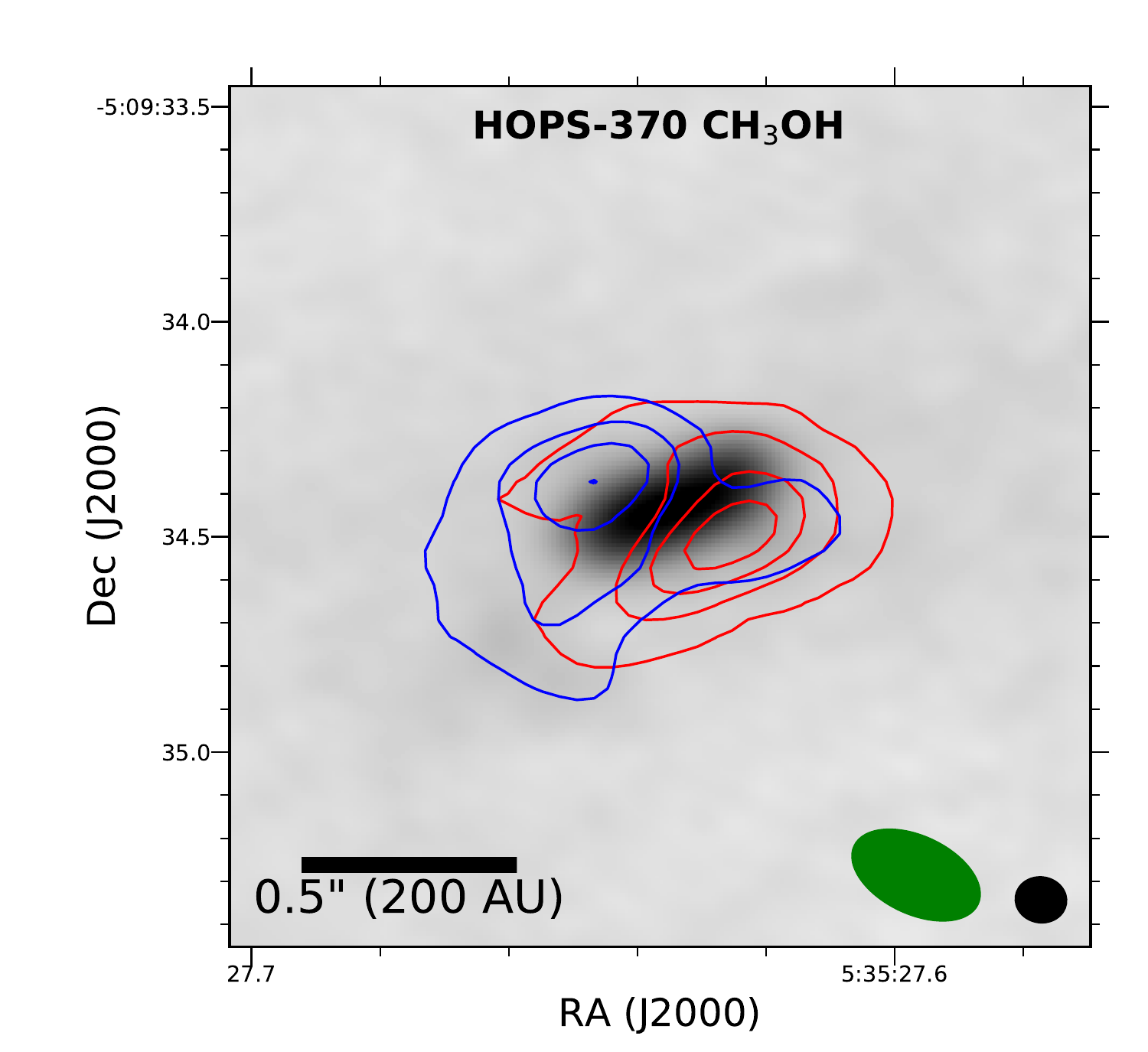}
\includegraphics[scale=0.375]{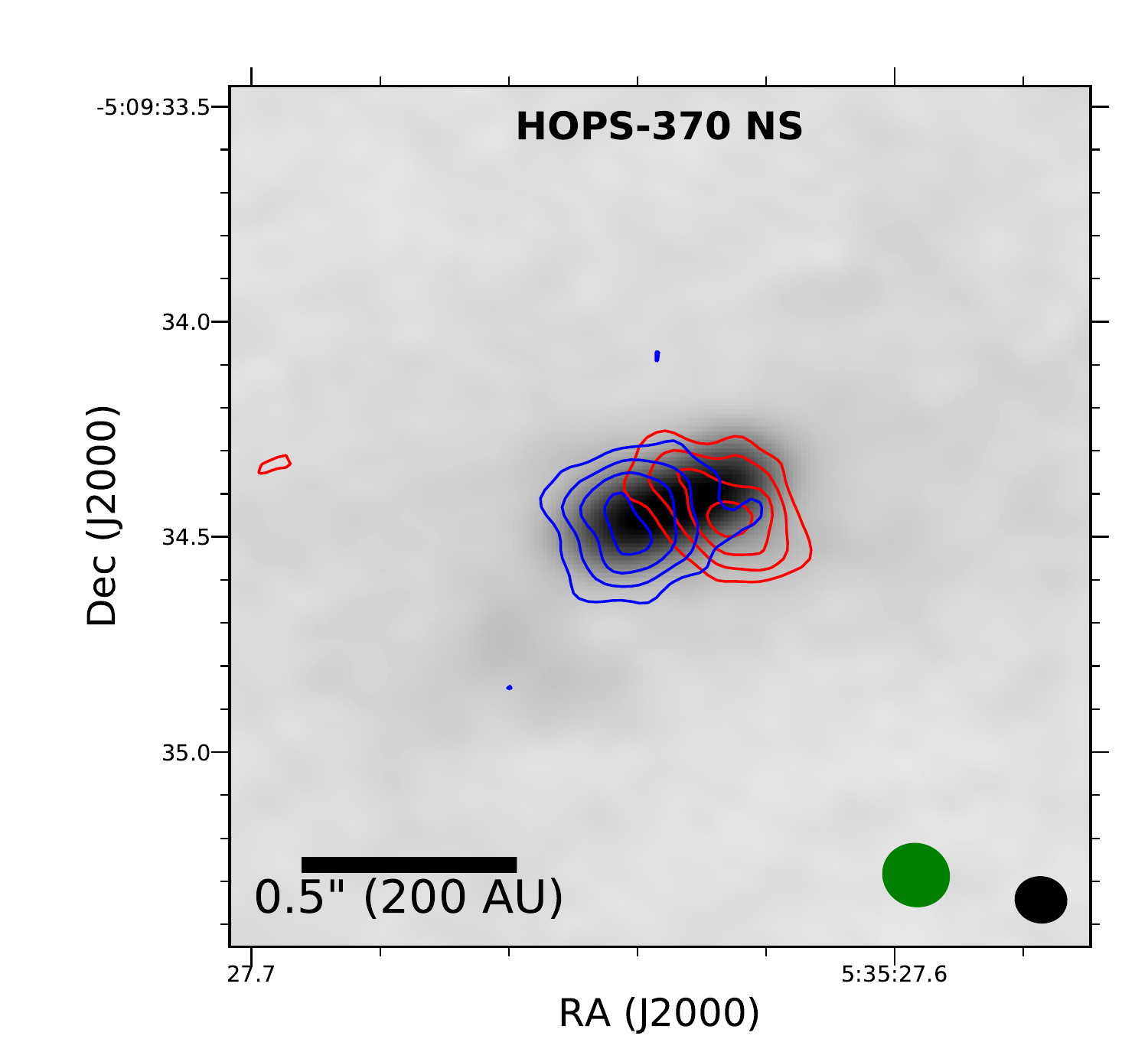}
\includegraphics[scale=0.375]{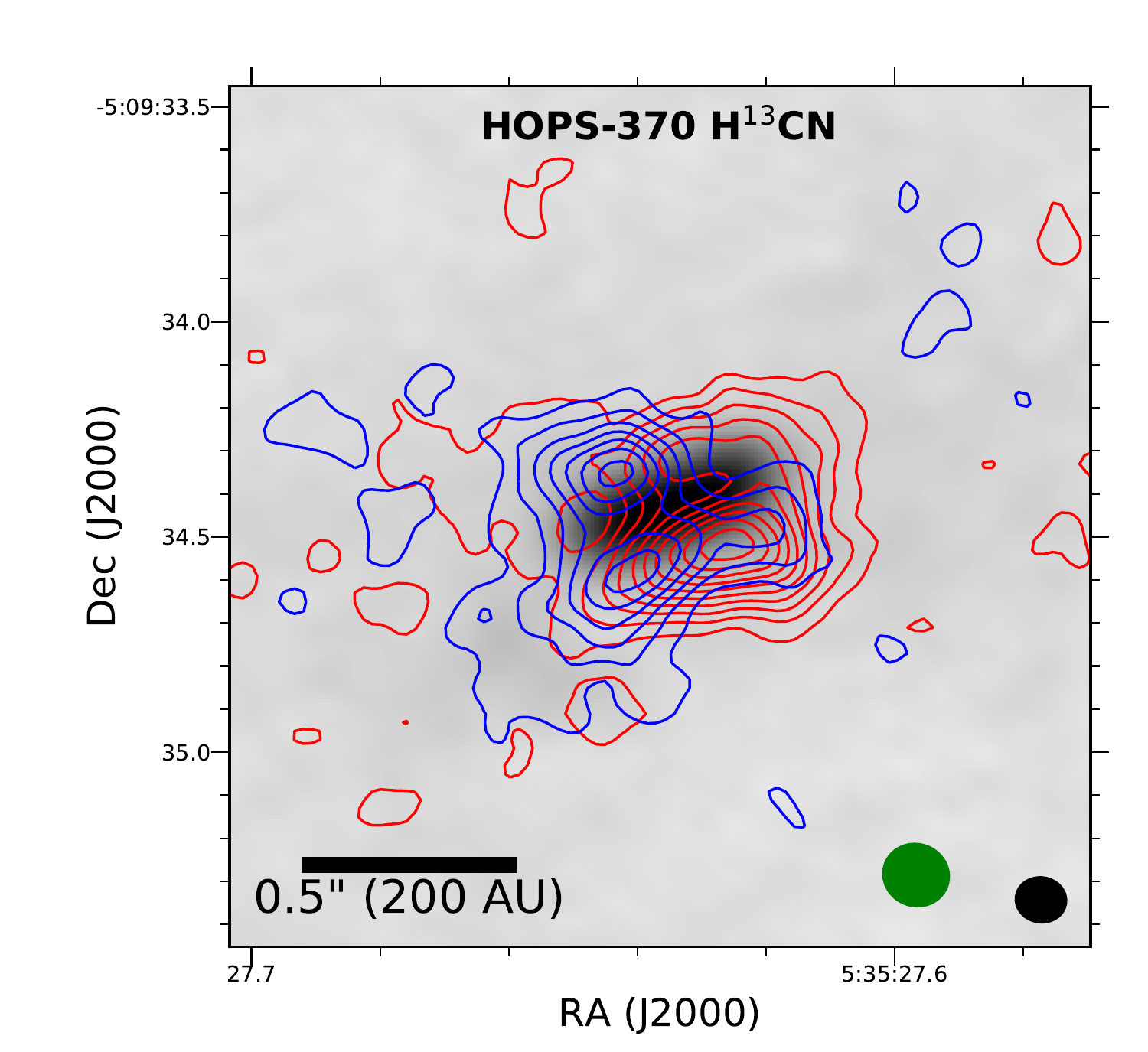}
\end{center}
\caption{Position-velocity (PV) diagrams of the molecular line emission
taken across the major axis of the disk, summed within a 0\farcs6 strip. 
The rotation signature of the disk is evident on 
scales less than $\pm$1\arcsec\ at blue- and red-shifted velocities. The emission on scales
$>$1\arcsec\ between velocities of $\sim$9.5 and 13~\kms\ corresponds to emission from the
cloud/envelope that is not well-recovered in these observations. The \cateo\ emission, however,
is dominated by an extended, low-velocity component and the higher-velocity disk emission
is only marginally-detected.
The white dashed line shows the source velocity at $\sim$11~\kms, and the white dotted line
is the Keplerian velocity curve for a 2.5~\msun\ protostar drawn for comparison.
The contours drawn start at and increase on 5$\sigma$ intervals, where $\sigma$ is 
$\sim$4$\times$ the image RMS from Table 1.
}
\label{pvdiagrams}
\end{figure}

\begin{figure}
\begin{center}
\includegraphics[scale=0.4]{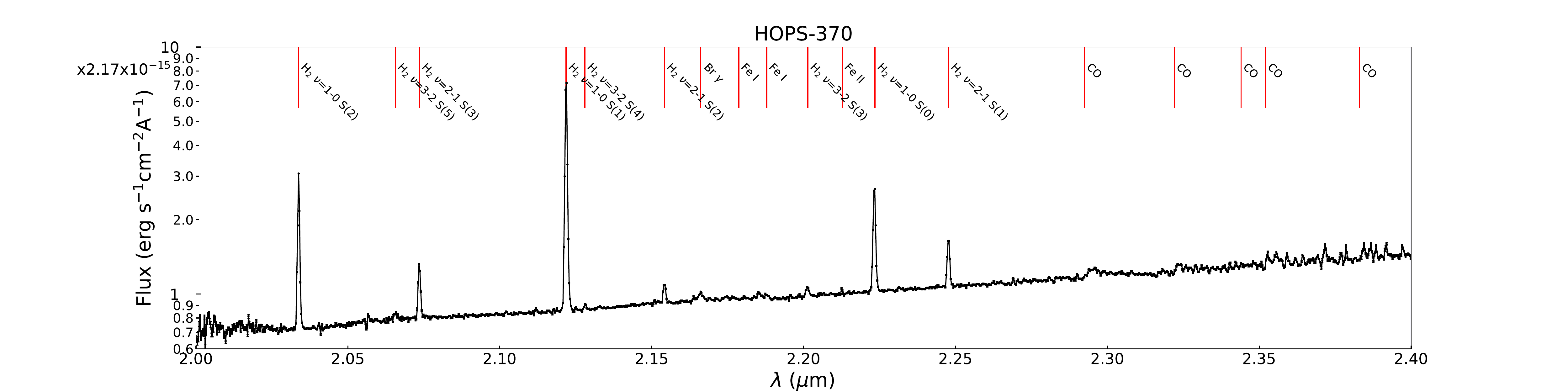}

\end{center}
\caption{Near-infrared spectrum of HOPS-370 from 2 to 2.4~\micron. The prominent spectral
features are molecular hydrogen emission lines that are presumably due to shock-heated
H$_2$ in the outflow from HOPS-370. Emission in Br~$\gamma$ and CO band heads are also 
well-detected, though less prominent. Absorption lines are absent in the spectrum of HOPS-370, indicating
that the spectrum is highly veiled.
}
\label{near-ir-spectrum}
\end{figure}

\begin{figure}
\begin{center}
\includegraphics[scale=0.65]{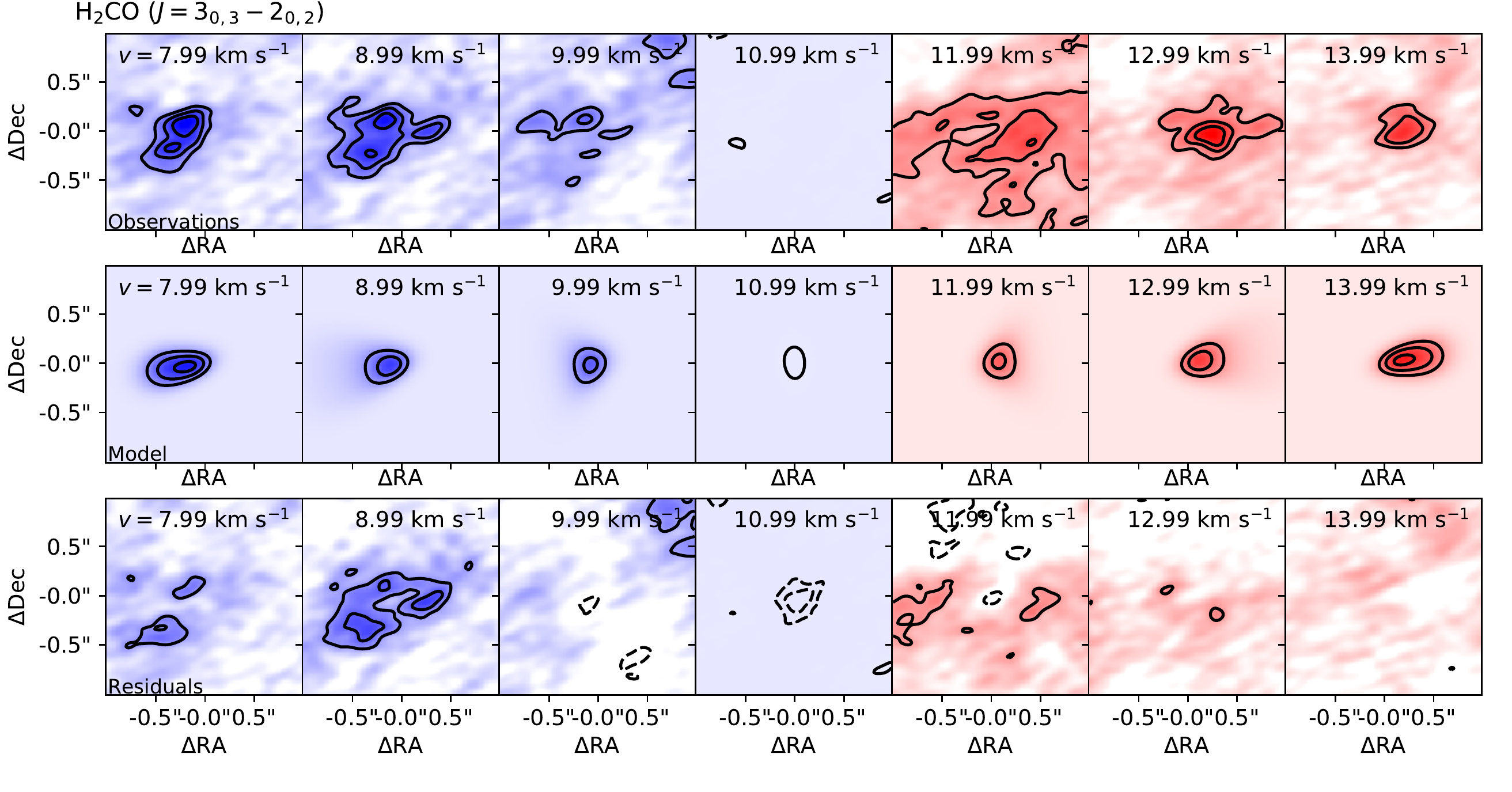}
\end{center}
\caption{Comparison of molecular line data to the kinematic model fits. The top row
shows the data, the middle row shows the model, and the bottom row shows the residuals.
We show H$_2$CO ($J=3_{0,3}\rightarrow2_{0,2}$) in Figure 7a, H$_2$CO ($J=3_{2,2}\rightarrow2_{2,1}$) in
Figure 7b, H$_2$CO ($J=3_{2,1}\rightarrow2_{2,0}$) in Figure 7c, and SO ($J_N = 6_5\rightarrow5_4$)
in Figure 7d, CH$_3$OH in Figure 7e, and NS in Figure 7f; the
RMS noise in each panel is 0.017, 0.013, 0.013, 0.016, 0.011 and, 0.021 mJy~beam$^{-1}$, respectively.
The contours in all panels start at and increase on 3$\sigma$ intervals, using the same measurement
of the RMS noise for the data, model, and residual. The color stretch is also identical for the data,
model, and residual for each molecule.
}
\label{linefit}
\end{figure}

\begin{figure}
\begin{center}
\figurenum{7b}
\includegraphics[scale=0.65]{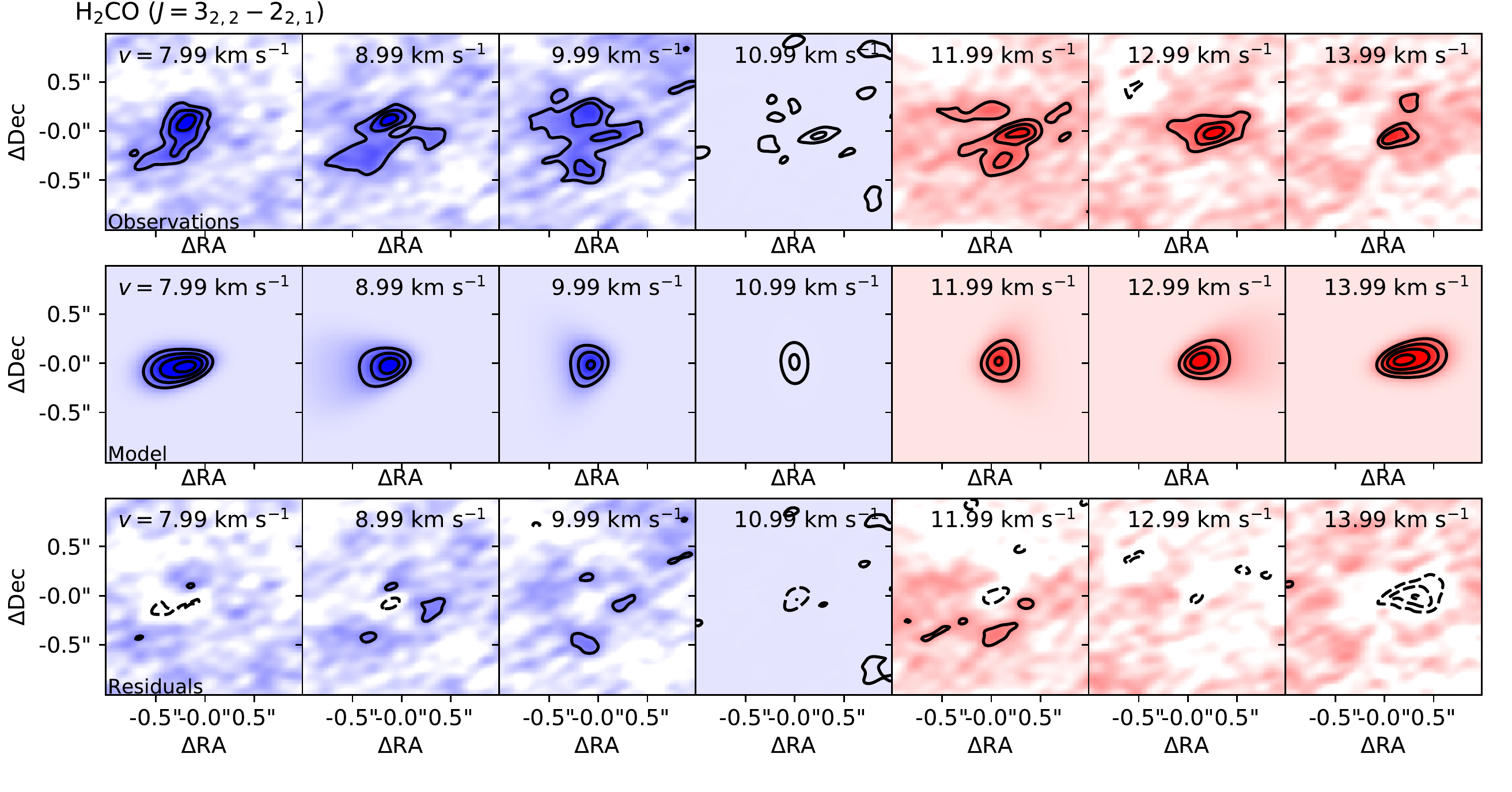}
\end{center}
\caption{Figure 7b.
}
\end{figure}

\begin{figure}
\begin{center}
\figurenum{7c}
\includegraphics[scale=0.65]{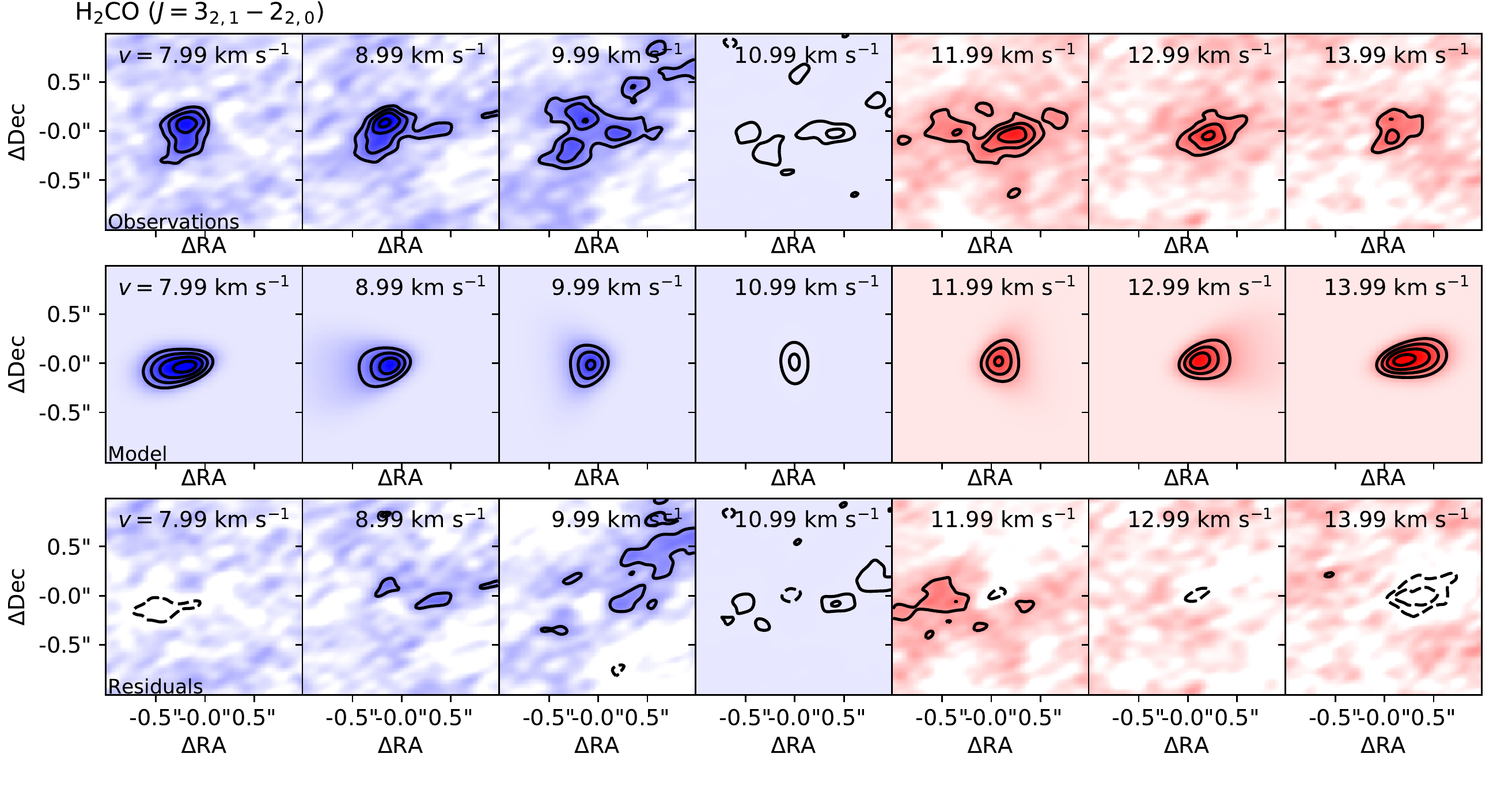}
\end{center}
\caption{Figure 7c.
}
\end{figure}

\begin{figure}
\begin{center}
\figurenum{7d}
\includegraphics[scale=0.65]{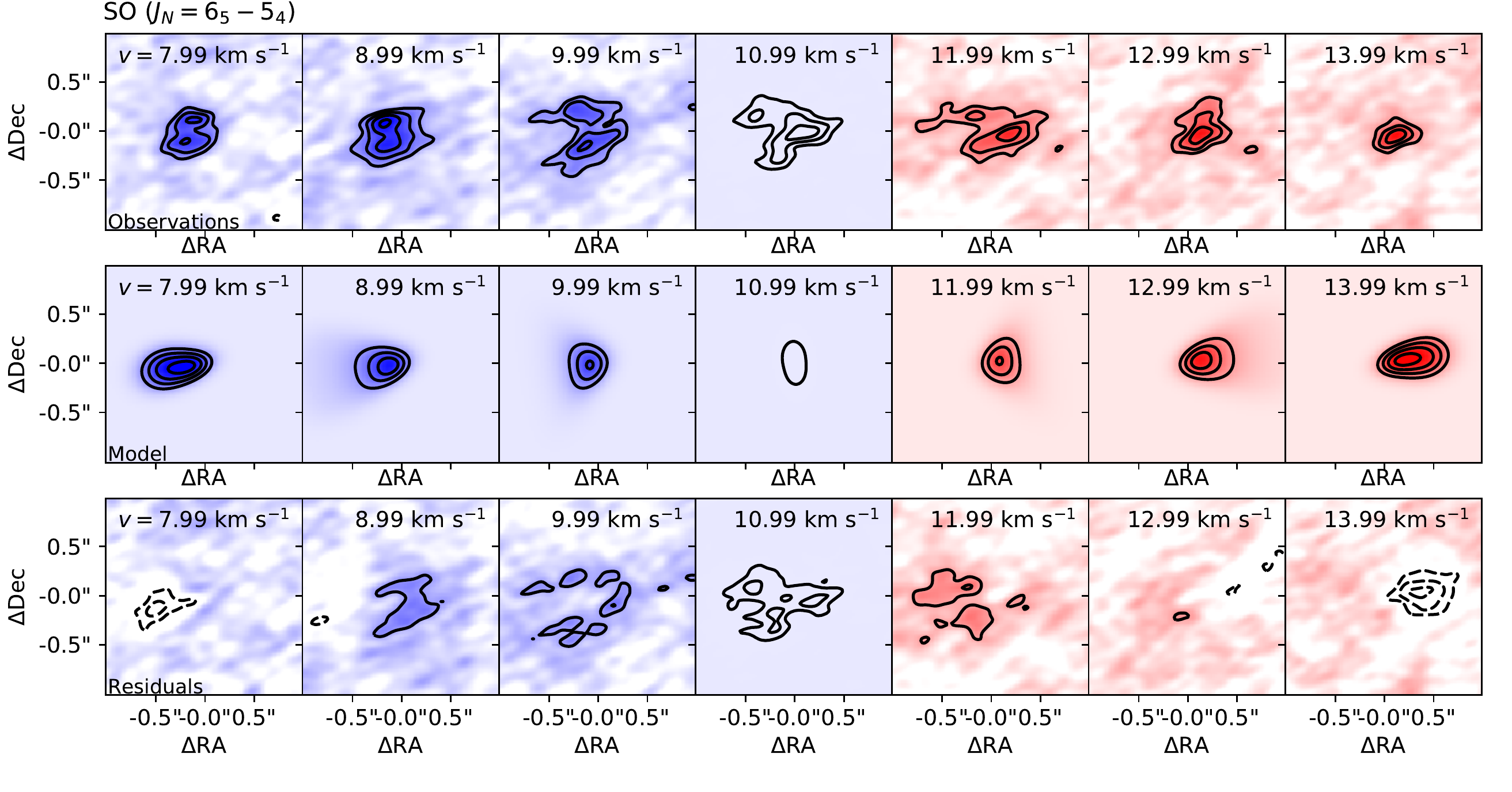}
\end{center}
\caption{Figure 7d.
}
\end{figure}

\begin{figure}
\begin{center}
\figurenum{7e}
\includegraphics[scale=0.65]{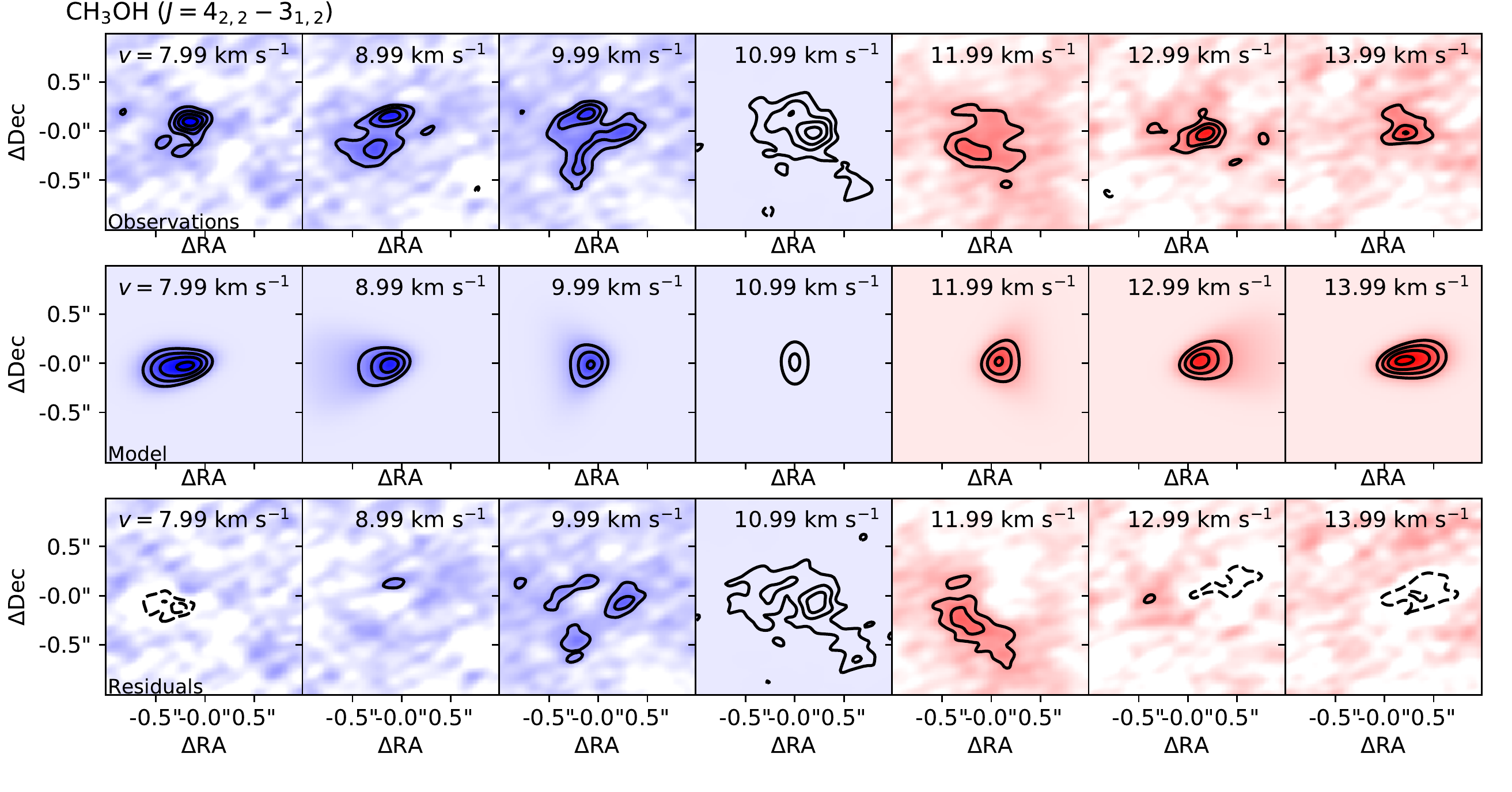}
\end{center}
\caption{Figure 7e.
}
\end{figure}

\begin{figure}
\begin{center}
\figurenum{7f}
\includegraphics[scale=0.65]{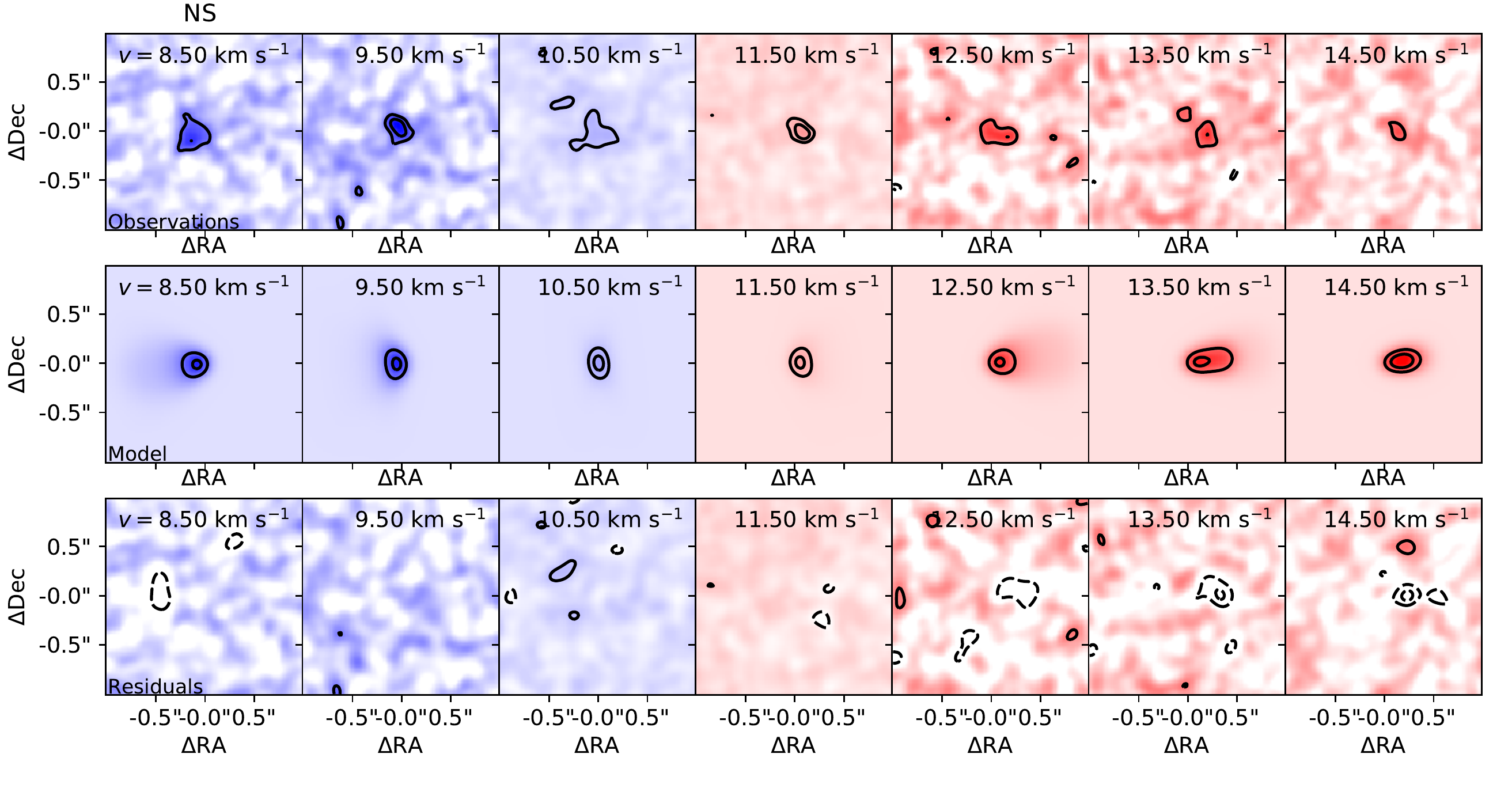}
\end{center}
\caption{Figure 7f.
}
\end{figure}

\begin{figure}
\begin{center}
\includegraphics[scale=0.575]{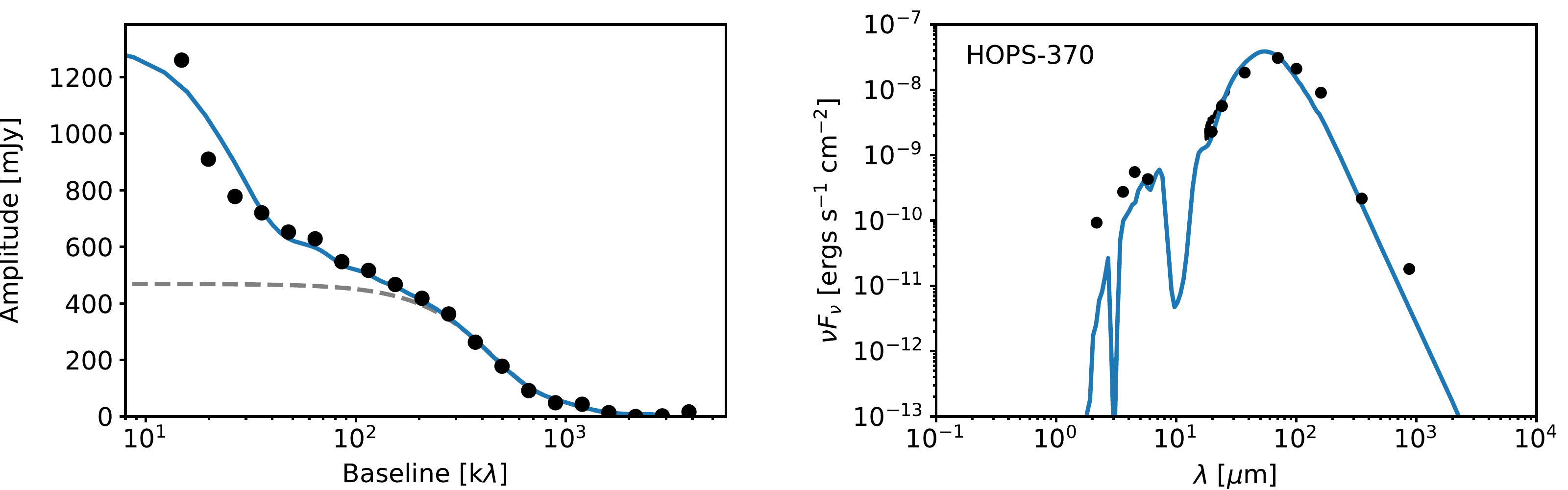}
\end{center}
\caption{Model fit visibilities and SED of HOPS-370. The 
circularly averaged visibility amplitudes 
at 0.87~mm is shown in the left panel and the SED is shown 
in the right panel. Note that the model fitting was performed
using the two dimensional visibility data and not the azimuthally-averaged
plots shown here. In all panels, the black points
show the data and the model fit is shown as the thick blue line; statistical 
uncertainties are smaller than the points shown.
The left panel also shows the contribution to the visibility amplitudes
from only the protostellar disk (gray dashed line), while the 
thick blue line shows the total contribution
from both the disk and envelope to the visibility amplitudes. The SED includes
photometry from \textit{Spitzer}, \textit{Herschel}, APEX (350 and 870~\micron) from \citet{furlan2016},
19 and 37~\micron\ flux densities from SOFIA \citep{adams2012}, 
and the SOFIA FORCAST spectrum (Karnath et al. in prep.).
}
\label{contvis}
\end{figure}

\begin{figure}
\begin{center}
\includegraphics[scale=0.575]{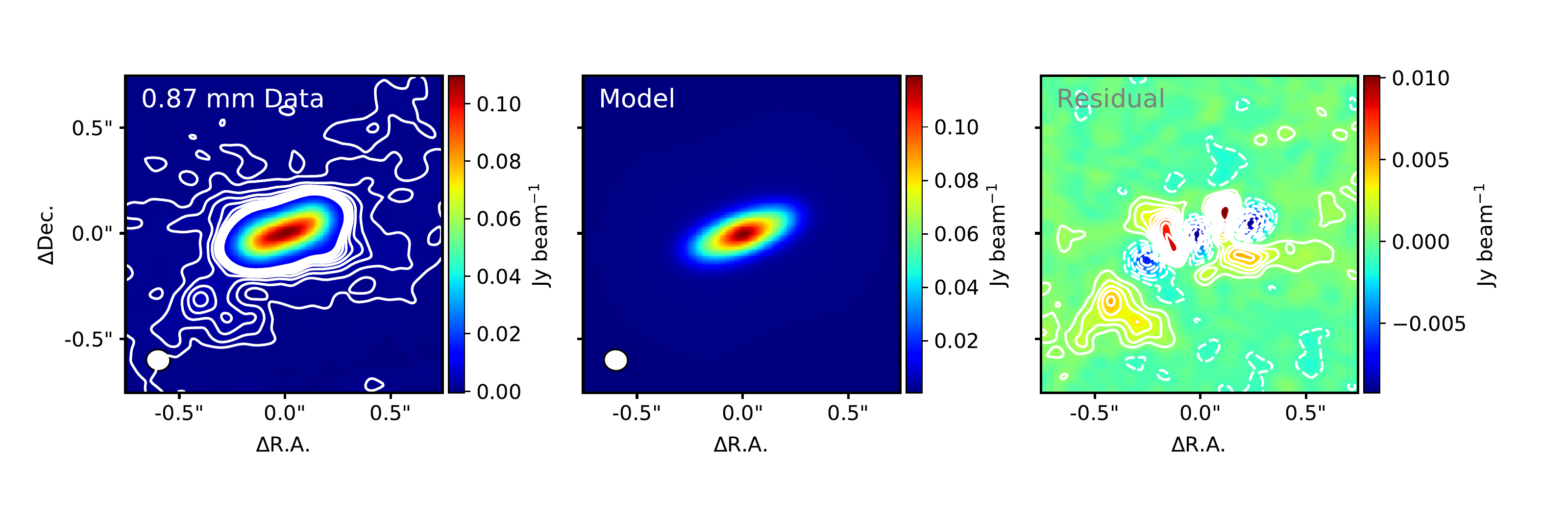}

\end{center}
\caption{Image comparison of the model fits to HOPS-370. The data are shown in
the left panel, the model is shown in the middle panel, and the residuals are shown
in the right panel. The residual image is generated from imaging the residual
visibility amplitudes and does not reflect an image-plane subtraction.
The contours in the data and residual image start at and increase on $\pm$3$\sigma$ intervals; negative residuals
are plotted as dashed contours, $\sigma$=0.31 mJy~beam$^{-1}$. The contours are shown
on the data to highlight the emission that is not shown due to the color scaling.
}
\label{contimage}
\end{figure}

\begin{figure}
\begin{center}
\includegraphics[scale=0.235]{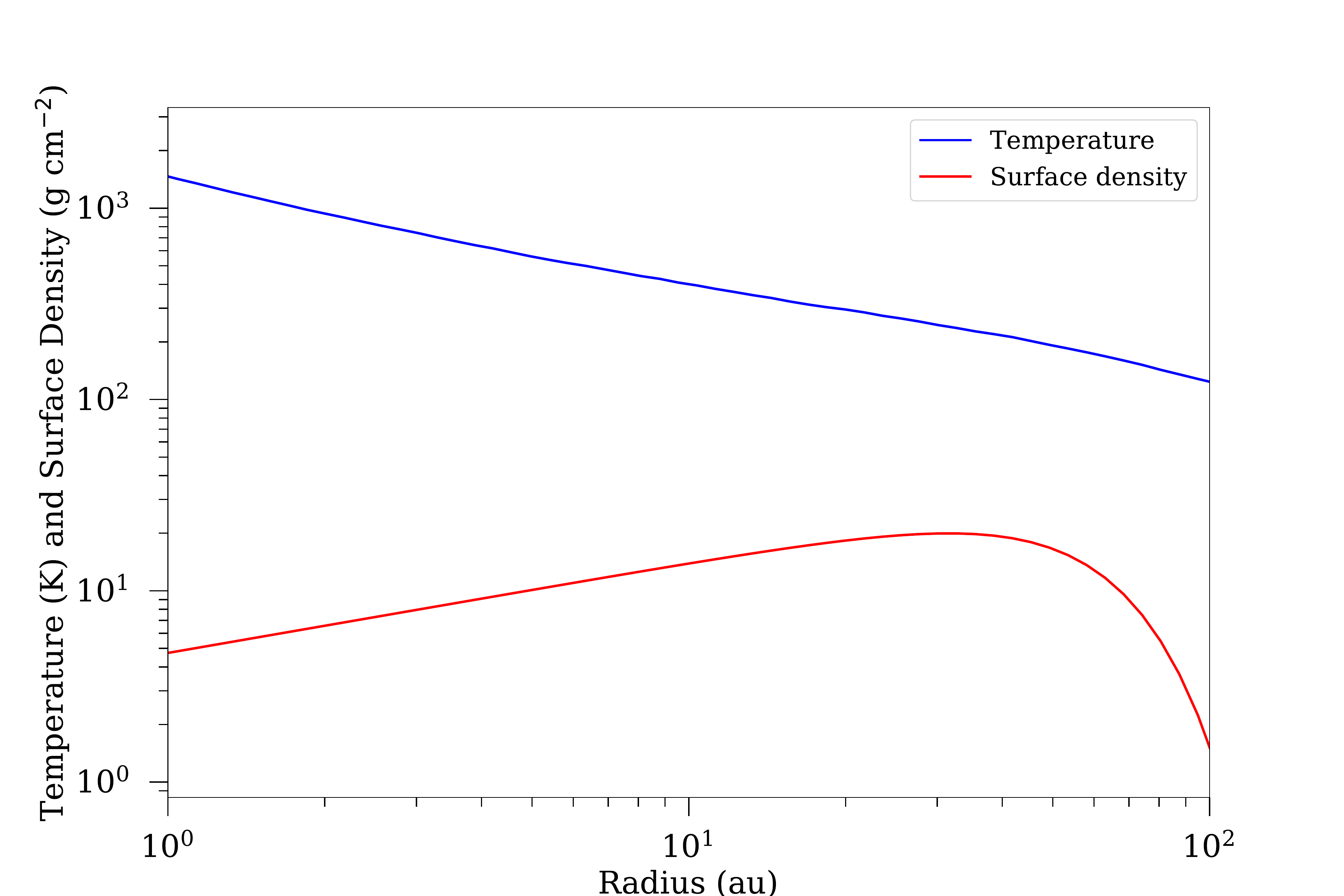}
\includegraphics[scale=0.235]{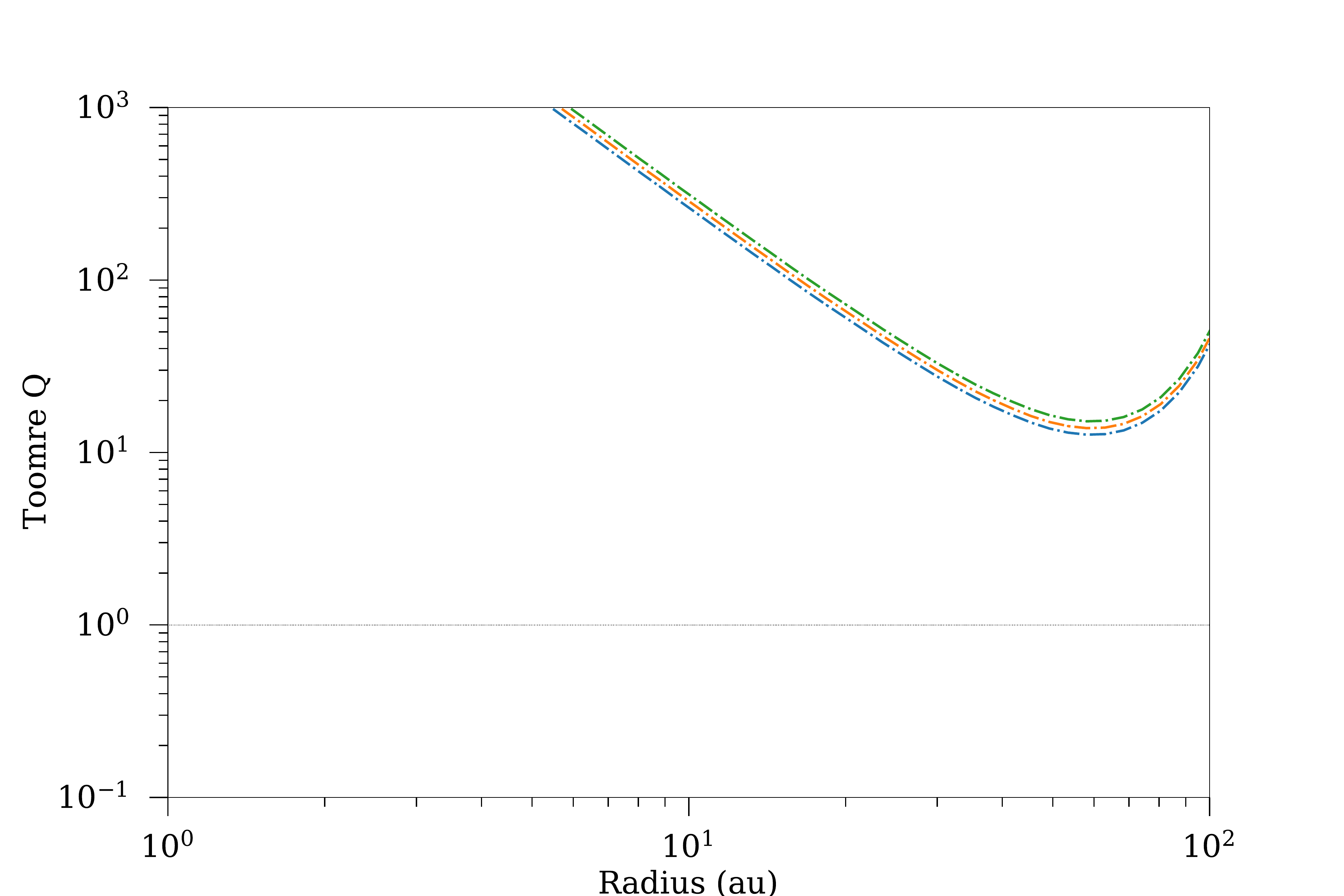}

\end{center}
\caption{The left panel shows the radial temperature and surface density profiles of the
disk of HOPS-370, derived from the best fitting radiative transfer model. The right panel
shows the inferred Toomre Q parameter calculated using Equation \ref{eq:qexact}. The three
lines correspond to three different assumptions of protostar mass: 2.1, 2.5, and 2.7~\msun\
in order from the lowest to highest lines. These results indicate that gravitational instability is
not significant within the disk of HOPS-370.
}
\label{sigmaTQ}
\end{figure}

\begin{figure}
\begin{center}
\includegraphics[scale=0.65]{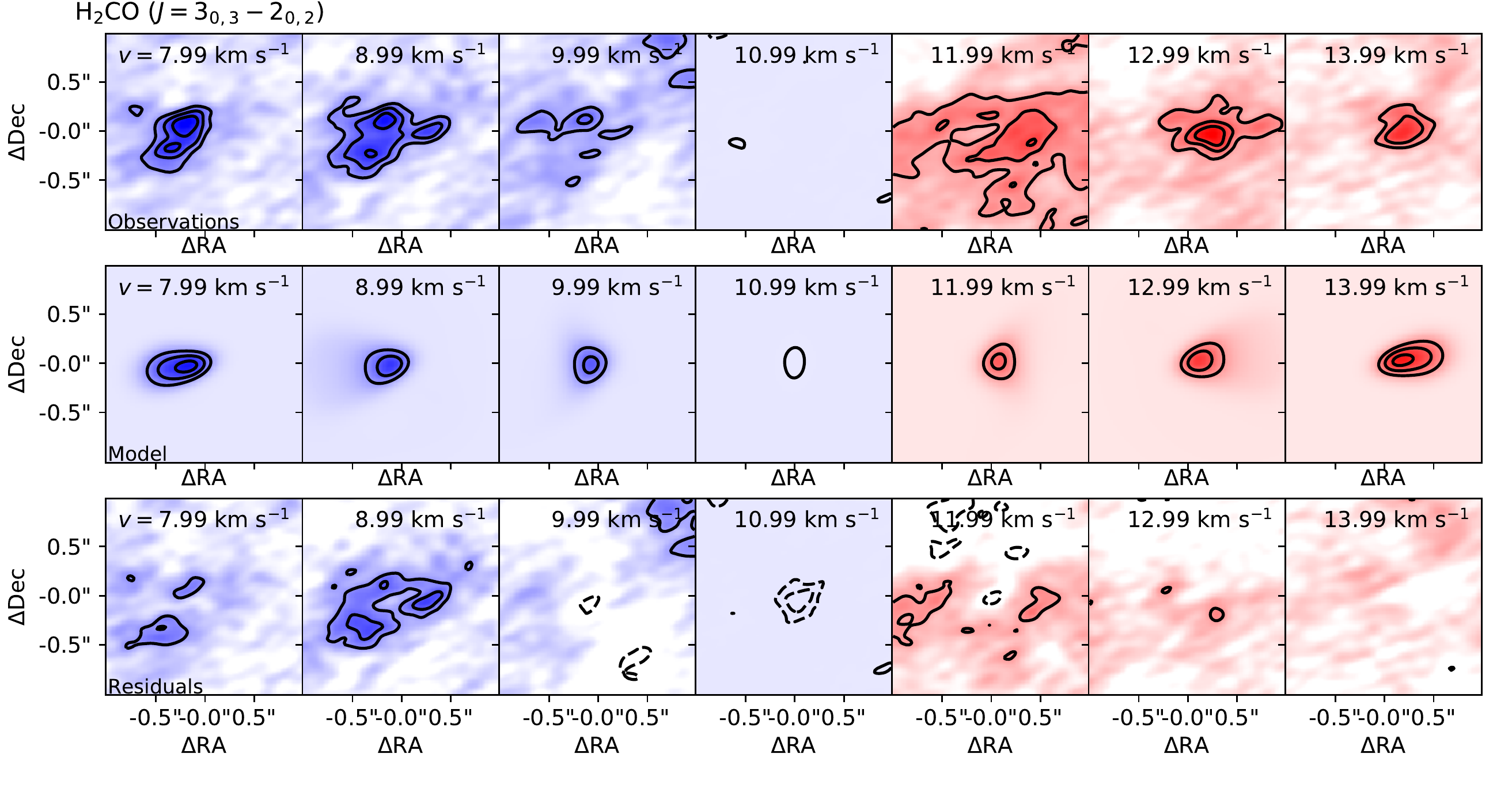}
\end{center}
\caption{Same as Figure 7, but for models that include an envelope in the fit.
}
\label{linefit-wenv}
\end{figure}

\begin{figure}
\begin{center}
\figurenum{11b}
\includegraphics[scale=0.65]{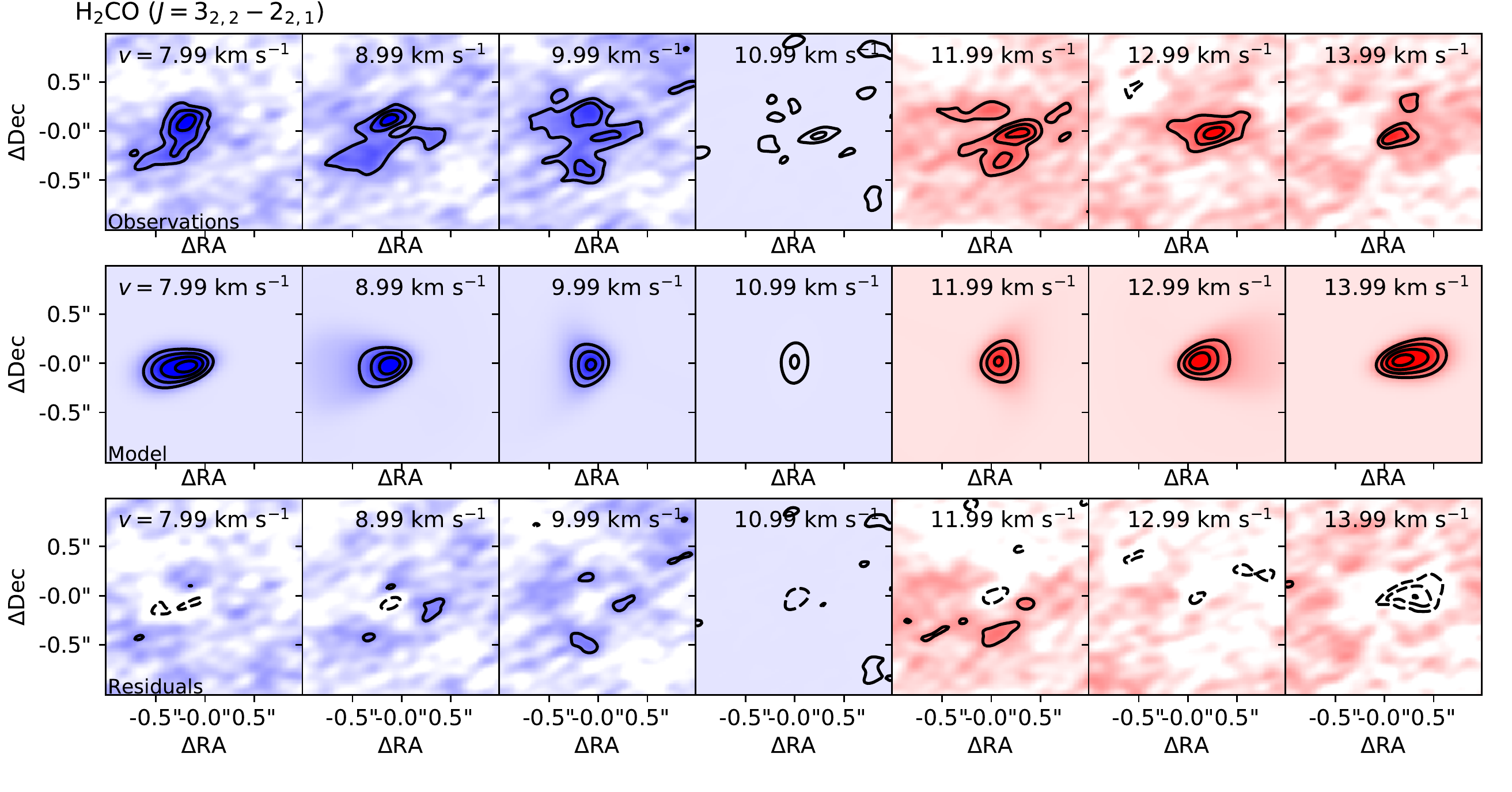}
\end{center}
\caption{Figure 11b.
}
\end{figure}

\begin{figure}
\begin{center}
\figurenum{11c}
\includegraphics[scale=0.65]{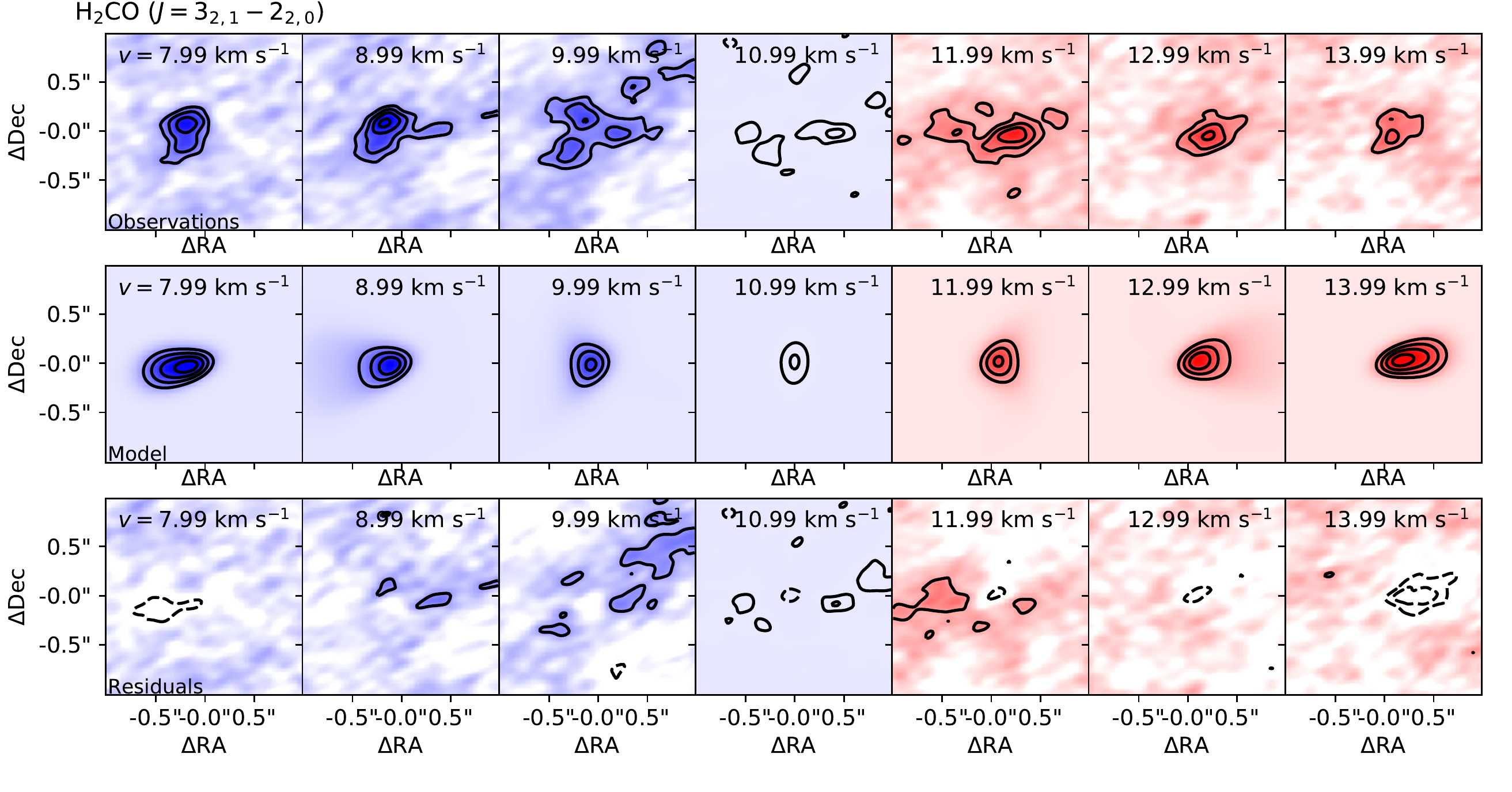}
\end{center}
\caption{Figure 11c.
}
\end{figure}

\begin{figure}
\begin{center}
\figurenum{11d}
\includegraphics[scale=0.65]{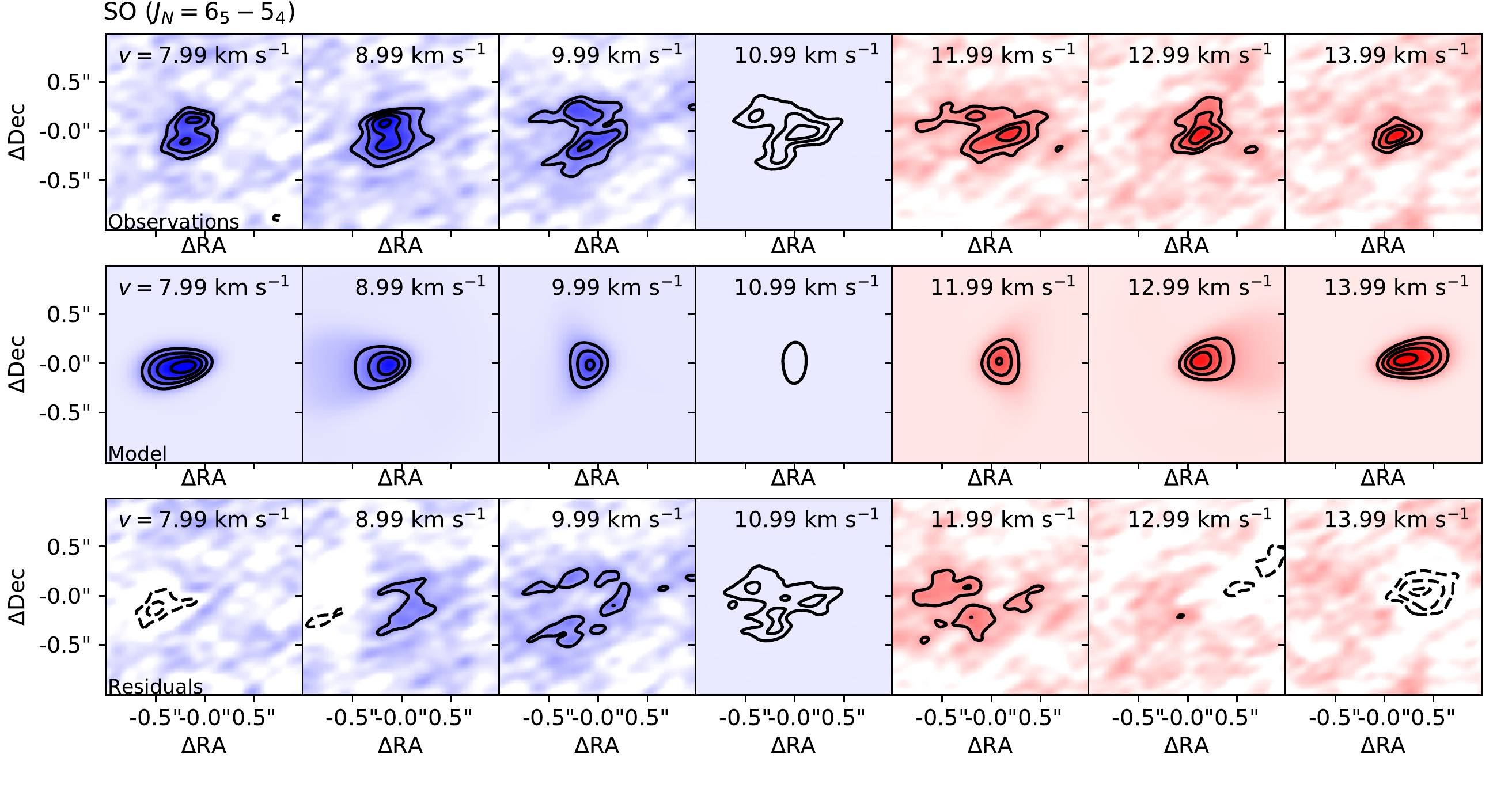}
\end{center}
\caption{Figure 11d.
}
\end{figure}

\begin{figure}
\begin{center}
\figurenum{11e}
\includegraphics[scale=0.65]{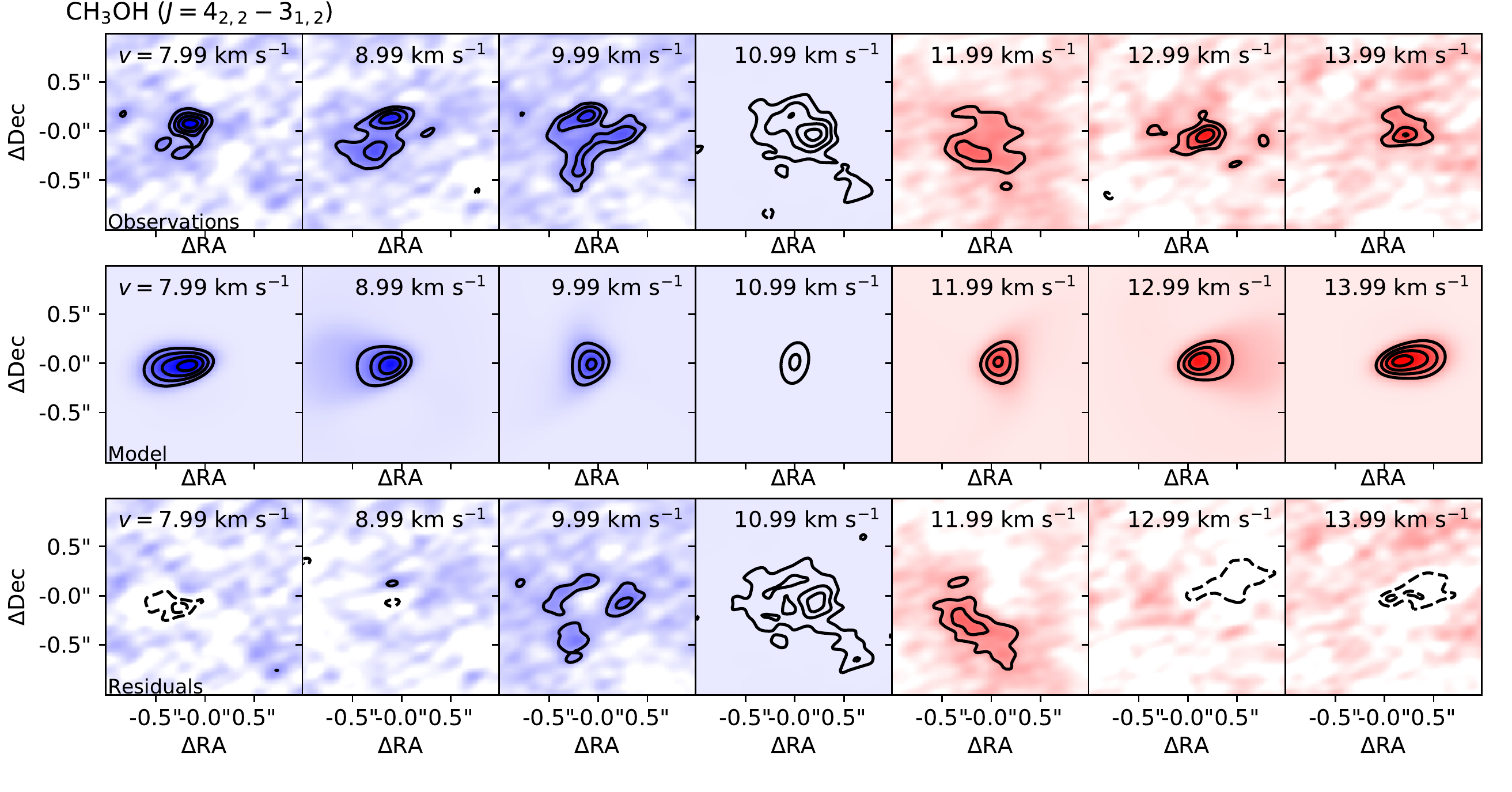}
\end{center}
\caption{Figure 11e.
}
\end{figure}

\begin{figure}
\begin{center}
\figurenum{11f}
\includegraphics[scale=0.65]{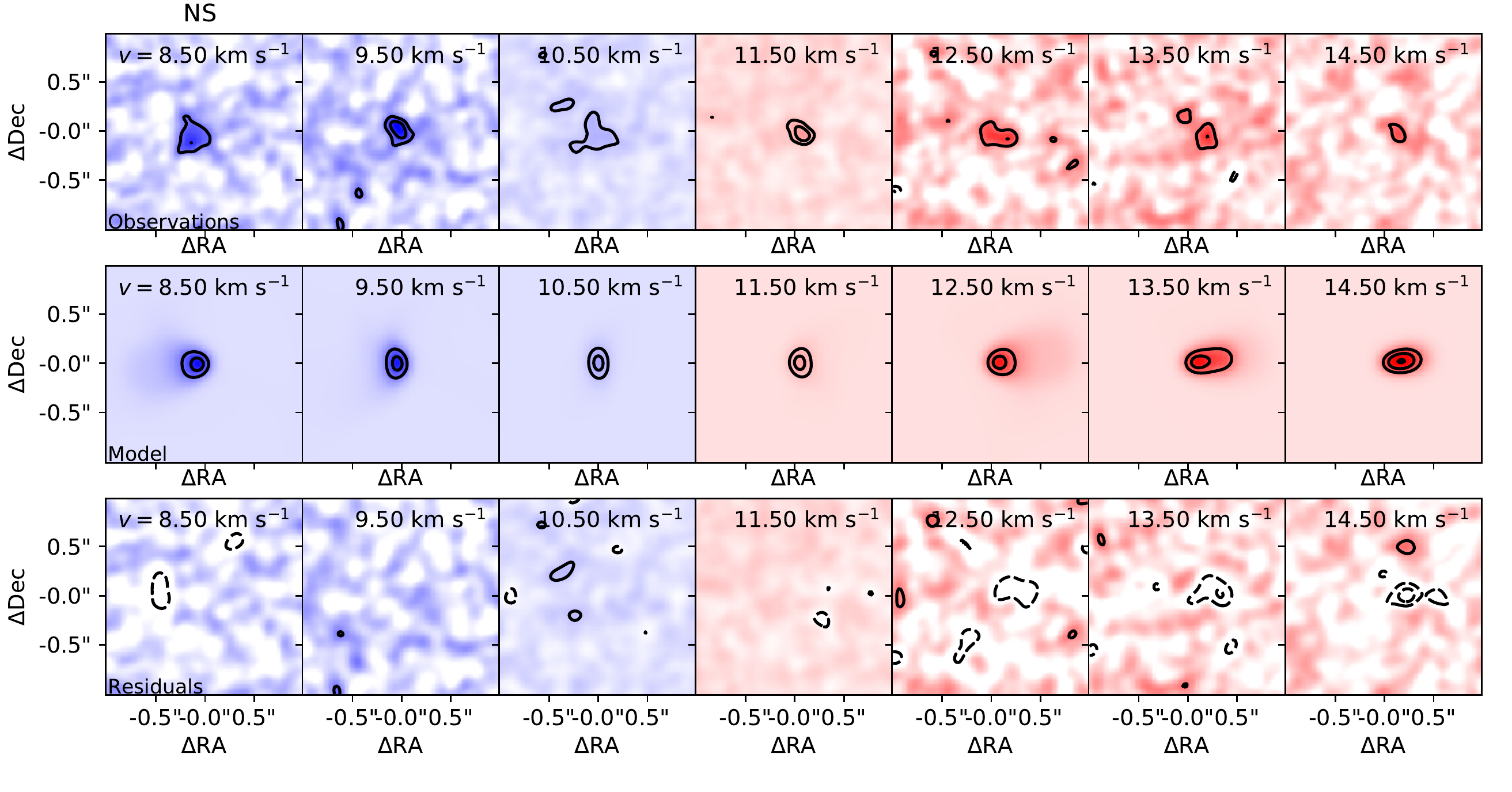}
\end{center}
\caption{Figure 11f.
}
\end{figure}

\clearpage

\input{tab1}

\input{tab2}

\input{tab3}

\input{tab4}

\input{tab5}
\input{tab6}

\input{tab7}
\end{document}

%% file: tab1.tex
\begin{deluxetable}{llllllll}
\tabletypesize{\scriptsize}
\tablewidth{0pt}
\tablecaption{Observational Setups}
\tablehead{\colhead{Telescope} & \colhead{Tracer} & \colhead{Frequency} & \colhead{Raw Channels} & \colhead{Map Channels} & \colhead{Image RMS} & \colhead{Beam} & \colhead{Detected?}\\
                                &                    & \colhead{(GHz)}     & \colhead{(kHz, km~s$^{-1}$)}      & \colhead{(km~s$^{-1}$)}           & \colhead{(mJy~beam$^{-1}$)} & \colhead{(\arcsec)} &
}
\startdata
VLA   & Continuum                                 & 29.0, 36.9 & 2000           & \nodata & 0.0072  & 0.08$\times$0.07 & Y\\
ALMA Band 6  & C$^{18}$O ($J=2\rightarrow1$)      & 219.560354 & 61.035, 0.083  & 0.5 & 11   & 0.32$\times$0.18 & Y\\
      & $^{13}$CO ($J=2\rightarrow1$)             & 220.398684 & 61.035, 0.083  & 0.25 & 23  & 0.32$\times$0.18 & Y\\
      & H$_2$CO ($J=3_{0,3}\rightarrow2_{0,2}$)   & 218.222192 & 122.07, 0.167  & 0.33 & 12  & 0.32$\times$0.18 & Y\\
      & H$_2$CO ($J=3_{2,2}\rightarrow2_{2,1}$)   & 218.475632 & 122.07, 0.167  & 0.33 & 12  & 0.32$\times$0.18 & Y\\
      & CH$_3$OH ($J=4_{2,2}\rightarrow3_{1,2}$)  & 218.440050 & 122.07, 0.167  & 0.33 & 12  & 0.32$\times$0.18 & Y\\
      & H$_2$CO ($J=3_{2,1}\rightarrow2_{2,0}$)   & 218.760066 & 122.07, 0.167  & 0.33 & 12  & 0.32$\times$0.18 & Y\\
      & SO ($J_N = 6_5\rightarrow5_4$)            & 219.949442 & 122.07, 0.167  & 0.33 & 14  & 0.32$\times$0.18 & Y\\
      & $^{12}$CO ($J=2\rightarrow1$)             & 230.538    & 282.23, 0.367  & 1.0 & 9.2  & 0.31$\times$0.17 & Y\\
      & N$_2$D$^+$ ($J=3\rightarrow2$)            & 231.321828 & 282.23, 0.367  & 0.367 & 13 & 0.30$\times$0.17 & N\\
      & $^{13}$CS ($J=5\rightarrow4$)             & 231.220685 & 282.23, 0.367  & 0.366 & 11 & 0.30$\times$0.17 & Y\\
      & Continuum                                 & 232.5      & 31250          & \nodata & 0.22 & 0.23$\times$0.13 & Y\\
ALMA Band 7      & Continuum                      & 333, 344   & 31250         & \nodata & 0.31 & 0.11$\times$0.1 & Y\\
      & NS ($J=15/2\rightarrow13/2$, $F=17/2\rightarrow15/2$)  & 346.220137  & 564.45, 0.489    & 0.5  & 20 & 0.20$\times$0.20 & Y\\
      & NS ($J=15/2\rightarrow13/2$, $F=15/2\rightarrow13/2$)  & 346.221163  & 564.45, 0.489    & 0.5  & 20 & 0.20$\times$0.20 & Y\\
      & H$^{13}$CN ($J=4\rightarrow3$)            & 345.3397693 & 564.45, 0.489  & 0.5  & 19 & 0.18$\times$0.17 & Y\\
      & SO$_2$ ($J=13_{2,12}\rightarrow12_{1,11}$) & 345.3385391 & 564.45, 0.489  & 0.5  & 19 & 0.18$\times$0.17 & Y\\
\enddata
\tablecomments{
The lines contained in the ALMA Band 7 data were detected within a spectral window
centered on the  $^{12}$CO ($J=3\rightarrow2$) transition. 
The spectral resolutions listed for the ALMA continuum bands are for a single channel while 
the full bandwidths were $\sim$1.875 GHz, also the full bandwidth for the VLA observations
was 8 GHz, spilt into 2, 4~GHz bands centered at the listed frequencies. The the raw channels columns
refers to the spectral resolution of the data itself, while the image channels refers to the spectral
resolution after averaging during the imaging process.
}

\end{deluxetable}

%% file: tab2.tex
\begin{deluxetable}{lllllll}
\tabletypesize{\scriptsize}
\tablewidth{0pt}
\tablecaption{HOPS-370 Continuum Flux density}
\tablehead{\colhead{Wavelength} & \colhead{Flux Density} & \colhead{Peak I$_{\nu}$}   & \colhead{RMS}               & \colhead{Decon. Size} & \colhead{Decon. PA} & \colhead{Reference}\\
           \colhead{(mm)}       & \colhead{(mJy)}        & \colhead{(mJy~beam$^{-1}$)}& \colhead{(mJy~beam$^{-1}$)} & \colhead{(\arcsec)}   & \colhead{(\degr)}                 
}           
\startdata
0.87 & 533.2$\pm$10.0 & 109.89  & 0.39     & 0.34$\times$0.11 & 109.7 & 1\\
1.3  & 207.4$\pm$3.0    & 95.7    & 0.22     & 0.32$\times$0.10 & 109.7 & 1\\
7.0 & 4.0$\pm$0.4       & \nodata &  0.015   &  0.24$\times$0.22 & 108 & 2\\
9.1 & 3.65$\pm$0.36    & \nodata &  \nodata &  0.0069  & \nodata & 1\\
9.1 (disk-only) & 0.732$\pm$0.075 & 0.122    & 0.0069 & 0.27$\times$0.09  & 112.4& 1\\
13.0 & 2.6$\pm$0.3      & \nodata &  0.010   &  0.52$\times$0.21 & 7.2 & 2\\
30.0 & 2.16$\pm$0.22    & \nodata &  0.009   &  2.8$\times$1.8\tablenotemark{a} & 30 & 2\\
50.0 & 1.76$\pm$0.18    & \nodata &  0.011   &  4.8$\times$2.6\tablenotemark{a} & 42 & 2\\
\enddata
\tablecomments{Integrated flux densities measured toward HOPS-370. The 
flux densities and peak intensities from this work are derived from 
image-plane Gaussian fitting of HOPS-370. The 9.1~mm
(disk dust-only) measurement is derived from the multi-component Gaussian fitting and this
is the component that fits the disk. The other 9.1~mm flux density reflects the
total integrated flux from the four component fit. Uncertainties reflect statistical uncertainties and
do not include the uncertainty in the absolute flux calibration for the 0.87~mm, 1.3~mm, and 9.1~mm
flux densities. The other measurements include a 10\% absolute calibration uncertainty.
References:
 (1) this work, and (2) \citep{osorio2017}.
}
\tablenotetext{a}{The source was unresolved at these wavelengths and the angular size reflects
the convolved size.}
\end{deluxetable}

%% file: tab3.tex
\begin{deluxetable}{lllllll}
\tabletypesize{\scriptsize}
\tablewidth{0pt}
\tablecaption{VLA 9~mm Gaussian Fit}
\tablehead{\colhead{Component} & \colhead{RA}      & \colhead{Dec.}    & \colhead{Int. Flux} & \colhead{Peak Intensity}     & \colhead{Deconvolved Size} & \colhead{PA}\\
                               &  \colhead{(ICRS)} & \colhead{(ICRS)}  & \colhead{(mJy)}      & \colhead{(mJy beam$^{-1}$)} & \colhead{\arcsec}         & \colhead{(\degr)} 
}
\startdata
Component 1 (Central Peak)  & 05:35:27.63360 & -05:09:34.408  & 1.44$\pm$0.013   & 1.41$\pm$0.012  & 0.016$\times$0.007  &  15.9 \\
Component 2 (Jet)  & 05:35:27.63422 & -05.09.34.302  & 7.54$\pm$0.046   & 2.11$\pm$0.01   & 0.285$\times$0.021  &  7.0 \\
Component 3 (Jet)  & 05:35:27.63438 & -05.09.34.282  & -6.06$\pm$0.043  & -1.93$\pm$0.01  & 0.250$\times$0.019  &  7.2 \\
Component 4 (disk)   & 05:35:27.63473 & -05.09.34.416  & 0.732$\pm$0.075 & 0.122$\pm$0.01  & 0.270$\times$0.090  &  112.4 \\
\enddata
\tablecomments{The individual components were fit non-interactively using the CASA task \textit{imfit}. We 
only provided initial guesses for each component. Component 1 was constrained to have an angular size 
equivalent to the beam, while Components 2 and 3 had initial parameter estimates elogated in their fitted
directions, but without restrictions on their angular size or position angle. Then Component 4 was provided
with an initial position angle and angular size consistent with the expected disk, but the Gaussian parameters
were also not fixed for Component 4. Component 4 is thus taken to provide an estimate of the flux density
from the dust emission from the disk at 9~mm. The fact that Component 2 has a large positive flux density
and Component 3 has a large negative flux density is not important, the combination of these two components
fits the jet morphology well. The sum of the flux densites from all componets agrees well the the total flux density
measured within a polygon and the flux density from \citet{osorio2017}.
}

\end{deluxetable}

%% file: tab4.tex
\begin{deluxetable}{lll}
\tabletypesize{\scriptsize}
\tablewidth{0pt}
\tablecaption{Molecular Line Modeling Parameters}
\tablehead{\colhead{Parameter Description} & \colhead{Parameter}    & \colhead{Parameter Range}}           
\startdata
 Stellar Mass                              &    $M_{*}$ (\msun)     &   0.1 - 10.0\\
 Disk Mass                                 &    $M_{disk}$ (\msun)  &    0.01 - 1.0\\
 Disk Outer Radius                         &    $r_{c}$ (au)     &    1.0 - 10000.0\\
 Disk Inner Radius                         &    $R_{in,disk}$   (au)     &    0.1\\
 Surface Density Power-law Index           &      $\gamma$          &     0.0 - 2.0\\
 Disk Vertical Density Profile             &      $h(r)$            &            HSEQ\\
 Temperature at 1 au                       &      $T_0$ (K)         &          10.0 - 1000.0\\
 Temperature Profile Power-law Index       &      $q$               &       0.35 (0.0 - 1.0)\\  
 Envelope Mass\tablenotemark{a}            &     $M_{env}$ (\msun)  &    0 - 10.0\\
 Envelope Outer Radius\tablenotemark{a}    &     $R_{env}$ (au)     &    1000.0 - 10000.0\\
 Envelope Inner Radius\tablenotemark{a}    &     $R_{in,env}$ (au)     &    0.1\\
 Centrifugal Radius\tablenotemark{a}       &     $R_{c}$ (au)     &    =$R_{disk}$\\
 Turbulent velocity Width                  &      $a$ (\kms)        &     0.001 - 0.13\\  
 Inclination                               &           $i$ (\degr)        &   72.2\\
 Density Scaling in Outflow Cavity         &     $f_{cav}$          &    0.5\\
 Outflow Cavity Shape power-law index      &         $\xi$          &     1.0\\
 Right Ascension Offset                    &       $\Delta$x (\arcsec)       &  -0.2 - 0.2\\
 Declination Offset                        &       $\Delta$y (\arcsec)        &  -0.2 - 0.2\\
 Distance                                  &       d (pc)           &  400.0\\
 Position Angle East of North              &           $pa$ (\degr) & 310 - 380\\
 System Velocity                           &           $v_{sys}$ (\kms) & 7.0 - 15.0\\
 H$_2$CO Abundnace                         &           (per H$_2$)   & 1.0$\times$10$^{-9}$\\
 SO Abundance                              &           (per H$_2$)   & 3.14$\times$10$^{-9}$ (1.0$\times$10$^{-8}$ - 1.0$\times$10$^{-10}$)\\
 CH$_3$OH Abundance                        &           (per H$_2$)   & 1.0$\times$10$^{-8}$ (1.0$\times$10$^{-7}$ - 1.0$\times$10$^{-10}$)\\
 NS Abundance                              &           (per H$_2$)   & 13.14$\times$10$^{-9}$ (1.0$\times$10$^{-8}$ - 1.0$\times$10$^{-10}$)\\
\enddata
\tablecomments{The disk and envelope parameters were varied (range) and fixed (single value)
for the molecular line modeling. For $q$ and the SO abundance, we explored the effects of
having these parameters fixed for most runs, but allowed them to vary in a few specific instances that
are described in the text: hence the single value and range of values in parentheses.
More explanation of the parameters and the model setup is 
provided in Section 4.1.
The parameter f$_{cav}$ is the reduction of 
envelope density within the outflow cavity, and the outflow cavity shape and opening 
angle are defined by the parameter $\xi$, translating to opening angle can be done 
by calculating 2tan$^{-1}$($\xi$).
}
\tablenotetext{a}{Note that only the models shown in the Appendix included an envelope. The models presented
in Figure 7 and Table 5 did not include an envelope.}
\end{deluxetable}

%% file: tab5.tex
\begin{deluxetable}{lllllllllllll}
\rotate
\tabletypesize{\scriptsize}
\tablewidth{0pt}
\tablecaption{HOPS-370 Molecular Line Models - Exponentially Tapered Disks without Envelope}
\tablehead{\colhead{Transition(s)} & \colhead{Mass} & \colhead{Disk Radius} & \colhead{Disk Mass}  & \colhead{V$_{lsr}$}     & \colhead{Pos. Angle} & \colhead{$\Delta$x} & \colhead{$\Delta$y} & \colhead{$\gamma$} & \colhead{T(1 au)} & \colhead{q} & \colhead{$log_{10}$(Mol./H$_2$)\tablenotemark{b}}\\
                                   & \colhead{(M$_{\sun}$)} & \colhead{(AU)} & \colhead{(M$_{\sun}$)} & \colhead{(km~s$^{-1}$)} & \colhead{\degr}      & \colhead{(\arcsec)} & \colhead{(\arcsec)} &                    &    \colhead{(K)}  &       & 
}
\startdata
  H$_2$CO ($J=3_{0,3}\rightarrow2_{0,2}$) & 2.72$\pm$0.03 & 82.28$\pm$0.7 & 0.31$\pm$0.004 & 11.07$\pm$0.02 & 353.86$\pm$0.34 & 0.004$\pm$0.002 & 0.022$\pm$0.002 & 1.08$\pm$0.01 & 995.92$\pm$5.59 & 0.35 & \nodata \\ 
                        H$_2$CO (2 lines) & 2.33$\pm$0.03 & 82.79$\pm$0.8 & 0.43$\pm$0.007 & 11.08$\pm$0.01 & 352.69$\pm$0.39 & 0.011$\pm$0.002 & 0.012$\pm$0.002 & 1.16$\pm$0.02 & 998.06$\pm$1.72 & 0.35 & \nodata \\ 
                        H$_2$CO (3 lines) & 2.52$\pm$0.02 & 82.99$\pm$1.5 & 0.54$\pm$0.012 & 11.09$\pm$0.01 & 352.64$\pm$0.25 & 0.006$\pm$0.001 & 0.014$\pm$0.001 & 1.10$\pm$0.02 & 998.82$\pm$1.63 & 0.35 & \nodata \\ 
 CH$_3$OH ($J=4_{2,2}\rightarrow3_{1,2}$) & 2.37$\pm$0.03 & 109.29$\pm$1.3 & 0.39$\pm$0.007 & 10.99$\pm$0.02 & 358.02$\pm$0.58 & 0.010$\pm$0.002 & 0.004$\pm$0.002 & 0.89$\pm$0.02 & 998.70$\pm$1.89 & 0.35 & \nodata \\ 
 CH$_3$OH\tablenotemark{a}, H$_2$CO (3 lines) & 2.50$\pm$0.02 & 90.36$\pm$1.0 & 0.49$\pm$0.007 & 11.06$\pm$0.01 & 353.45$\pm$0.24 & 0.007$\pm$0.001 & 0.013$\pm$0.001 & 1.05$\pm$0.01 & 998.38$\pm$1.09 & 0.35 & -0.88$\pm$0.013  \\ 
           SO ($J_N = 6_5\rightarrow5_4$) & 2.51$\pm$0.05 & 104.78$\pm$2.1 & 0.19$\pm$0.005 & 11.42$\pm$0.03 & 345.10$\pm$0.74 & 0.023$\pm$0.002 & 0.041$\pm$0.002 & 0.80$\pm$0.03 & 999.21$\pm$1.11 & 0.35 & \nodata \\ 
         SO\tablenotemark{a} and CH$_3$OH & 2.37$\pm$0.04 & 120.53$\pm$0.9 & 0.29$\pm$0.005 & 11.03$\pm$0.03 & 353.00$\pm$0.58 & 0.011$\pm$0.002 & 0.029$\pm$0.002 & 0.67$\pm$0.02 & 998.44$\pm$1.85 & 0.35 & -0.88$\pm$0.005  \\ 
   SO\tablenotemark{a}, H$_2$CO (3 lines) & 2.55$\pm$0.02 & 99.29$\pm$0.3 & 0.42$\pm$0.003 & 11.12$\pm$0.01 & 351.46$\pm$0.24 & 0.009$\pm$0.001 & 0.018$\pm$0.001 & 0.93$\pm$0.01 & 999.63$\pm$0.47 & 0.35 & -2.84$\pm$0.203  \\ 
                                       NS & 3.63$\pm$0.16 & 79.77$\pm$20.8 & 0.01$\pm$0.006 & 11.05$\pm$0.09 & 350.56$\pm$1.69 & -0.007$\pm$0.004 & 0.014$\pm$0.004 & 0.31$\pm$0.24 & 776.35$\pm$246.00 & 0.35 & \nodata \\ 
                  NS, SO\tablenotemark{a} & 1.83$\pm$0.03 & 82.87$\pm$1.3 & 0.02$\pm$0.001 & 10.98$\pm$0.03 & 352.76$\pm$0.54 & 0.016$\pm$0.002 & 0.062$\pm$0.001 & 1.06$\pm$0.02 & 998.44$\pm$1.23 & 0.35 & -0.88$\pm$0.011  \\ 
            NS\tablenotemark{a}, CH$_3$OH & 2.44$\pm$0.03 & 109.43$\pm$1.7 & 0.38$\pm$0.008 & 11.00$\pm$0.02 & 356.56$\pm$0.56 & 0.008$\pm$0.002 & 0.006$\pm$0.002 & 0.89$\pm$0.02 & 996.65$\pm$3.85 & 0.35 & -0.88$\pm$0.018  \\ 
   NS\tablenotemark{a}, H$_2$CO (3 lines) & 2.56$\pm$0.02 & 88.86$\pm$1.1 & 0.50$\pm$0.008 & 11.09$\pm$0.01 & 352.46$\pm$0.27 & 0.007$\pm$0.001 & 0.014$\pm$0.001 & 1.06$\pm$0.01 & 998.87$\pm$1.23 & 0.35 & -0.88$\pm$0.012  \\ 
                 H$_2$CO (3 lines, q fit) & 2.37$\pm$0.02 & 69.69$\pm$0.6 & 1.00$\pm$0.003 & 11.13$\pm$0.01 & 353.76$\pm$0.31 & 0.001$\pm$0.001 & 0.009$\pm$0.001 & 1.39$\pm$0.01 & 997.92$\pm$2.28 & 0.001$\pm$0.001 & \nodata \\ 
\enddata
\tablecomments{The results from each model lists the molecular line(s) fit with the \textit{pdspy} models.
For H$_2$CO, `3 lines', refers to all three ($J=3_{n,n}\rightarrow2_{n,n}$) transitions, while `2 lines' refers to the higher
excitation ($J=3_{2,2}\rightarrow2_{2,1}$) and ($J=3_{2,1}\rightarrow2_{2,0}$) transitions (see Table 1). The parameter
$q$, the power-law index of the disk 
temperature profile is fixed for all models except when `q fit' is listed in the description.
The Mass column refers to the protostar mass, $\Delta x$ and $\Delta y$ refer to the offset of the
model center of mass with respect to the image phase center, $\gamma$ is the power-law index of the 
disk surface density profile, and SO Abund. refers to the SO abundance adopted or fit by the models.}
\tablenotetext{a}{Abundance of denoted molecule was allowed to vary as part of the fitting process to enable
better convergence.}
\tablenotetext{b}{Abundance provided is for the molecule in the first column that have a superscript a.}
\end{deluxetable}

%% file: tab6.tex
\begin{deluxetable}{lll}
\tabletypesize{\scriptsize}
\tablewidth{0pt}
\tablecaption{HOPS-370 Continuum Modeling Results}
\tablehead{\colhead{Parameter Description} & \colhead{Parameter}    & \colhead{Fit Value}}           
\startdata
 System Luminosity                         &    $L_{*}$ (\lsun)     &   $279.9^{+14.1}_{-13.1}$\\
 Disk Mass                                 &    $M_{disk}$ (\msun)  &    $0.035^{+0.005}_{-0.003}$\\
 Disk Outer Radius                         &    $R_{disk}$ (au)     &    $61.9^{+ 0.7}_{- 1.2}$\\
 Disk Inner Radius                         &    $R_{in,disk}$ (au)  &    $0.54^{+0.27}_{-0.09}$\\
 Envelope Mass                             &     $M_{env}$ (\msun)  &    $0.12^{+0.01}_{-0.01}$\\
 Envelope Radius                           &     $R_{env}$ (au)     &    $1881^{+78}_{-137}$\\
 Envelope Inner Radius                     &     $R_{in,env}$ (au)  &    =$R_{in,disk}$\\
 Envelope Centrifugal Radius               &     $R_{c}$ (au)       &    =$R_{disk}$\\
 Surface Density Power-law Index           &      $\gamma$          & $-0.47^{+0.02}_{-0.01}$\\
 Flaring Power-law Index                   & $\beta$        & $0.66^{+0.02}_{-0.03}$\\
 Scale Height at 1 au                      & $h_0$ (au)             &    $0.132^{+0.009}_{-0.008}$\\
 Inclination                               &           $i$          &   $74.0^{+ 0.2}_{- 0.2}$\\
 Density Scaling in Outflow Cavity         &     $f_{cav}$          &    $0.33^{+0.01}_{-0.02}$\\
 Outflow Cavity Shape power-law index      &         $\xi$          &     $1.12^{+0.06}_{-0.05}$\\
 Maximum dust grain size                   &    $a_{max}$ (\micron) &  $432^{+31}_{-47}$\\
 Dust grain size distribution power-law index &           $p$       &  $2.63^{+0.04}_{-0.06}$\\
 Position Angle East of North              &       $pa$ (\degr)     & $109.7^{+ 0.1}_{- 0.2}$\\
 Distance                                  &       d (pc)           &  400.0\\
\enddata
\tablecomments{The disk and envelope parameters derived from continuum modeling are reported here.
More explanation of the parameters is provided in Sections 4.1 and 4.2. However, we explain
a few that are less intuitive. 
The parameter f$_{cav}$ is the reduction of 
envelope density within the outflow cavity, and the outflow cavity shape and opening 
angle are defined by the parameter $\xi$, translating to opening angle can be done 
by calculating 2tan$^{-1}$($\xi$), corresponding to a full opening angle of $\sim$98\degr.
The maximum dust size and grain size distribution are defined by a$_{max}$ and p, where 
the dust grain size distribution follows a power-law n(a)$^{-p}$. Absolute flux calibration
uncertainty was not taken into account in the modeling.
}
\end{deluxetable}

%% file: tab7.tex
\begin{deluxetable}{lllllllllllllll}
\rotate
\tabletypesize{\tiny}
\tablewidth{0pt}
\tablecaption{HOPS-370 Molecular Line Models - Exponentially Tapered Disks with Envelope}
\tablehead{\colhead{Transition(s)} & \colhead{Mass} & \colhead{Disk Radius} & \colhead{Disk Mass} & \colhead{Envelope Radius} & \colhead{Envelope Mass}  & \colhead{V$_{lsr}$}     & \colhead{Pos. Angle} & \colhead{$\Delta$x} & \colhead{$\Delta$y}  & \colhead{$\gamma$} & \colhead{T(1 au)} & \colhead{q} & \colhead{$log_{10}$(Mol./H$_2$)\tablenotemark{b}}\\
                                   & \colhead{(M$_{\sun}$)} & \colhead{(AU)} & \colhead{(M$_{\sun}$)} & \colhead{(AU)} & \colhead{(M$_{\sun}$)} & \colhead{(km~s$^{-1}$)} & \colhead{\degr}      & \colhead{(\arcsec)} & \colhead{(\arcsec)} &                              &       (K)     &                 &
}
\startdata
  H$_2$CO ($J=3_{0,3}\rightarrow2_{0,2}$) & 2.81$\pm$0.03 & 32.97$\pm$3.3 & 0.77$\pm$0.070 & 8345.89$\pm$8536.0 & 0.548$\pm$0.660 & 11.19$\pm$0.02 & 353.35$\pm$0.34 & -0.000$\pm$0.002 & 0.015$\pm$0.001 & 1.32$\pm$0.02 & 997.43$\pm$3.76 & 0.35 & \nodata  \\ 
                        H$_2$CO (2 lines) & 2.46$\pm$0.03 & 36.07$\pm$4.1 & 0.85$\pm$0.075 & 1081.80$\pm$182.1 & 0.049$\pm$0.010 & 11.21$\pm$0.02 & 352.00$\pm$0.44 & 0.011$\pm$0.002 & 0.004$\pm$0.002 & 1.37$\pm$0.03 & 998.69$\pm$1.66 & 0.35 & \nodata  \\ 
                        H$_2$CO (3 lines) & 2.66$\pm$0.02 & 62.99$\pm$3.1 & 0.69$\pm$0.034 & 1076.49$\pm$21.0 & 0.071$\pm$0.004 & 11.20$\pm$0.01 & 352.39$\pm$0.28 & 0.006$\pm$0.001 & 0.009$\pm$0.001 & 1.16$\pm$0.02 & 999.52$\pm$0.66 & 0.35 & \nodata  \\ 
 CH$_3$OH ($J=4_{2,2}\rightarrow3_{1,2}$) & 2.46$\pm$0.03 & 81.51$\pm$5.6 & 0.51$\pm$0.035 & 1009.67$\pm$20.1 & 0.037$\pm$0.004 & 11.07$\pm$0.02 & 357.55$\pm$0.54 & 0.012$\pm$0.002 & -0.001$\pm$0.002 & 1.08$\pm$0.04 & 998.72$\pm$1.91 & 0.35 & \nodata  \\ 
 CH$_3$OH\tablenotemark{a}, H$_2$CO (3 lines) & 2.60$\pm$0.02 & 74.65$\pm$2.5 & 0.57$\pm$0.019 & 1013.96$\pm$29.7 & 0.054$\pm$0.002 & 11.14$\pm$0.01 & 353.58$\pm$0.22 & 0.007$\pm$0.001 & 0.008$\pm$0.001 & 1.09$\pm$0.02 & 999.09$\pm$0.85 & 0.35 & -7.99$\pm$0.005  \\ 
           SO ($J_N = 6_5\rightarrow5_4$) & 2.63$\pm$0.04 & 90.16$\pm$4.1 & 0.23$\pm$0.011 & 1013.35$\pm$15.2 & 0.013$\pm$0.001 & 11.53$\pm$0.02 & 344.85$\pm$0.61 & 0.024$\pm$0.002 & 0.036$\pm$0.002 & 1.00$\pm$0.05 & 999.36$\pm$0.73 & 0.35 & \nodata  \\ 
         SO\tablenotemark{a} and CH$_3$OH & 2.35$\pm$0.04 & 113.69$\pm$2.9 & 0.30$\pm$0.012 & 1067.64$\pm$35.0 & 0.023$\pm$0.003 & 11.12$\pm$0.03 & 352.97$\pm$0.65 & 0.013$\pm$0.001 & 0.026$\pm$0.002 & 0.65$\pm$0.04 & 993.91$\pm$3.08 & 0.35 & -8.70$\pm$0.007  \\ 
   SO\tablenotemark{a}, H$_2$CO (3 lines) & 2.63$\pm$0.02 & 85.17$\pm$2.4 & 0.50$\pm$0.015 & 5466.58$\pm$202.3 & 0.320$\pm$0.026 & 11.19$\pm$0.01 & 351.38$\pm$0.23 & 0.009$\pm$0.001 & 0.015$\pm$0.001 & 1.02$\pm$0.02 & 999.23$\pm$0.88 & 0.35 & -8.76$\pm$0.005  \\ 
                                       NS & 3.83$\pm$0.15 & 55.96$\pm$26.9 & 0.03$\pm$0.018 & 5889.31$\pm$3114.6 & 0.046$\pm$0.067 & 11.01$\pm$0.07 & 350.24$\pm$1.52 & -0.010$\pm$0.004 & 0.013$\pm$0.003 & 0.55$\pm$0.35 & 424.31$\pm$38.63 & 0.35 & \nodata  \\ 
                  NS, SO\tablenotemark{a} & 1.85$\pm$0.03 & 14.17$\pm$2.2 & 0.06$\pm$0.004 & 4671.31$\pm$117.8 & 0.006$\pm$0.002 & 11.10$\pm$0.03 & 353.53$\pm$0.55 & 0.018$\pm$0.002 & 0.060$\pm$0.002 & 1.57$\pm$0.02 & 998.54$\pm$1.37 & 0.35 & -7.50$\pm$0.002  \\ 
            NS\tablenotemark{a}, CH$_3$OH & 2.52$\pm$0.04 & 88.72$\pm$6.8 & 0.46$\pm$0.035 & 1074.55$\pm$180.6 & 0.029$\pm$0.007 & 11.07$\pm$0.03 & 356.83$\pm$0.66 & 0.009$\pm$0.003 & 0.004$\pm$0.002 & 1.04$\pm$0.05 & 991.41$\pm$4.22 & 0.35 & -9.77$\pm$0.018  \\ 
   NS\tablenotemark{a}, H$_2$CO (3 lines) & 2.67$\pm$0.02 & 78.74$\pm$2.7 & 0.55$\pm$0.021 & 5475.02$\pm$481.8 & 0.532$\pm$0.066 & 11.17$\pm$0.01 & 352.43$\pm$0.31 & 0.006$\pm$0.001 & 0.010$\pm$0.001 & 1.07$\pm$0.02 & 998.17$\pm$1.80 & 0.35 & -9.82$\pm$0.018  \\ 
                 H$_2$CO (3 lines, q fit) & 2.57$\pm$0.02 & 52.98$\pm$0.9 & 0.99$\pm$0.011 & 1054.51$\pm$15.0 & 0.18$\pm$0.005 & 11.33$\pm$0.01 & 352.71$\pm$0.26 & 0.004$\pm$0.001 & 0.003$\pm$0.001 & 1.27$\pm$0.01 & 998.85$\pm$1.46 & 0.0003$\pm$0.0005 & \nodata  \\ 
\enddata
\tablecomments{The results from each model lists the molecular line(s) fit with the \textit{pdspy} models.
For H$_2$CO, `3 lines', refers to all three ($J=3_{n,n}\rightarrow2_{n,n}$) transitions, while `2 lines' refers to the higher
excitation ($J=3_{2,2}\rightarrow2_{2,1}$) and ($J=3_{2,1}\rightarrow2_{2,0}$) transitions (see Table 1). The parameter
$q$, the power-law index of the disk 
temperature profile is fixed for all models except when `q fit' is listed in the description.
The Mass column refers to the protostar mass, $\Delta x$ and $\Delta y$ refer to the offset of the
model center of mass with respect to the image phase center, $\gamma$ is the power-law index of the 
disk surface density profile, and SO Abund. refers to the SO abundance adopted or fit by the models.}
\tablenotetext{a}{Abundance of denoted molecule was allowed to vary as part of the fitting process to enable
better convergence.}
\tablenotetext{b}{Abundance provided is for the molecule in the first column that have a superscript a.}
\end{deluxetable}